\documentclass[fleqn,usenatbib]{mnras}
\usepackage{newtxtext,newtxmath}

\usepackage[T1]{fontenc}
\usepackage{ae,aecompl}
\usepackage{macros}
\usepackage{pdflscape}

\usepackage{geometry}
\usepackage{fancyhdr}
\usepackage{amsmath,amssymb}
\usepackage{upgreek}
\usepackage{graphicx} 
\usepackage[caption=false]{subfig}
\usepackage{wasysym}
\usepackage{lastpage}
\usepackage{etoolbox}
\usepackage{threeparttable}
\usepackage{rotating}
\usepackage{hyperref}

\graphicspath{{./}{Figures/}}

\title[Prograde metal-poor stars]{The origin of metal-poor stars on prograde disk orbits in FIRE simulations of Milky Way-mass galaxies}

\author[Santistevan et al.]{
Isaiah B. Santistevan$^{1}$\thanks{E-mail: ibsantistevan@ucdavis.edu},
Andrew Wetzel$^{1}$,
Robyn E. Sanderson$^{2,3}$,
Kareem El-Badry$^{4}$,
\newauthor
Jenna Samuel$^{1}$,
Claude-Andr{\'e} Faucher-Gigu{\`e}re$^{5}$
\\
$^{1}$Department of Physics \& Astronomy, University of California, Davis, CA 95616, USA\\
$^{2}$Department of Physics \& Astronomy, University of Pennsylvania, Philadelphia, PA 19104, USA\\
$^{3}$Center for Computational Astrophysics, Flatiron Institute, New York, NY 10010, USA\\
$^{4}$Department of Astronomy, Theoretical Astrophysics Center, University of California Berkeley, Berkeley, CA 94720, USA\\
$^{5}${Department of Physics and Astronomy and CIERA, Northwestern University, 2145 Sheridan Road, Evanston, IL 60208, USA}\\
}

\date{Accepted XXX. Received YYY; in original form ZZZ}

\pubyear{2020}

\begin{document}
\label{firstpage}
\pagerange{\pageref{firstpage}--\pageref{lastpage}}
\maketitle

\begin{abstract}
In hierarchical structure formation, metal-poor stars in and around the Milky Way (MW) originate primarily from mergers of lower-mass galaxies.
A common expectation is therefore that metal-poor stars should have isotropic, dispersion-dominated orbits that do not correlate strongly with the MW disk.
However, recent observations of stars in the MW show that metal-poor ($\FeH\lesssim-2$) stars are preferentially on prograde orbits with respect to the disk.
Using the FIRE-2 suite of cosmological zoom-in simulations of MW/M31-mass galaxies, we investigate the prevalence and origin of prograde metal-poor stars.
Almost all (11 of 12) of our simulations have metal-poor stars on preferentially prograde orbits today and throughout most of their history: we thus predict that this is a generic feature of MW/M31-mass galaxies.
The typical prograde-to-retrograde ratio is $\sim2:1$, which depends weakly on stellar metallicity at $\FeH\lesssim-1$.
These trends predicted by our simulations agree well with MW observations.
Prograde metal-poor stars originate largely from a single LMC/SMC-mass gas-rich merger $7-12.5\Gyr$ ago, which deposited existing metal-poor stars and significant gas on an orbital vector that sparked the formation of and/or shaped the orientation of a long-lived stellar disk, giving rise to a prograde bias for all low-metallicity stars.
We find sub-dominant contributions from in-situ stars formed in the host galaxy before this merger, and in some cases, additional massive mergers.
We find few clear correlations between any properties of our MW/M31-mass galaxies at $z=0$ and the degree of this prograde bias as a result of diverse merger scenarios.
\end{abstract}



\section{Introduction} %
\label{sec:intro}      %

The populations of stars in different regions of a galaxy, such as the disk, the bulge, and the halo, can reveal its formation history.
In the standard picture of galaxy formation, unenriched gas collapses in dark matter (DM) halos to form stars and galaxies \citep[e.g.][]{Ostriker75, Rees77, White78, Fall80}.
Many works have studied the hierarchical nature of galaxy formation, such as determining the ages of globular clusters throughout the MW's stellar halo, and the formation of disk galaxies within their DM halos \citep[e.g.][]{Searle78, White91, Mo98}.
\citet{Eggen62} first suggested that old, metal-poor stars orbit the MW in elliptical orbits compared to metal-rich stars, which move in more circular orbits, a picture that broadly holds today \citep[e.g.][]{Chiba00}.
Many authors have explored how metal-poor stars were deposited into the stellar halo of the MW by mergers of low-mass dwarf galaxies \citep[e.g.][]{Bullock01, Bullock05, Ibata94, Newberg03, Nissen10}.
We now understand that most of the mass in the stellar halo came from a few massive mergers, which is a natural consequence of the steepness of the stellar mass halo mass relation.
\citet{Deason16} showed that only $1 - 2$ satellite galaxies with $M_{\rm star} \sim 10^8 - 10^{10} \Msun$ deposited a majority of the stars in the halo, and dwarf galaxies with $M_{\rm star} \sim 10^5 - 10^8 \Msun$ contributed the bulk of metal-poor ($\FeH \lesssim -2$) stars.
A recent observational study by \citet{Naidu20} suggests that at distances greater than $\sim 15 \kpc$ from the galactic center, $\sim 80$ per cent of halo stars came from 2 mergers: Gaia-Enceladus \citep{Helmi18, Belokurov18} and Sagittarius \citep{Ibata94, Newberg03, Majewski03}.

Many studies focused on the properties and origins of these old and/or metal-poor stars, such as their spatial and elemental distributions within the MW, the masses of the accreted dwarf galaxies they came from, and the formation of the oldest Population III stars \citep[e.g.][]{Scannapieco06, Brook07, Deason16, Griffen18}.
Spectroscopic surveys that have observed stars with $\FeH \lesssim -2$, such as RAVE, GALAH, LAMOST, and APOGEE, provide the elemental abundances and ages of stars in different regions of the MW and now allow for an investigation into the early formation period \citep{RAVE, GALAH, LAMOST, APOGEE}.
The Pristine survey also has pushed metallicity measurements down to $\FeH < -3$, with the potential of reaching the `ultra metal-poor' ($\FeH < -4$) population \citep{PRISTINE}.
Finally, the recent Hectochelle in the Halo at High Resolution (H3) Survey is targeting stars outside of the disk and bulge regions of the MW, and is reaching metallicities down to $\FeH \sim -3$ \citep{H3SURVEY}.

Recent simulation work on the spatial distributions of old, metal-poor stars within a MW-mass galaxy suggests that they are less centrally concentrated than younger stars.
For example, \citet{Scannapieco06} used DM-only simulations of MW-mass halos and concluded that both unenriched stars, and their descendants, occupy a wide range of radii throughout the MW.
In a similar study, \citet{Brook07} used cosmological baryonic simulations of 4 MW-mass galaxies and found that metal-free stars are distributed throughout the MW's halo, though the authors suggest that the oldest stars are more centrally concentrated.
Using a sample of old ($z_{\rm form} > 5$), metal-poor ($\FeH < -2$) stars in 3 of the FIRE-2 simulations that we examine here, \citet{ElBadry18} proposed that this spatially extended distribution of metal-poor stars in a MW-mass galaxy results from a combination of kinematic heating of stars to larger orbits in the host galaxy from a time-varying galactic potential and from early mergers that deposited these stars on radial orbits.
However, they did not examine any preference for metal-poor stars to be on prograde orbits aligned with the disk.
Similarly, using the APOSTLE simulations, \citet{Starkenburg17} found that the fraction of stars with $\FeH < -2.5$, older than $\gtrsim 13 \Gyr$, increases with increasing distance from the center of the galaxy.
Although these studies agree that the number density of old stars is largest near the central bulge region, the overabundance of younger stars can make finding older stars difficult.

The origin of the MW's stellar halo and its disk are likely connected, particularly for the inner stellar halo and the thick disk.
\citet{Gallart19} showed that stellar populations in the inner halo and thick disk partially overlap in kinematic space.
The thick disk extends $\gtrsim 1 \kpc$ from the midplane and is comprised of primarily older stars across a wide range of iron abundances \citep[$-2 \lesssim \FeH \lesssim -0.5$; e.g.][]{Gilmore83, Freeman02}.
Several authors have proposed different formation mechanisms for the thick disk such as the kinematic heating of thin disk stars to larger vertical or radial distances, and the disruption and accretion of satellite galaxies \citep[e.g.][]{Walker96, Villalobos08, Loebman11, Haywood18}
However, many works now show that the disk formed thick and settled over time.
\citet{Bird13} used the cosmological MW-mass simulation `Eris' \citep{ERIS} and concluded that the MW's disk formed `upside-down', where older stars formed farther from the midplane with larger vertical velocity dispersions compared to younger stars, which formed in the thin disk with smaller velocity dispersions.
More recent cosmological simulations strongly support this idea \citep[e.g.][]{Grand16, Bird21}, including \citet{Ma17}, who analyzed one of the simulated galaxies in our sample (m12i).
The authors concluded that older stars in the thick disk formed during the early, bursty phase of the MW when the Galaxy was gas-rich, while younger stars formed during the more calm period of MW growth and that the disk becomes thinner over time.

A great deal of work is dedicated to understanding the existence of net rotation in the stellar halo, and its origins.
In an early study, \citet{Norris89} found that, in a population of $\sim 500$ halo stars, the majority showed net retrograde motion.
We now know that the halo is comprised of two components, an inner and outer component, each with different elemental abundances and kinematics \citep{Zinn93, Carney96, Bonaca17}.
\citet{Deason11} found that metal-rich stars in the halo ($\FeH > -2$) favored prograde motion and were likely accreted from a satellite, while metal-poor stars ($\FeH < -2$) were more retrograde on average and could have been part of the early-forming halo.
They also noted that many globular clusters in the MW halo orbit in the prograde direction.
Another study by \citet{Bonaca17} concluded that metal-rich stars ($\FeH > -1$) near the MW have prograde orbits, while metal-poor stars have no average orbit direction.
This suggests the metal-rich halo stars formed in-situ in the MW galaxy, while the metal-poor stars likely were accreted.
The authors found similar features in one of the simulated galaxies in our sample, m12i.
Finally, recent work focused on kinematically, or elementally, coherent structures throughout the MW suggest that many previous mergers currently comprise most of the retrograde population in the MW halo \citep{Myeong19, Naidu20}.

Recently, \citet{Sestito19} analyzed all known MW stars with $\FeH < -4$ and concluded that 11 of the 42 stars (26 per cent of their sample) have orbits that are confined to within $\pm 3 \kpc$ of the disk mid-plane.
Of these stars on disk-like orbits, 10 are prograde ($J_{\phi} > 0$), and only one shows retrograde motion ($J_{\phi} < 0$).
They determined that the remaining 31 metal-poor stars in their sample have halo-like orbits, bringing them far away from the disk, which is typical for accretion-based origin.
Subsequently, \citet{Sestito20} selected stars with $\FeH \lesssim -2$ from both the Pristine survey \citep{PRISTINE} and LAMOST DR3 \citep{Li18}, increasing their sample size from 42 to 1027 (a factor of $\sim24$).
They again found that stars in the prograde region of action space outnumber those in the retrograde region, and $\sim 31$ per cent of the sample have prograde orbits that are confined to $\pm 3 \kpc$ of the disk mid-plane.
New observational studies continue to support the results from \citet{Sestito19, Sestito20}, particularly the discovery of metal-poor stars near the plane of the MW \citep[e.g.][]{Venn20,DiMatteo20}.
\citet{Sestito19} posit 3 possible explanations as to the origins of these metal-poor, disk-like stars: (i) they formed in-situ within the MW disk itself and remained on prograde, near-circular orbits; (ii) another galaxy merged into the MW on an orbit that was aligned with the (pre-existing) MW disk, depositing its stars on prograde orbits; or (iii) these stars formed in one or more progenitor galaxies that merged into the MW as it was forming.

Our goal in this paper is to test the origin and frequency of metal-poor stars on prograde disk orbits, using 12 MW/M31-mass galaxies from the FIRE-2 suite of cosmological zoom-in simulations.
The main questions that we address are:
\renewcommand{\labelenumi}{\roman{enumi})}
\begin{enumerate}
\item How commonly do MW/M31-mass galaxies show a preference for prograde motions in their metal-poor stars, as observed in the MW, and what is the range in strength of this prograde bias?

\item How does this prograde bias depend on the metallicity, distance, and/or age of the stars?

\item What process(es) cause this prograde bias among metal-poor stars?

\item Do any properties of the MW/M31-mass galaxy or their galaxy mergers correlate with this prograde bias?
\end{enumerate}


\section{Methods}   %
\label{sec:methods} %

\subsection{FIRE-2 Simulations} %
\label{sec:sims}                %

\begin{table*}
\centering
\begin{threeparttable}
\caption{
	Properties of the 12 MW/M31-mass galaxies in the FIRE-2 simulation suite that we analyze, ordered by decreasing prograde-to-retrograde ratio for metal-poor stars.
	Simulations with `m12' names are isolated hosts from the Latte suite, while the others are from the ELVIS on FIRE suite of LG-like galaxies.
	We measure all masses and mass fractions at $z = 0$, using only metal-poor ($\FeH < -2.5$) stars (except for the total stellar mass of the galaxy, which uses all stars).
	Columns: name; 
	$M_{\rm star, 90}$ is the host's stellar mass within $R_{\rm star,90}$, the disk radius enclosing 90 per cent of the stellar mass within 20 kpc;  
	$M_{\rm star,pro} / M_{\rm star,ret}$ is the prograde bias, defined as the mass ratio of prograde to retrograde stars;
	$t_{\rm lb,merger,1}$ is the lookback time of the merger that contributed the most prograde stars; 
	$f_{\rm merger,1}$ is the fraction of all prograde stars that came from the primary (most contributing) merger; 
	$f_{\rm merger, \ top \ 3}$ is the fraction of prograde stars from the 3 most contributing mergers; $f_{\rm in-situ}$ is the fraction of prograde stars that formed in-situ (within 15 kpc);
	$f_{\rm merger,1} + f_{\rm in-situ}$ is the sum;
	$f_{\rm merger,\ top \ 3} + f_{\rm in-situ}$ is the sum of fractions from the 3 most contributing mergers and the fraction that formed in-situ.
	}
	\begin{tabular}{|c|c|c|c|c|c|c|c|c|}
		\hline
		\hline
		Name & $M_{\rm star,90}$ & $M_{\rm star,pro} / M_{\rm star,ret}$ & $t_{\rm lb,merger,1}$ & $f_{\rm merger,1}$ & $f_{\rm merger, \ top\ 3}$ & $f_{\rm in-situ}$ & $f_{\rm in-situ} + f_{\rm merger,1}$ & $f_{\rm merger,\ top \ 3} + f_{\rm in-situ}$ \\
		 & [$10^{10} \Msun$] & & [Gyr] & & & & & \\
		\hline
        m12m$^{\rm a}$ & 10.0 & 9.14 & 9.56 & 0.44 & 0.74 & 0.21 & 0.65 & 0.95 \\
        m12w$^{\rm b}$ & 4.8 & 3.02 & 8.17 & 0.20 & 0.50 & 0.13 & 0.32 & 0.63 \\
        m12c$^{\rm c}$ & 5.1 & 2.67 & 8.99 & 0.48 & 0.75 & 0.21 & 0.70 & 0.96 \\
        Romeo$^{\rm c}$ & 5.9 & 2.35 & 12.52 & 0.24 & 0.38 & 0.19 & 0.43 & 0.57 \\
        m12b$^{\rm c}$ & 7.3 & 2.29 & 2.84 & 0.27 & 0.59 & 0.06 & 0.34 & 0.65 \\
        Juliet$^{\rm c}$ & 3.3 & 2.01 & 12.25 & 0.14 & 0.28 & 0.15 & 0.30 & 0.43 \\
        Thelma$^{\rm c}$ & 6.3 & 1.96 & 8.61 & 0.28 & 0.57 & 0.10 & 0.38 & 0.67 \\
        Romulus$^{\rm d}$ & 8.0 & 1.56 & 7.60 & 0.53 & 0.77 & 0.05 & 0.58 & 0.82 \\
        m12f$^{\rm e}$ & 6.9 & 1.48 & 11.91 & 0.20 & 0.35 & 0.16 & 0.36 & 0.51 \\
        Remus$^{\rm d}$ & 4.0 & 1.42 & 12.02 & 0.23 & 0.45 & 0.10 & 0.34 & 0.55 \\
        Louise$^{\rm c}$ & 2.3 & 1.28 & 10.75 & 0.15 & 0.34 & 0.31 & 0.46 & 0.65 \\
        m12i$^{\rm f}$ & 5.5 & 0.98 & 10.50 & 0.24 & 0.57 & 0.24 & 0.48 & 0.81 \\
		\hline
		\hline
\end{tabular}
\label{tab:hosts}
\begin{tablenotes}
\item \textit{Note:} Simulation introduced in: a: \citet{Hopkins18}, b: \citet{Samuel20}, c: \citet{GarrisonKimmel19a}, d: \citet{GarrisonKimmel19b}, e: \citet{GarrisonKimmel17}, f: \citet{Wetzel16}.
\end{tablenotes}
\end{threeparttable}
\end{table*}

We use cosmological zoom-in baryonic simulations of MW/M31-mass galaxies from the Feedback In Realistic Environments (FIRE) project\footnote{FIRE project web site: \url{http://fire.northwestern.edu}} \citep{Hopkins18}.
We ran all of the simulations using the same $N$-body gravitational plus hydrodynamics code \textsc{Gizmo} \citep{Hopkins15}, with the mesh-free finite-mass (MFM) hydrodynamics method and the FIRE-2 physics model \citep{Hopkins18}.
Across a temperature range of $10 - 10^{10} \K$, the FIRE-2 model includes several radiative cooling and heating processes for gas such as free-free emission, photoionization/recombination, Compton scattering, photoelectric, metal-line, molecular, fine-structure, dust-collisional, and cosmic-ray heating, including the spatially uniform, redshift-dependent cosmic UV background from \cite{FaucherGiguere09}, where HI reionization occurs at $z_{\rm reion} \sim 10$.
The simulations self-consistently generate and track 11 elemental abundances (H, He, C, N, O, Ne, Mg, Si, S, Ca, Fe), including the sub-grid diffusion/mixing of these elements in gas via turbulence \citep{Hopkins16, Su17, Escala18}.

Because these FIRE simulations do not model the initial metal enrichment from Pop III stars, we initialize gas elements in the simulations with a metallicity floor of $\FeH = -4$.
We choose this metallicity floor as a rough model for the level of enrichment we expect from Population III stars, which also corresponds to the lowest metallicity in one of the observational studies we compare our results to, \citet{Sestito19}.
As we will show, the strength of the prograde bias does not depend on metallicity at low metallicities, so the choice of the metallicity floor is unimportant.
Stars form in gas that is self-gravitating, Jeans unstable, molecular \citep[following][]{Krumholz11}, and dense ($n_{\rm H}$ > 1000 cm$^{-3}$).
A star particle inherits the mass and elemental abundances from its progenitor gas particle, and represents a single stellar population, assuming a \cite{Kroupa01} initial mass function.
It then evolves along stellar population models from \textsc{STARBURST99 v7.0} \citep{Leitherer99}.
Furthermore, the FIRE-2 simulations include the following feedback processes: core-collapse and Ia supernovae, stellar winds, radiation pressure, photoionization, and photo-electric heating.

We generated the cosmological zoom-in initial conditions for each simulation embedded within periodic cosmological boxes of lengths $70.4 - 172 \Mpc$ using the code \textsc{MUSIC} \citep{Hahn11}, at $z \approx 99$.
Each simulation saves of 600 snapshots down to $z = 0$, spaced every $\approx 25 \Myr$, and all assume flat $\Lambda$CDM cosmology with the following range in cosmological parameters, consistent with \citet{Planck18}: $\Omega_{\rm m} = 0.266 - 0.31$, $\Omega_{\rm b} = 0.0449 - 0.048$, $\sigma_{\rm 8} = 0.801 - 0.82$, $n_{\rm s} = 0.961 - 0.97$, and $h = 0.68 - 0.71$.

We analyze the same 12 MW/M31-mass galaxies in \citet{Santistevan20}, and Table~\ref{tab:hosts} lists their stellar masses at $z = 0$.
Half of our sample comes from the Latte suite of isolated MW/M31-mass galaxies first introduced in \cite{Wetzel16}.
The Latte suite has gas and initial star particle masses of $7100 \Msun$, but because of stellar mass loss, the average star particle mass is $\approx 5000 \Msun$.
The mass of DM particles in the zoom-in region is $3.5 \times 10^4 \Msun$.
The gravitational softening lengths for star and DM particles are fixed at 4 and 40 pc (Plummer equivalent), comoving at $z > 9$ and physical thereafter, while gas elements use adaptive force softening, equal to their hydrodynamic smoothing, which adapts down to 1 pc.
We chose the halos in the Latte suite to have $M_{\rm 200m} = 1 - 2 \times 10^{12} \Msun$ and have no other similar-mass halos within $5 \times R_{\rm 200m}$, but we imposed no additional selection criteria beyond this.

The other half of our sample comes from the `ELVIS on FIRE' suite of LG-like MW+M31 pairs, introduced in \citet{GarrisonKimmel19b, GarrisonKimmel19a}, which have approximately $2 \times$ better mass resolution, with gas and initial star particle masses of $3500 - 4000 \Msun$.
Each pair of halos was chosen based on their masses (each with $M_{\rm 200m} = 1 - 3 \times 10^{12} \Msun$ and total LG mass between $2 - 5 \times 10^{12} \Msun$), current separation ($600 - 1000\kpc$) and radial velocities at $z = 0$ ($\rm \upsilon_{rad} < 0$), and that they have no other massive halos within $2.8 \Mpc$ of either host center.

We selected all halos without prior knowledge of their formation/merger histories or satellite populations, except m12w, which we selected to have an LMC-mass satellite analog at $z \sim 0$ \citep{Samuel20}.
Therefore, our galaxy sample should reflect random/typical formation histories.
The host galaxies in our sample reflect a class of morphologies \citep{GarrisonKimmel18, ElBadry18_gas}, which host stellar-to-halo mass relations \citep{Hopkins18}, disk morphologies and metallicity gradients \citep{Ma17, Sanderson18_gaia}, and stellar halos \citep{Bonaca17, Sanderson18_halo} akin to that of the MW and M31.
Furthermore, each host galaxy possesses realistic satellite galaxy populations, with stellar masses and velocity dispersions \citep{Wetzel16, GarrisonKimmel19b}, radial and 3-D distributions \citep{Samuel20, Samuel20_Planes}, and star-formation histories \citep{GarrisonKimmel19a} that broadly agree with observations of the LG and the local Universe.

\subsection{ROCKSTAR Halo Catalogs \& Merger Trees} %
\label{sec:rockstar}  %

We generate (sub)halo catalogs using only DM particles, at all 600 snapshots, using the \textsc{ROCKSTAR} 6-D halo finder \citep{Behroozi13a}.
To generate merger trees, we use \textsc{CONSISTENT-TREES} \citep{Behroozi13b}.
All (sub)halos that we examine have zero contamination by low-resolution DM particles, because of the large zoom-in region that we generate for each host.

We assign star particles to (sub)halos in post-processing, which we outline below, but refer the reader to \citet{Samuel20} for more details.
We identify star particles with positions within $0.8 \times R_{\rm halo}$ (out to a maximum distance of $30 \kpc$) and velocities within $2 \times V_{\rm circ,max}$ of a (sub)halo's center-of-mass velocity.
Following these initial criteria, we keep star particles if they are within $1.5 \times R_{\rm star,90}$ (the radius enclosing 90 per cent of the stellar mass) of the current member star particle's center-of-mass and halo center position.
This guarantees that the centers-of-mass of both the galaxy and (sub)halo are consistent with one another.
We then keep star particles within $2 \times \sigma_{\rm vel,star}$ (the velocity dispersion of current member star particles) of the center-of-mass velocity of member star particles, and we iterate on both criteria until the (sub)halo's stellar mass converges to within 1 per cent.

We use two publicly available analysis packages: \textsc{HaloAnalysis}\footnote{\url{https://bitbucket.org/awetzel/halo\_analysis}} \citep{HaloAnalysis} for assigning star particles to halos and for reading and analyzing halo catalogs/trees, and  \textsc{GizmoAnalysis}\footnote{\url{https://bitbucket.org/awetzel/gizmo\_analysis}} \citep{GizmoAnalysis} for reading and analyzing particles from Gizmo snapshots.

\subsection{Selecting \& Tracking Metal-poor Stars} %
\label{sec:sample}                                 %

For each MW/M31-mass host galaxy, we define the `prograde bias' as the ratio of the total stellar mass of prograde stars to that of retrograde stars, $M_{\rm star,pro} / M_{\rm star,ret}$.
To measure the prograde bias, we first select star particles in the simulations at $z = 0$ based on the following two criteria, motivated by \cite{Sestito20}: (1) within $\pm 3 \kpc$ vertically of the disk midplane and within $4 - 12 \kpc$ radially from the center of the galaxy, which ensures that the sample is not contaminated significantly from bulge stars; (2)  with iron abundance $\FeH < -2.5$.
These conditions define our fiducial sample, however, we explore how our results vary with $\FeH$, and with how we spatially and kinematically select star particles in Section~\ref{sec:metals}.
We then define prograde or retrograde motion based on the star particle's action variables, $J_{\phi}$ and $J_z$.
$J_{\phi}$ is equal to the z-component of angular momentum, $L_z$.
We use the approximation $J_z \approx |z \, \upsilon_z|$, where $z$ is a star particle's vertical position and $\upsilon_z$ is its vertical velocity, relative to the host's disk.
Our choice is motivated by the definition for specific action (per unit mass),
\begin{equation}
J_z \equiv \oint \upsilon_z dz
\end{equation}
where the integral is over one orbit in z.
In the epicyclic approximation ($J_z \ll |J_{\phi}|$),
\begin{equation}
    J_z \approx \frac{|\upsilon_z| z_{\rm max}}{2}.
\end{equation}
Because $z$ follows simple harmonic motion,  $\langle |z| \rangle = z_{\rm max} / 2$, where $\langle |z| \rangle$ is the time-averaged height.
Therefore, for a population of orbits described by the epicyclic approximation, trading time-averaging for spatial-averaging, we get $J_z \approx |v_z \, z|$ for the population.
Even though this approximation is good only for stars near the plane of the disk on near-circular orbits, our geometric selection of stars avoids stars that orbit more than 3 kpc above or below the disk at $R = 8 \kpc$.
As we show in Section 3.1 the strength of the prograde bias does not change much for the different kinematic selections that we investigate.

\begin{figure}
\centering
\begin{tabular}{c}
\includegraphics[width = 0.95 \linewidth]{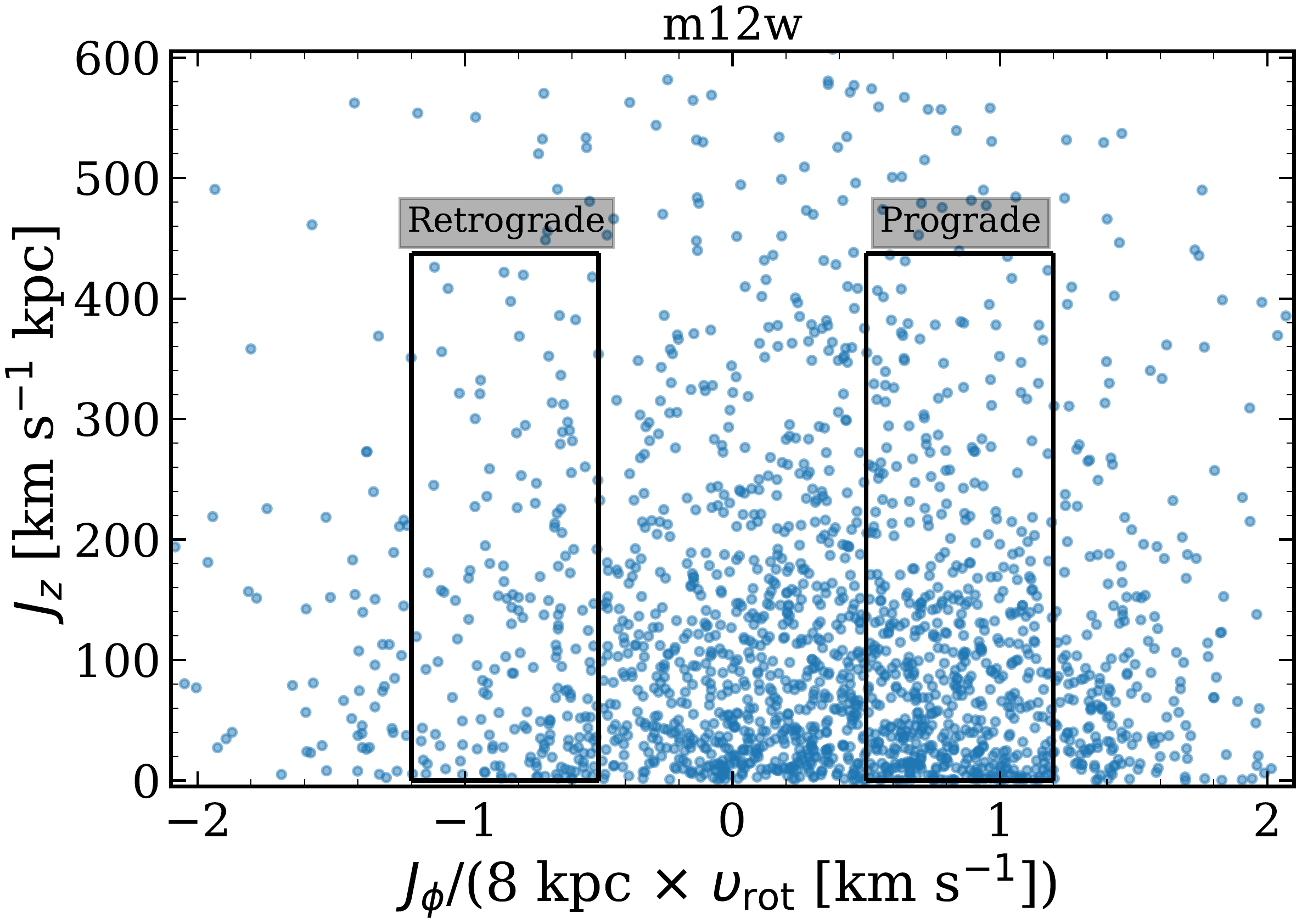}
\end{tabular}
\vspace{-2 mm}
\caption{
Metal-poor ($\FeH < -2.5$) stars in the disk ($\rm 4\leq R \leq 12 \kpc, |Z| \leq 3 \kpc$) of the m12w simulation, in action space.
Blue dots show individual star particles, and the right and left boxes show regions in action space that we select for prograde and retrograde orbits, respectively.
We adopt a similar selection region as in \citet{Sestito20} for the Pristine survey, which focused on metal-poor stars ($\FeH < -2.5$) in the MW.
We normalize $J_{\phi}$ by what `solar' values would be in the simulations, where $\upsilon_{\rm rot}$ is the rotational velocity of the galaxy at $8 \kpc$.
Metal-poor stars in m12w have a strong preference for prograde orbits, with $M_{\rm star,pro} / M_{\rm star,ret} \approx 3$.
}
\label{fig:selection}
\end{figure}

\citet{Sestito20} defined prograde stars with $0.5 < J_{\phi} / J_{\phi\odot} < 1.2$ and $J_z / J_{z\odot} < 1250$, and retrograde stars with $-1.2 < J_{\phi} / J_{\phi\odot} < -0.5$ and $J_z / J_{z\odot} < 1250$, where the solar values are $J_{\phi\odot} = 2009 \kmsi \kpc$ and $J_{z\odot} = 0.35 \kmsi \kpc$.
Our simulated MW/M31-mass galaxies have a range of rotational velocities ($\upsilon_{\rm rot}\sim 130-230\kmsi$), so we normalize $J_{\phi}$ by $8 \times \upsilon_{\rm rot} \kmsi \kpc$, where $\upsilon_{\rm rot}$ is the rotational velocity of the host galaxy, for stars at a similar galactocentric distance as the Sun.
Given that we approximate the $J_z$ action variable, we use the same MW values to normalize $J_z$ as in \citet{Sestito20}, and keep stars with $|J_z| < 438 \kmsi \kpc$.
As we will show below, our qualitative results do not depend on the details of our selection window, if for example, we select all prograde ($J_{\phi}>0$) and all retrograde ($J_{\phi}<0$) stars, or if we vary the $J_z$ cut between $\sim1/2 - 2\times$ our fiducial value of $|J_z| < 438 \kmsi \kpc$.

Figure~\ref{fig:selection} shows the action space coordinates of all metal-poor ($\FeH < -2.5$) star particles within our selection region, for one of our simulated galaxies, m12w.
The black rectangles show our fiducial selection windows to define prograde and retrograde stars.
An overabundance of metal-poor stars lie in the prograde region (right box) compared to the retrograde region (left box), similar to the MW results in \citet{Sestito20}.

We track these stars, which we select at $z = 0$, back in time to track their origin, specifically, to determine whether they formed (a) in-situ, within $15 \kpc$ physical of the most massive progenitor of the host or (b) in another galaxy that merged into the host.
At early times, this distance may encompass other nearby low-mass progenitor galaxies, so we tested how the in-situ fractions changed using stars within 10 and 20 kpc, and we found that they varied by approximately $\pm 4$ per cent on average, therefore, our results do not qualitatively change based on this selection.
To assign membership to another progenitor galaxy, we require that a star particle must have been a member of that galaxy for at least 3 consecutive snapshots.
We also tested requiring 2, 4, or 5 consecutive snapshots, and we found no significant differences in our results.


\section{Results}   %
\label{sec:results} %

\subsection{How does the prograde bias depend on stellar metallicity, distance, \& age?} %
\label{sec:metals}                                       %

\subsubsection{Comparison with the Pristine survey}
\label{sec:pristine}

\begin{figure}
\centering
\begin{tabular}{c}
\includegraphics[width=0.94\linewidth]{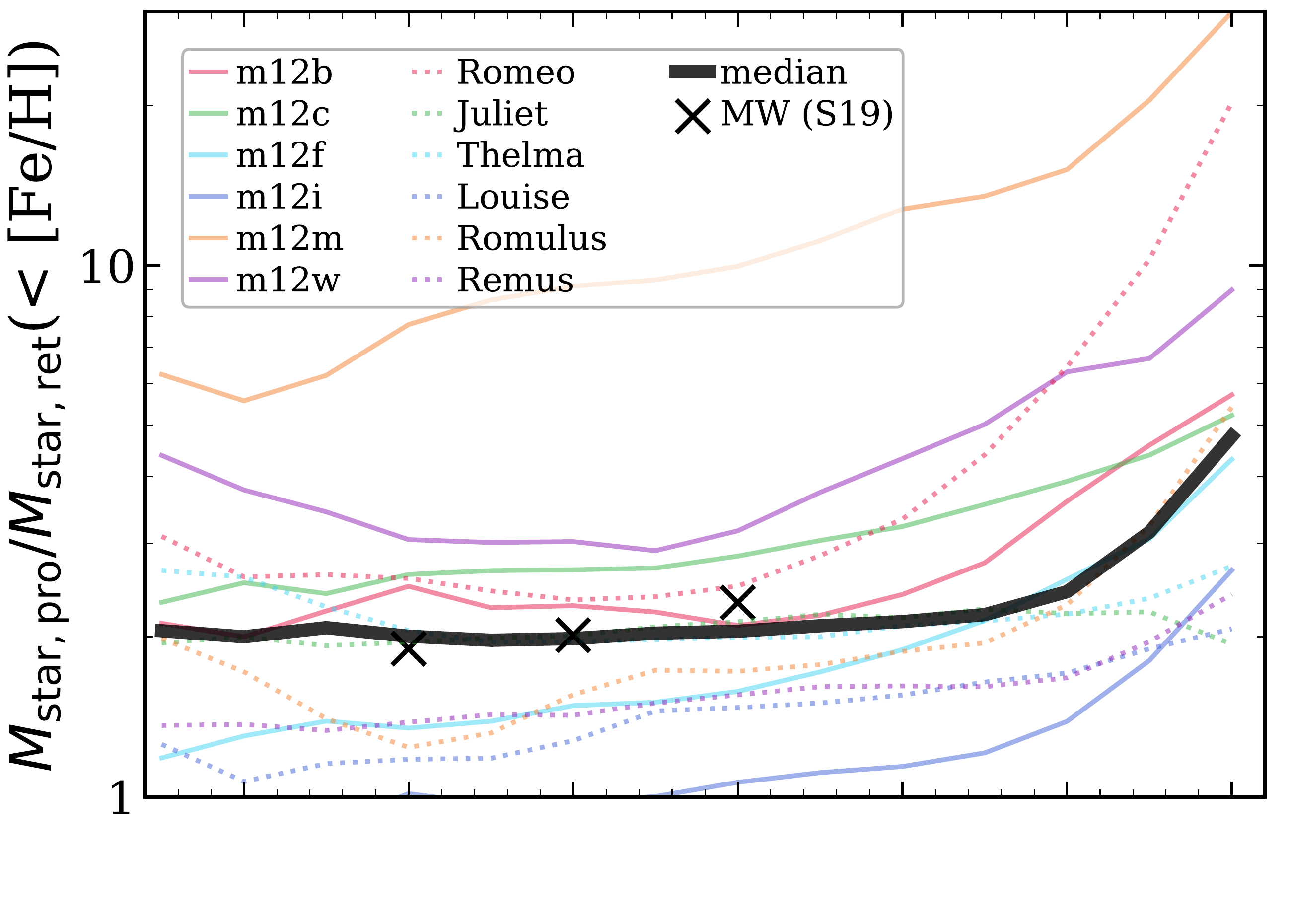} \vspace{-5 mm}\\
\includegraphics[width=0.94\linewidth]{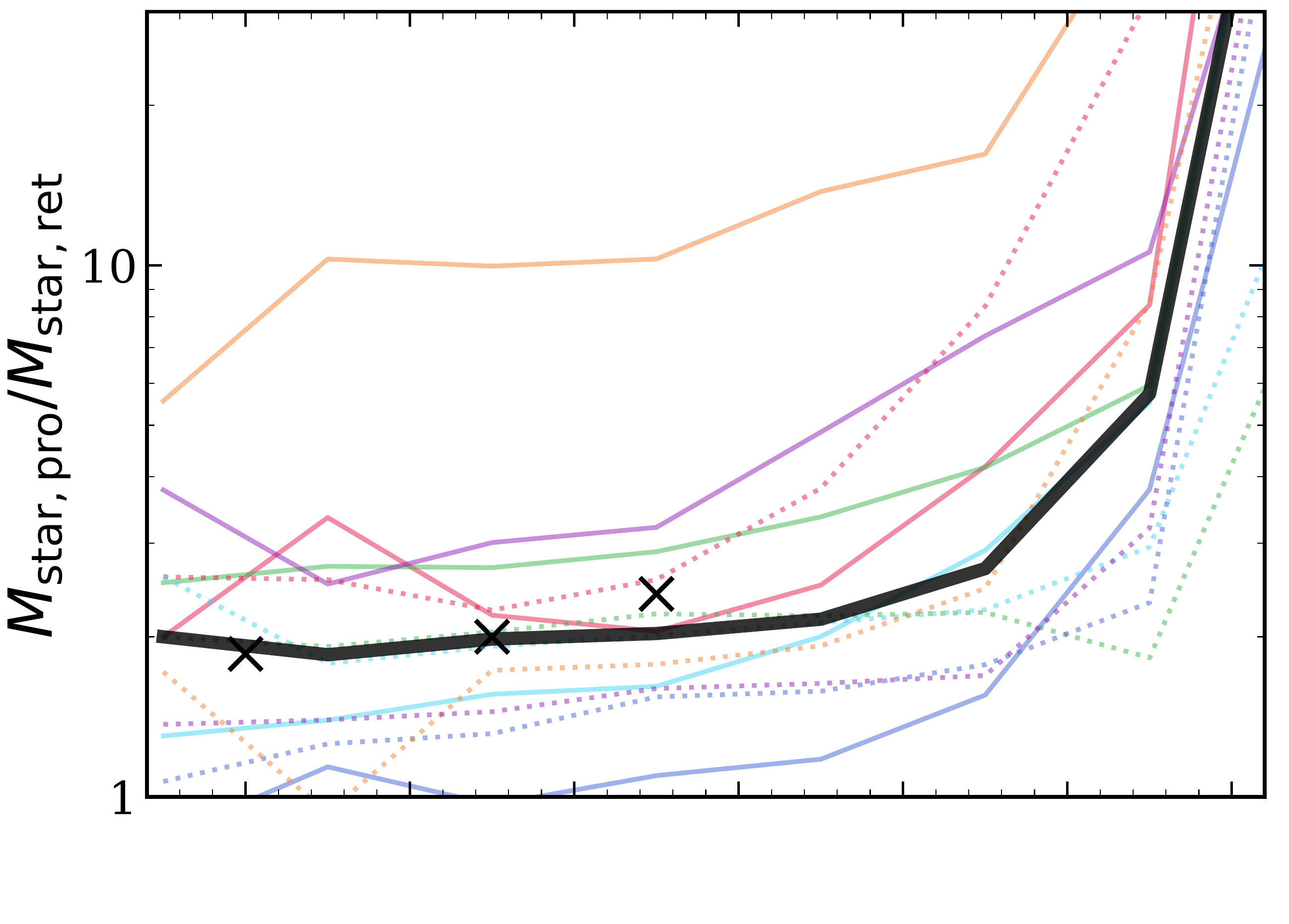} \vspace{-5 mm}\\
\includegraphics[width=0.94\linewidth]{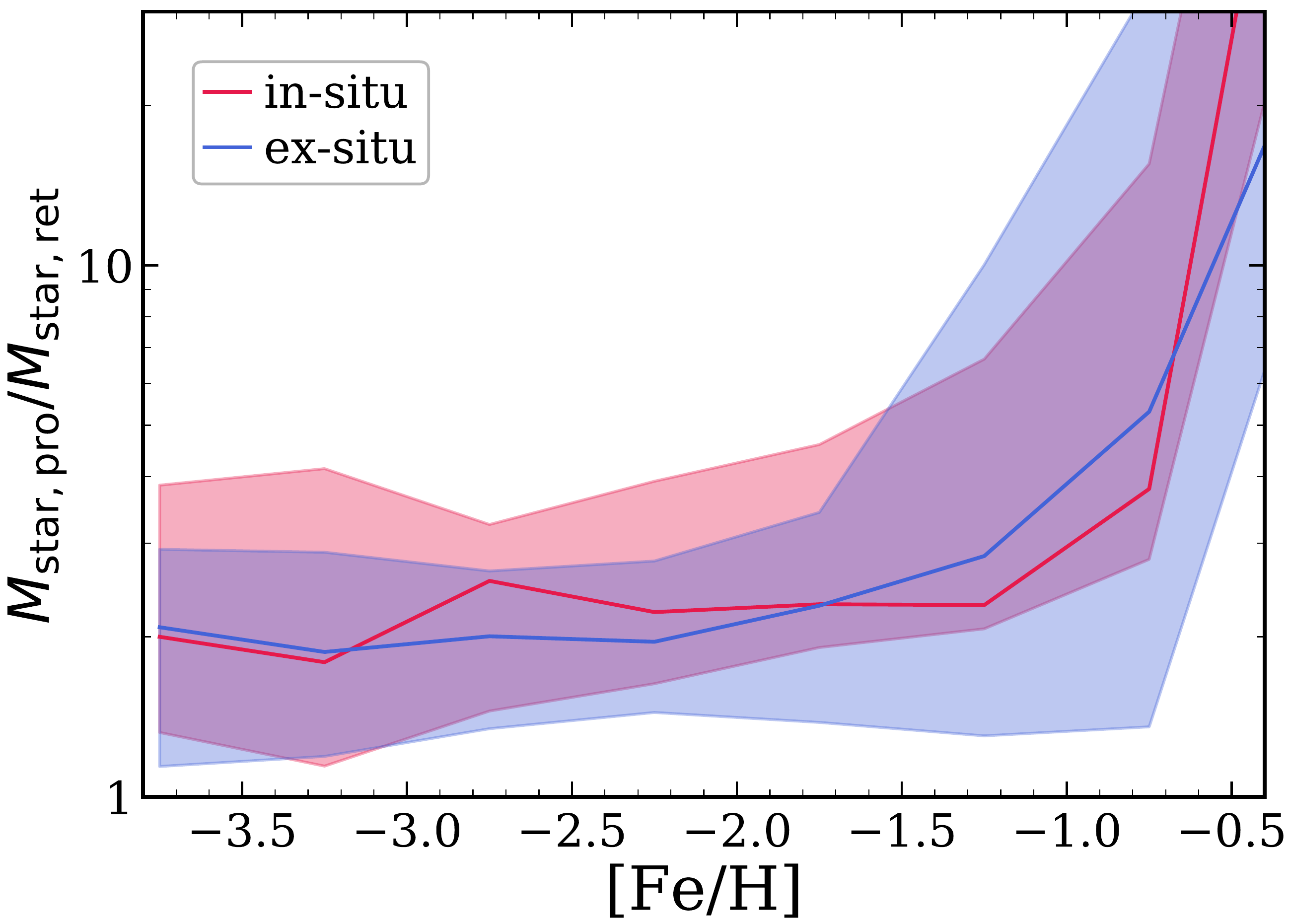}
\end{tabular}
\vspace{-2 mm}
\caption{
The prograde bias, defined as the mass ratio of prograde to retrograde stars, $M_{\rm star,pro}$ / $M_{\rm star,ret}$, versus stellar iron abundance for our 12 MW/M31-mass galaxy simulations at $z = 0$.
\textbf{Top and middle panels}:
the prograde bias as a function of cumulative $\FeH$ and binned $\FeH$, respectively.
Solid lines show the 6 isolated hosts, while dotted lines show the 6 LG-like hosts.
The thick black lines show the medians across all 12 hosts, which are consistent with MW observations from the Pristine survey \citep[][S19]{Sestito19}, which we show with black crosses.
11 of the 12 galaxies in our sample show a significant prograde bias ($> 1$) in metal-poor ($\FeH < -2.5$) stars.
For most hosts, the prograde bias is nearly constant with iron abundance at $\FeH \lesssim -1.5$, and it rises rapidly at higher $\FeH$.
\textbf{Bottom}: The binned prograde bias, split into in-situ (red) and ex-situ (blue) stars, where we select `in-situ' stars as formed within $15 \kpc$ of the host.
The lines show the medians, and the shaded regions show the 68 per cent scatter across 12 hosts.
Ex-situ and in-situ stars have similar prograde bias at $\FeH \lesssim -1.7$, while ex-situ stars show a slightly higher prograde bias at $-1.7 \lesssim \FeH \lesssim -0.7$.
As we will show, a single SMC/LMC-mass merger typically drives this trend.
At higher $\FeH$, the in-situ component quickly rises in prograde bias, as stars transition to the thin-disk component.
}
\label{fig:metal}
\end{figure}

We first examine how the present prograde bias varies with iron abundance, $\FeH$.
Figure~\ref{fig:metal} shows the cumulative prograde bias for all stars below a given $\FeH$, and the prograde bias for stars within 0.5 dex bins of $\FeH$, for all of our hosts, as a function of $\FeH$ at $z = 0$.
The thick black lines show the median across all 12 hosts, and we overlay Pristine observations of the MW from \citet{Sestito19} in black crosses labeled `MW (S19)'.
Solid lines show our 6 isolated hosts, while dashed line show our 6 hosts in LG-like pairs.
We show stars only up to $\FeH = -0.5$ to avoid the metal-rich stars in the thin disk that are on overwhelmingly prograde orbits, and we remind the reader of the metallicity floor of $\FeH = -4$ in the simulations.

The median prograde bias is fairly flat at $\FeH < -1$ for the cumulative plot (top panel) and at $\FeH < -1.75$ for the binned (middle panel) population.
A few hosts (such as m12b) show a slight rise at $\FeH \lesssim -3$. The prograde bias rapidly rises at higher iron abundance, as the stellar population quickly transitions to the more metal-rich, younger, thin disk (and thus highly prograde) component of each host.
The observed values from \citet{Sestito19} are close to our median lines, so the MW is both consistent and typical compared with our simulation suite.

Figure~\ref{fig:metal} shows that nearly all (11 of 12) of our hosts have significant prograde bias at all $\FeH$.
The key exception is m12i, which does not have any significant prograde bias at $\FeH \lesssim -2.5$.
However, m12i does have a prograde bias at higher $\FeH$, so this feature is ubiquitous (within our suite) at all $\FeH \gtrsim -2$.

Isolated and LG-like hosts differ slightly in Figure~\ref{fig:metal}.
Isolated hosts typically have higher prograde bias curves than LG-like hosts, with median values of 2.48 and 1.76 at $\FeH = -2.5$, respectively.
The differences in the prograde biases between isolated and LG-like galaxies is of order $\sim 1$, which is small compared to the host-to-host scatter, which is of order $\sim10$.
There are differences in the in-situ formation times between LG-like and isolated galaxies, with LG-like galaxies forming around $z\sim4.9$ ($t_{\rm lb}\sim12.5\Gyr$ ago) and isolated galaxies forming later, around $z\sim2.7$ ($t_{\rm lb}\sim11.3\Gyr$ ago), so the lack of strong differences in prograde bias may be surprising \citep{Santistevan20}.
However, as we discuss below, the prograde bias is primarily driven by mergers and sub-dominantly by in-situ formation; we do not explore these differences further throughout the rest of the paper.

Figure~\ref{fig:metal} (bottom panel) shows the same binned prograde bias, but we now split the stars into two origins: in-situ (formed within 15 kpc of the host's main progenitor), and ex-situ stars that formed in another galaxy that merged into the host.
The lines show the median values across all 12 hosts, and the shaded regions show the 68 per cent scatter.
One might naively expect that the in-situ population would have a larger prograde bias, to the degree that one expects that in-situ stars always formed on disk-like prograde orbits, and because a merger's orbit need not align itself with the host's existing disk.
However, we find only weak differences between in-situ and ex-situ stars, which are well within the host-to-host scatter, at all $\FeH < -0.5$.
At $\FeH \lesssim -1.75$, in-situ stars tend to be a little more prograde, but at higher $\FeH$ up to $\sim -0.75$, ex-situ stars in fact show a slightly stronger preference for prograde orbits.
At $\FeH > -0.75$, in-situ stars quickly become the more prograde population, because we now transition to stars that primarily formed in the host's thin disk.

Figure~\ref{fig:metal} shows that the prograde bias depends only weakly on iron abundance at $\FeH \lesssim -2$.
Thus, to maintain consistency with the selection in \citet{Sestito20}, we present results in Section~\ref{sec:fractions} and after (and Table~\ref{tab:hosts}) using $\FeH < -2.5$ as our fiducial selection for metal-poor stars.
Our results do not change significantly if we use a slightly different threshold in $\FeH$.

The results in Figure~\ref{fig:metal} do not qualitatively change for different values of the $J_z$ cut as well.
When we increase the selection to include stars with $|J_z| < 876 \kmsi \kpc$, double our fiducial value, the $z= 0$ prograde biases for stars with $\FeH < -2.5$ change by $\sim 3-6$ per cent.
Decreasing the selection by a half to only include stars with $|J_z| < 219 \kmsi \kpc$ changes the prograde biases by $\sim3-8$ per cent.
Neither of these fractional changes in the prograde bias are significant enough to completely erase the prograde signal, so, we continue to implement our fiducial $J_z$ cut from hereon, unless stated otherwise.

In our calculation of the prograde biases in Table~\ref{tab:hosts}, the number of prograde star particles range from 341 - 1964, with a mean of $\sim923$, and the range of retrograde star particles is 149 - 841, with a mean of $\sim462$.
These numbers correspond to fractional Poisson uncertainties that range from 2 - 5 per cent for prograde metal-poor star particles, and 3 - 8 per cent for retrograde metal-poor star particles.
The mean fractional Poisson uncertainties are 4 and 5 per cent for prograde and retrograde stars, respectively.
The typical prograde bias is $\sim2:1$, therefore, these uncertainties are not strong enough to erase this signal.
However, the host with the smallest prograde bias, m12i, is well within these uncertainties of having an equal number of prograde and retrograde stars, and its value of 0.98 should not be interpreted as having a `retrograde bias'.

Similarly, the Poisson uncertainties for each bin of $\FeH$ are not large enough to significantly change our results.
The mean fractional Poisson uncertainty in the number of prograde stars ranges from a min of 0.1 per cent to a max of 9.4 per cent for $\FeH = [-0.5, 0]$ and $\FeH = [-4, -3.5]$ respectively.
The uncertainty increases for decreasing $\FeH$ because there are fewer stars in lower metallicity bins.
Likewise, the mean fractional Poisson uncertainty in the number of retrograde stars ranges from 0.7 per cent for $\FeH = [-1, -0.5]$ to 13.3 per cent for $\FeH = [-4, -3.5]$.
The mean number of retrograde stars in the two highest metallicity bins, $\FeH = [-1, -0.5]$ and $\FeH = [-0.5, 0]$, are similar ($\sim90,000$), however, the more metal-rich bin has a wider range, which is why it does not have a smaller mean fractional uncertainty.

\subsubsection{Comparison with the H3 survey}
\label{sec:h3}

\begin{figure}
\centering
\begin{tabular}{c}
\includegraphics[width=0.95\linewidth]{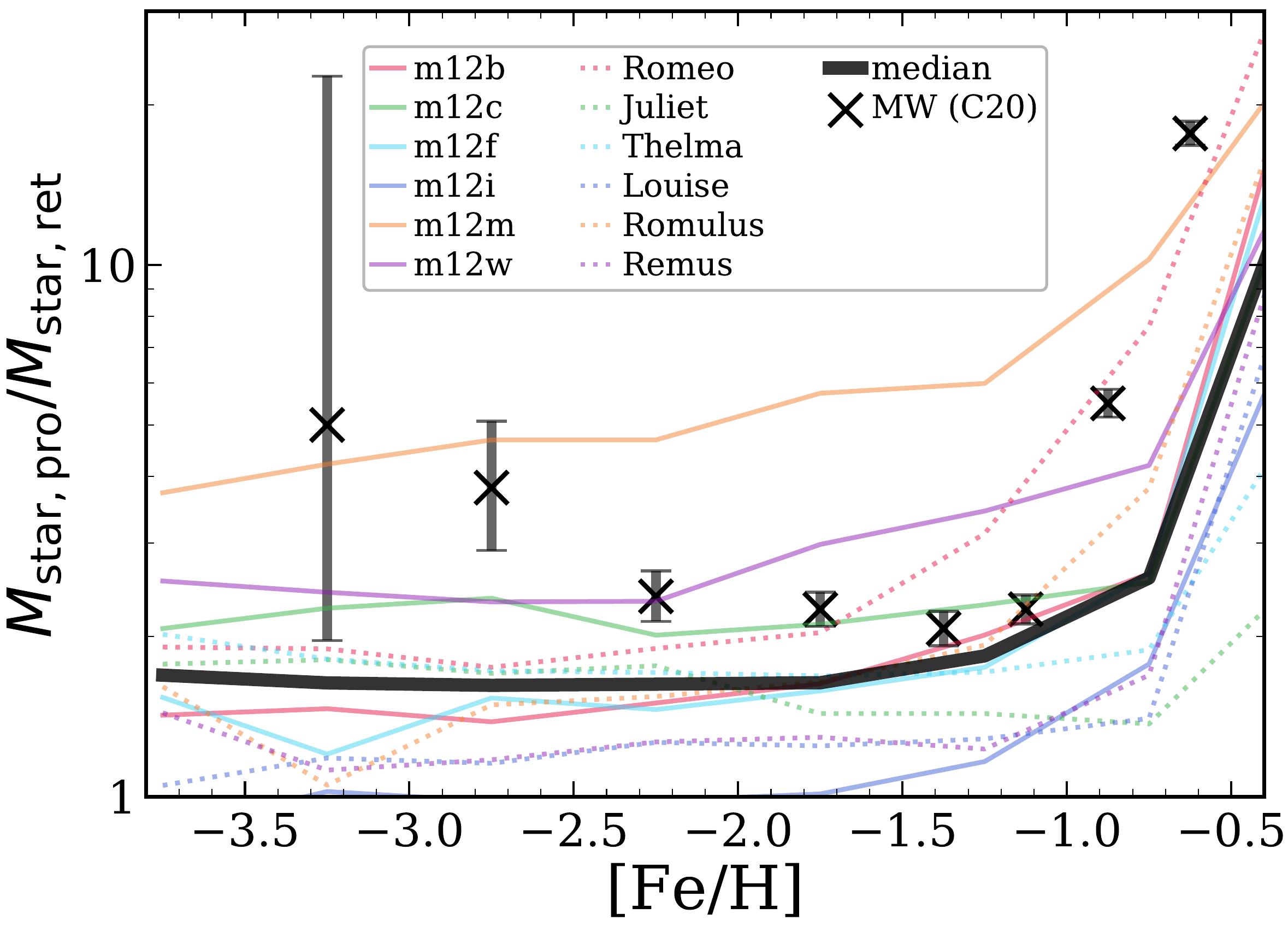}
\end{tabular}
\vspace{-3 mm}
\caption{
The prograde bias versus stellar iron abundance, similar to Figure~\ref{fig:metal} (middle panel), but using using a `stellar halo' selection function to mimic the H3 survey ($1 < |Z| < 3 \kpc$ and $|b| > 30^{\circ}$) and computing the prograde bias as the mass ratio of stars with $J_{\phi} > 0$ to those with $J_{\phi} < 0$.
Crosses show MW observations from the H3 survey \citep[][C20]{Carter20}.
Despite the differing selection function and prograde metric, the trends in our simulations remain similar to those in Figure~\ref{fig:metal}.
Our simulations qualitatively agree with the MW observations from C20, though now the prograde bias at $\FeH < -2.5$ is lower in most simulations than the MW, and most simulations do not show as strong an increase as the MW at $\FeH \gtrsim -1$, although typically at least 2 simulations have prograde biases at least as strong as these MW observations at any iron abundance.
}
\label{fig:h3}
\end{figure}

Recently, the H3 Spectroscopic Survey \citep{H3SURVEY} also examined the orbits of metal-poor ($\FeH < -2$) stars in the MW's halo \citep[out of the disk;][]{Carter20}.
They also found a strong prograde bias, with nearly 70 per cent of their stars on prograde orbits (defined as $J_{\phi} > 0$), in general agreement with \citep{Sestito20}, though \citet{Carter20} found some evidence that the prograde bias increases at lower $\FeH$ as well.
Figure~\ref{fig:h3} (crosses with uncertainties) shows their results.

Figure~\ref{fig:h3} also shows the results from our simulation suite when applying a selection window and prograde bias metric similar to \citet{Carter20}.
Specifically, we select star particles above the disk, at $1 < |Z| < 3 \kpc$, and at high Galactic latitude, $|b| > 30^{\circ}$.
Following \citet{Carter20}, we calculate the prograde bias as simply the ratio of the mass of star particles with $J_{\phi} > 0$ to those with $J_{\phi} < 0$.
As in Figure~\ref{fig:metal}, Figure~\ref{fig:h3} shows each of our hosts separately, and the thick black line shows the median.

The results in Figure~\ref{fig:h3} are similar to those in Figure~\ref{fig:metal} (middle).
The median prograde bias is flat across a wider range of iron abundance, $-3.75 < \FeH < -0.75$, before rising at higher $\FeH$.
The median prograde bias is consistently lower, with a value of $\sim 1.6$ at $\FeH = -2.5$ compared to $\sim 2$ in Figure~\ref{fig:metal}.
By selecting stars at higher vertical distance from the disk and without imposing a selection on $J_z$, we now select stars with orbits farther from the disk, which leads to a somewhat smaller prograde bias.

Again, our simulation suite broadly agrees with MW observations from \citet{Carter20}, although less well than with the results of \citet{Sestito20}.
At $-2.5 < \FeH < -1$, the observations are consistent with being constant, but they show an upward trend at lower $\FeH$.
The values below $\FeH < -2.5$ are higher than most of the simulated sample, but the uncertainties encompass much of the region spanned by our simulations.
The rise of the MW at $\FeH > -1$ is sharper than most of the simulations, but the simulations show similar qualitative trends.

To investigate further how this metallicity dependence changes with how we kinematically select star particles, we implemented the H3 kinematic selection ($J_{\phi} > 0$ versus $J_{\phi} < 0$) using the Pristine spatial selection window in Section~\ref{sec:pristine} and saw nearly identical results to Figure~\ref{fig:h3}.
The medians are essentially the same up to $\FeH \sim -0.5$, where the H3 selection has a slightly smaller value of $\sim 10$ compared to $\sim 15$ when using the Pristine window.
This suggests that implementing a more stringent spatial selection and ignoring the disk midplane, as in the H3 survey, does not change the prograde bias results when compared to our \textit{entire} disk selection.
As a final check, we examined how the prograde bias varies using the stellar populations solely above versus below the disk: both populations have nearly identical behavior.

\subsubsection{Prograde bias of the stellar halo}

\begin{figure}
\centering
\begin{tabular}{c}
\includegraphics[width = 0.95 \linewidth]{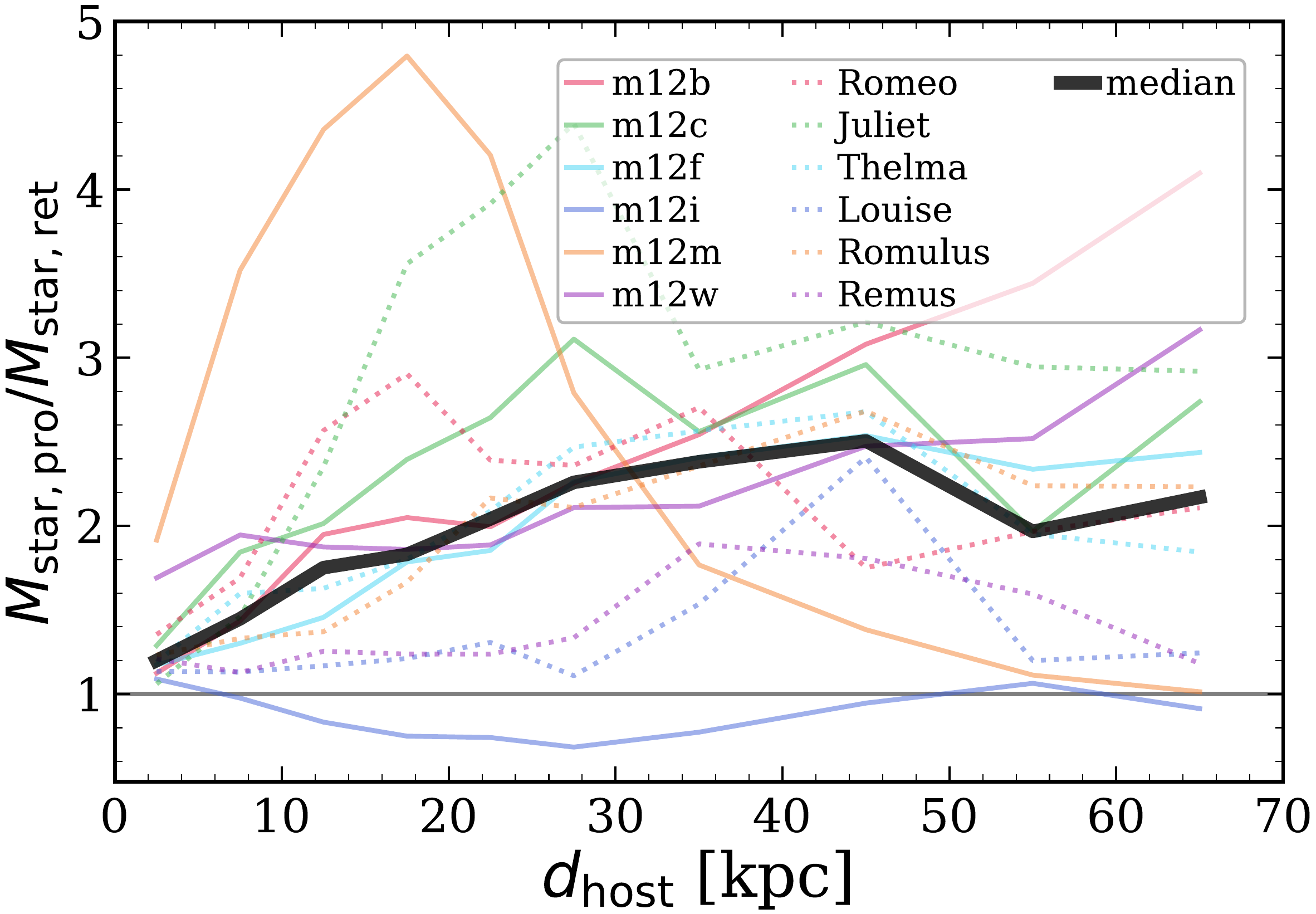}
\end{tabular}
\vspace{-2 mm}
\caption{
The prograde bias of the stellar halo.
We bin metal-poor ($\FeH < -2.5$) stars at $z = 0$ based on galacto-centric distance, and we compute the prograde bias as the mass ratio of stars with $J_{\phi} > 0$ to those with $J_{\phi} < 0$, relative to the host disk.
The black horizontal line at $M_{\rm star,pro} / M_{\rm star,ret} = 1$ denotes no asymmetry, that is, equal mass in prograde and retrograde.
Our most extreme hosts, m12m, shows a peak prograde bias of $\sim 4.8$ at $\sim 17.5\kpc$ before decreasing to 1 at $65\kpc$, and m12i again shows no prograde bias now in its stellar halo.
Similar to m12m, Juliet shows a peak of $\sim 4.5$ at $\sim 28 \kpc$, but the prograde bias remains constant at 3, instead of declining.
Both Louise and Remus have no appreciable prograde bias until $d \gtrsim 30 \kpc$, then they show peaks at $\sim2 - 2.5$ between $35 - 45\kpc$, and no prograde bias at $\gtrsim 65 \kpc$.
The median prograde bias increases with distance until $d_{\rm host} \sim 45 \kpc$, staying roughly constant at $\sim 2.5$ out to $100\kpc$. 
}
\label{fig:halo}
\end{figure}

Many studies within the literature debate the possible rotation of the MW's stellar halo, with some arguing for prograde motion and others arguing for retrograde motion.
To investigate the possible rotation of the stellar halos in the simulations, we select all metal-poor stars ($\FeH < -2.5$) and bin them based on their galacto-centric distance, with no additional geometric or kinematic selection.
We define the prograde bias in the same way as in the previous section: the stellar mass ratio of stars with $J_{\phi} > 0$ to those with $J_{\phi} < 0$.

Figure~\ref{fig:halo} shows how the prograde bias for this `halo' selection varies as a function of galacto-centric distances, for stars at $z = 0$.
The median prograde bias increases with distance to $\sim 2.5$ at $\sim 45 \kpc$, and the signal flattens at larger distances, up to $100\kpc$, though we show only out to $70 \kpc$.
In many hosts, the prograde bias increases with distance up to a certain point, then the prograde bias stays roughly constant.
In m12b and m12w, the prograde bias monotonically increases with distance, indicating that stars in the outer regions of these stellar halos are heavily biased towards prograde motion.
Our most asymmetric host, m12m, has a peak in its prograde bias at $\sim 4.8$ for stars at $15 - 20\kpc$, but then the signal decreases and eventually disappears.
In our least asymmetric host, m12i, we do not see a prograde bias anywhere, but instead it dips below 1, indicating a retrograde bias at $\sim 5 - 50\kpc$.
Finally, Louise and Remus exhibit virtually no prograde bias until $\sim 30 \kpc$, then they increase to $2 - 2.5$ around $\sim 40\kpc$ and fall back to 1 at $65\kpc$.

In Figure~\ref{fig:halo}, at the solar circle, $d_{\rm host} \sim 8 \kpc$, the median prograde bias is $\sim 1.5$, slightly lower than in Figure~\ref{fig:h3}, where the prograde bias is $\sim 1.8$.
Thus, using more stringent selection criteria closer to the disk results in a slightly larger prograde bias.
To ensure that this net prograde motion is not driven by stars in the plane of the disk, we also examined a similar trend as Figure~\ref{fig:halo} but only selecting stars at $|Z| > 3\kpc$ at all distances, that is, excluding the disk plane.
We find nearly similar results as in Figure~\ref{fig:halo}.
The prograde bias increases until $\sim 45 \kpc$ to a value of 2.5, then remains constant.
Thus, we find that the details of geometric selection do not influence the results of Figure~\ref{fig:halo} significantly.

Using a similar definition to denote metal-poor stars, $\FeH < -2$, \citet{Deason11} suggest that the metal-poor and metal-rich ($\FeH>-2$) components are rotating in a retrograde and prograde sense, respectively, but the retrograde signal may be a reflection about the adopted value of the local standard of rest.
Similarly, \citet{Bonaca17} suggest that metal-rich halo stars ($\FeH > -1$) have a net prograde rotation, while metal-poor stars show no net rotation.
Although we do not look at a complimentary metal-rich halo population in our analysis ($\FeH > -2.5$), Figure~\ref{fig:halo} shows a bias for prograde rotation ($\sim 2-2.5 : 1$) in our halo selection.
\citet{Deason11} conclude that the metal-rich stars likely were accreted from a satellite, while the metal-poor stars formed in the early host galaxy.
This qualitatively agrees with the curves for m12m, Juliet, Louise, and Remus in Figure~\ref{fig:halo}, given that the prograde bias peaks at a given distance and subsequently decreases afterward, suggesting possible substructure with net prograde rotation.
\citet{Bonaca17} conclude the opposite, metal-rich stars formed in-situ and metal-poor stars were accreted, which also qualitatively agrees with our results in Section~\ref{sec:fractions}: mergers dominate the origin of the prograde metal-poor stars.
Finally, more recent studies, such as \citet{Myeong19} and \citet{Naidu20}, suggest that the retrograde stellar halo population are a consequence of mergers, though we do not investigate the origins of the retrograde populations in our simulations.

\subsubsection{Dependence on stellar age}

\begin{figure}
\centering
\begin{tabular}{c}
\includegraphics[width = 0.95 \linewidth]{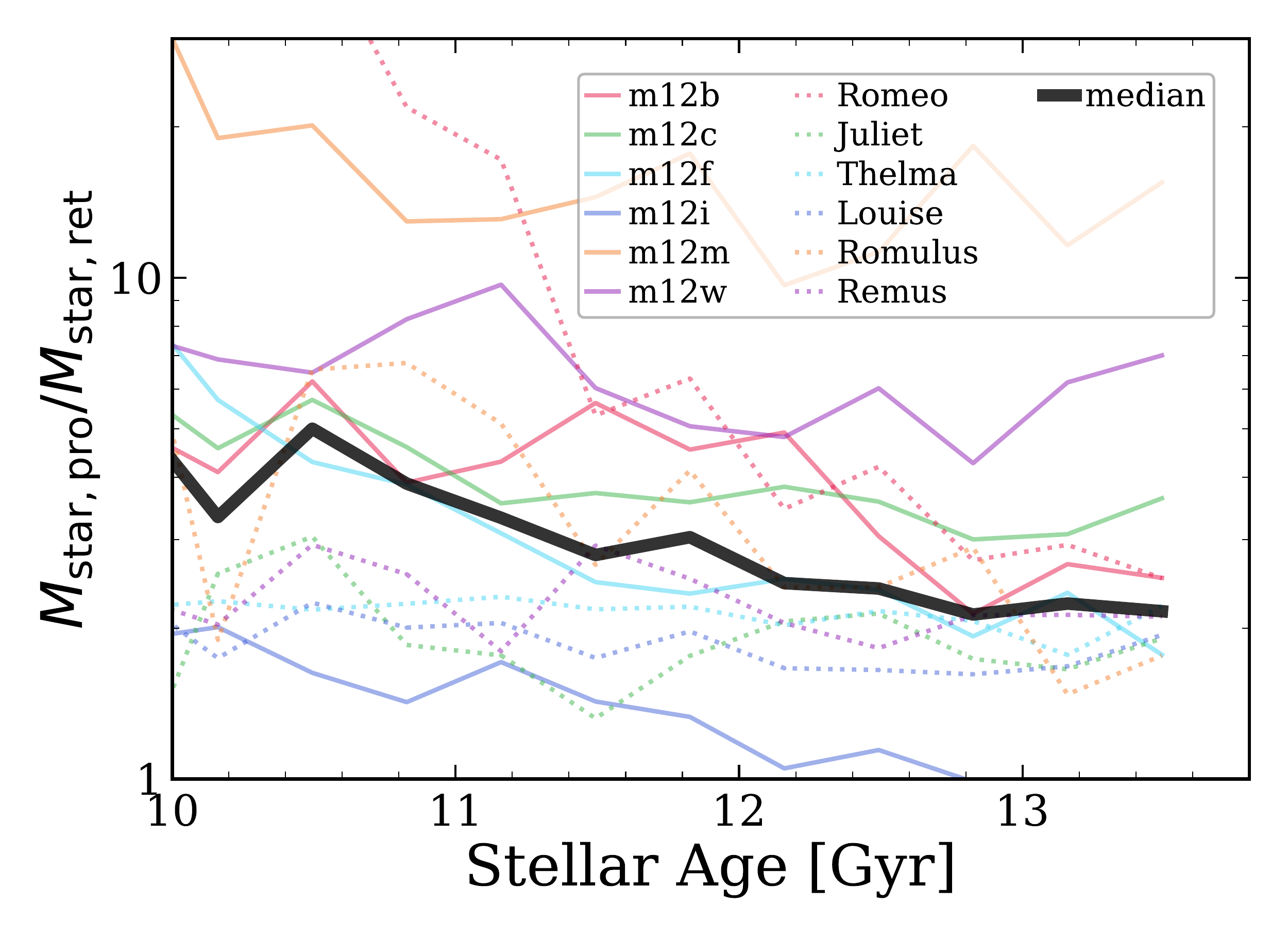}
\end{tabular}
\vspace{-2 mm}
\caption{
The prograde bias versus stellar age (regardless of iron abundance) at $z = 0$.
Solid lines show the 6 isolated hosts, dotted lines show the 6 LG-like hosts, and the thick black line shows the median across all 12 hosts.
The prograde bias decreases with stellar age, such that younger stars tend to be on more prograde orbits, consistent with the expectation that younger stars typically form on diskier orbits.
However, even stars in our oldest age bin ($\gtrsim 13.5 \Gyr$) have a median prograde bias of $\sim 2.2$.
All but one host (m12i) shows significant prograde bias to arbitrarily old stellar ages, and all hosts show prograde bias $> 1$ at stellar ages $\lesssim 12.5 \Gyr$.
Thus, the strong preference for prograde orbits in our simulations extends to arbitrarily old and/or arbitrarily metal-poor stars.
}
\label{fig:age}
\end{figure}

In general, a star's metallicity correlates with its age, though with significant scatter.
Thus we investigate how the prograde bias at $z = 0$ depends on stellar age, with no selection based on metallicity.
We once again select stars as in Section~\ref{sec:pristine}, following the approach of \citet{Sestito20} from the Pristine survey, though we find broadly similar results following the methodology of \citet{Carter20} from the H3 Survey.

Figure~\ref{fig:age} shows the the prograde bias at $z = 0$ as a function of stellar age, using age bins 0.5 Gyr wide.
The prograde bias increases for younger stars, with a median prograde bias of $\sim 4.5$ at $10 \Gyr$ of age, as stars transition increasingly to thin-disk formation in each host.
This trend is more dramatic for even younger stars (not shown) where the prograde bias reaches $\sim100$ for stars of age $\sim7\Gyr$.
All host galaxies except m12i have a median prograde bias of $\sim 2.2$ even at the oldest ages ($\gtrsim 13.5 \Gyr$), and m12i does show prograde bias starting at $\sim 12.5 \Gyr$ age.
We thus conclude that the prograde bias at $z = 0$ is nearly ubiquitous for star at all ages, including arbitrarily old stars.
While this result might be surprising, as we will show below, it can be understood as arising from the fact that the prograde bias arises largely from a single major merger, which brings in old stars with it.

The oldest stars ($t_{\rm form}>13\Gyr$) have a wide range in iron abundance, $\FeH = [-3.87, -0.52]$, with a median value of $\FeH = -2.5$.
This is broadly consistent with \citet{ElBadry18}, who used 3 of the simulations in our sample and found that stars that form at $z_{\rm form}>5$ ($t_{\rm lb} > 12.6 \Gyr$) have a range of $\FeH = [-3, -1]$, while stars with $\FeH \sim -2$ formed at $1.5 \lesssim z \lesssim 8$ ($9.4\Gyr \lesssim t_{\rm lb} \lesssim 13.1\Gyr$).

\subsection{Which galaxies contributed to the prograde bias?}
\label{sec:fractions}

We next investigate the origin of the prograde bias in metal-poor stars, where we define `metal-poor' as $\FeH < -2.5$ hereon.
Using the tracking method in Section~\ref{sec:sample}, we first attribute all star particles as either formed in-situ (formation within $15 \kpc$ from the host) or ex-situ.
For ex-situ star particles, we track which progenitor galaxy they were a member of prior to merging into the host, and we track the total stellar mass that each progenitor contributed to the prograde metal-poor population of each host.
We examine the three progenitors that contributed the most mass to the prograde metal-poor stars (or, put differently, the three progenitors that contributed the largest number of prograde metal-poor stars), and we refer to the galaxy that contributed the most, second most, and third most as the `primary', `secondary', and `tertiary' merger, respectively.
These mergers, however, are not always the top three most massive mergers to have ever contributed prograde metal-poor stars to the host.
Table~\ref{tab:hosts} lists values for the in-situ fractions, $f_{\rm in-situ}$, primary merger fractions, $f_{\rm merger,1}$, and fraction of stars contributed from the primary, secondary, and tertiary mergers combined, $f_{\rm merger, top 3}$.
Perhaps surprisingly, none of these fractions show any clear correlation with the strength of the prograde bias.

For all hosts except Louise and Juliet, the primary merger fractions are higher than the in-situ fractions, with median values of $f_{\rm merger, 1} \approx 0.24$ and $f_{\rm in-situ} \approx 0.16$.
\textit{Thus, the primary merger generally was responsible for more prograde metal-poor stars than the (most massive progenitor of the) host galaxy.}
The primary merger fractions range from $\approx 15 - 55$ per cent: the merger in Romulus contributed the most, and the mergers in Juliet and Louise contributed the least.
The in-situ fractions range from $\approx 5 - 30$ per cent.
Furthermore, on average, the primary merger contributed more prograde stars ($\sim24$ per cent) than retrograde stars ($\sim16$ per cent) to the host galaxy.

\begin{figure}
\centering
\begin{tabular}{c}
\includegraphics[width = 0.95 \linewidth]{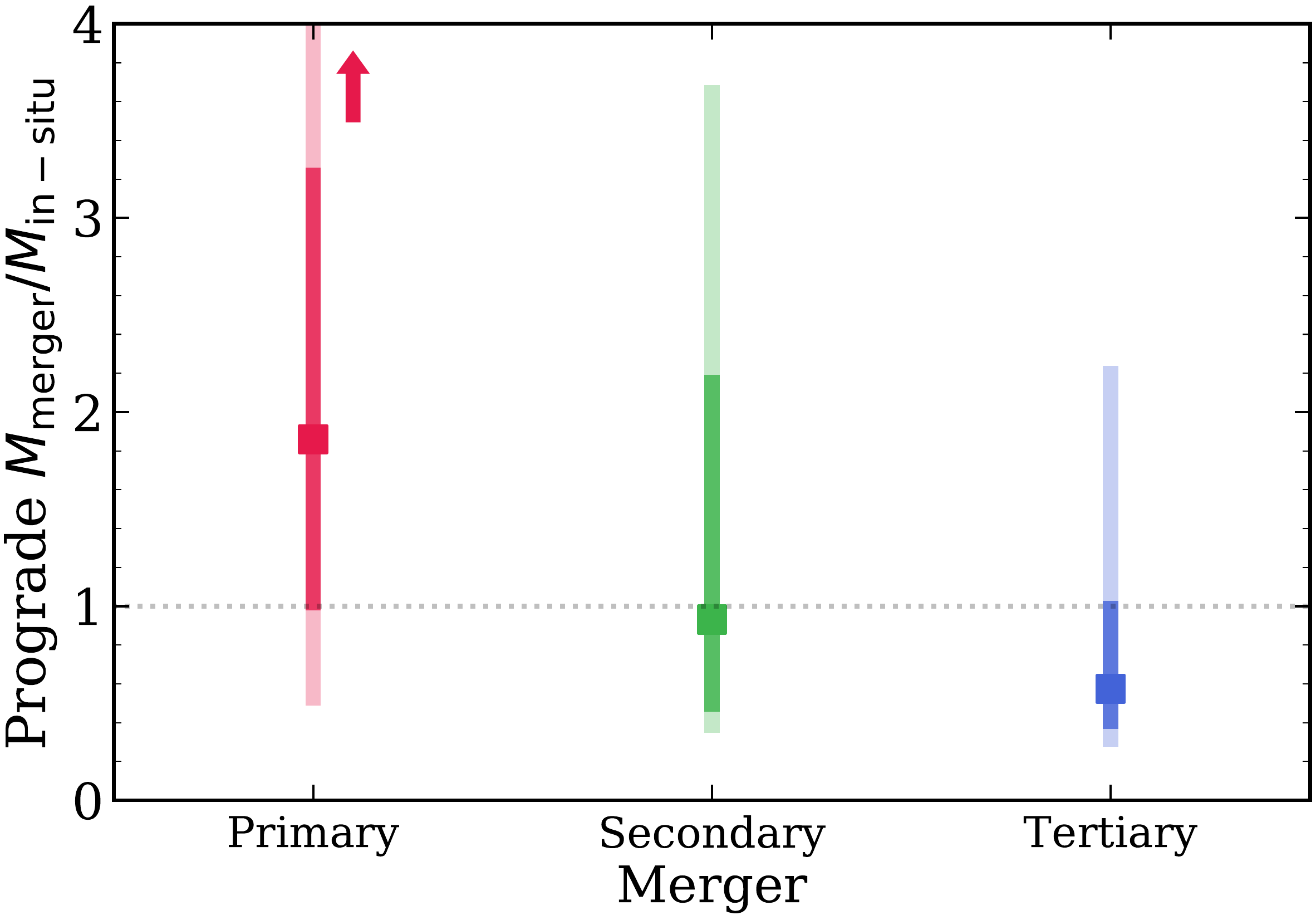}
\end{tabular}
\vspace{-2 mm}
\caption{
The ratio of the mass of prograde metal-poor stars contributed from a merger to that from in-situ formation, for the primary (red), secondary (green), and tertiary (blue) galaxy mergers, defined as the 3 progenitor galaxies that deposited the most mass to the prograde metal-poor population.
Squares show the median values across all 12 hosts, and the dark and light vertical bars indicate the 68 per cent scatter and full range, respectively.
The red arrow indicates that the full range of values for the primary-to-in-situ ratios extend beyond the axis (to $\sim11$).
Relative to in-situ stars, the primary merger typically contributed $\approx 2 \times$ more stars, the secondary merger was comparable to in-situ, and the tertiary merger contributed only $\sim 1/2$ as much as in-situ.
The sum of these 4 components nearly always accounts for the majority of prograde metal-poor stars (see also Table~\ref{tab:hosts}), and any single merger beyond the tertiary merger is typically negligible.
Thus, the primary source of prograde metal-poor stars at $z = 0$ is from a single galaxy merger.
}
\label{fig:merger_vs_insitu}
\end{figure}

Figure~\ref{fig:merger_vs_insitu} shows the ratio of the stellar masses contributed by the primary (red), secondary (green), and tertiary (blue) mergers, relative to in-situ formation, of prograde metal-poor stars at $z = 0$.
The square points show the median values across our 12 hosts, and the dark and light vertical bands show the 68 per cent scatter and full range, respectively.
The full scatter extends up to $\approx 10$ in the primary merger to in-situ ratio, though we truncate the plot for clarity.
The primary merger contributed twice as many stars as in-situ formation, so \textit{the primary merger is the single most important contributor to prograde metal-poor stars}.
Furthermore, the secondary merger contributed comparable stellar mass as in-situ formation.
Finally, the tertiary merger contributed typically only $\sim 1/2$ that of in-situ formation.
We emphasize that most metal-poor stars formed early, $\gtrsim 12 \Gyr$ ago; this includes both stars that form in-situ and ex-situ.
Almost all metal-poor stars formed prior to these 3 mergers ($\gtrsim93$ per cent), thus, these mergers did not induce in-situ formation that contributed to the prograde bias.

While Figure~\ref{fig:merger_vs_insitu} shows the masses of the mergers and in-situ stars that comprise the prograde population, we also investigated the total stellar masses of these mergers relative to each other.
Specifically, we computed the ratio of the peak stellar masses (across their histories) between the primary to the secondary, and the secondary to the tertiary.
We found that the peak stellar mass ratio between the secondary to the tertiary was $\sim 2$, consistent with Figure~\ref{fig:merger_vs_insitu}, but the ratio between the primary to the secondary was $\sim 6$, greater than in Figure~\ref{fig:merger_vs_insitu}.
This implies that the primary merger contributes a smaller \textit{fraction} of its original stars to the prograde metal-poor population compared to the secondary merger, even though it contributes more prograde stars in total.

Table~\ref{tab:hosts} shows that in-situ formation together with the primary merger account for $30 - 70$ per cent of all prograde metal-poor stars.
Considering the in-situ fraction along with the top 3 mergers, this accounts for $43 - 96$ per cent.
Juliet has the smallest summed fraction, so an unusually large number of additional mergers contributed significantly to its prograde metal-poor stars.
In m12m, m12c, and Romulus, the combined fractions reach $\approx 82$ per cent or higher, and the top 3 mergers account for roughly $\approx 75$ per cent.
We also find no clear differences in these fractions for isolated versus paired hosts; m12c (isolated) and Romulus (LG-like) have the 2 highest primary merger fractions, and m12b (isolated) and the Romulus have the 2 lowest in-situ fractions.
Furthermore, m12m, which has the highest prograde bias, has a relatively high primary merger fraction and average in-situ fraction, while m12i, which has the smallest (no) prograde bias, has average values for both.
These results highlight the diversity of merger/growth histories that lead to a prograde bias for metal poor stars at $z = 0$, and that the fractions of such stars from in-situ formation or the primary merger do not predict the strength of this kinematic feature.

We thus conclude that mergers dominate the origin of prograde metal-poor stars, and the primary merger dominates over any other merger.
We thus focus most of our subsequent analysis on the properties of the primary merger.

\subsection{How did the prograde bias evolve over time?} %
\label{sec:asym_evo}                           %

\begin{figure*}
\centering
\begin{tabular}{c @{\hspace{-0.1ex}} c}
    \includegraphics[width=0.48\linewidth]{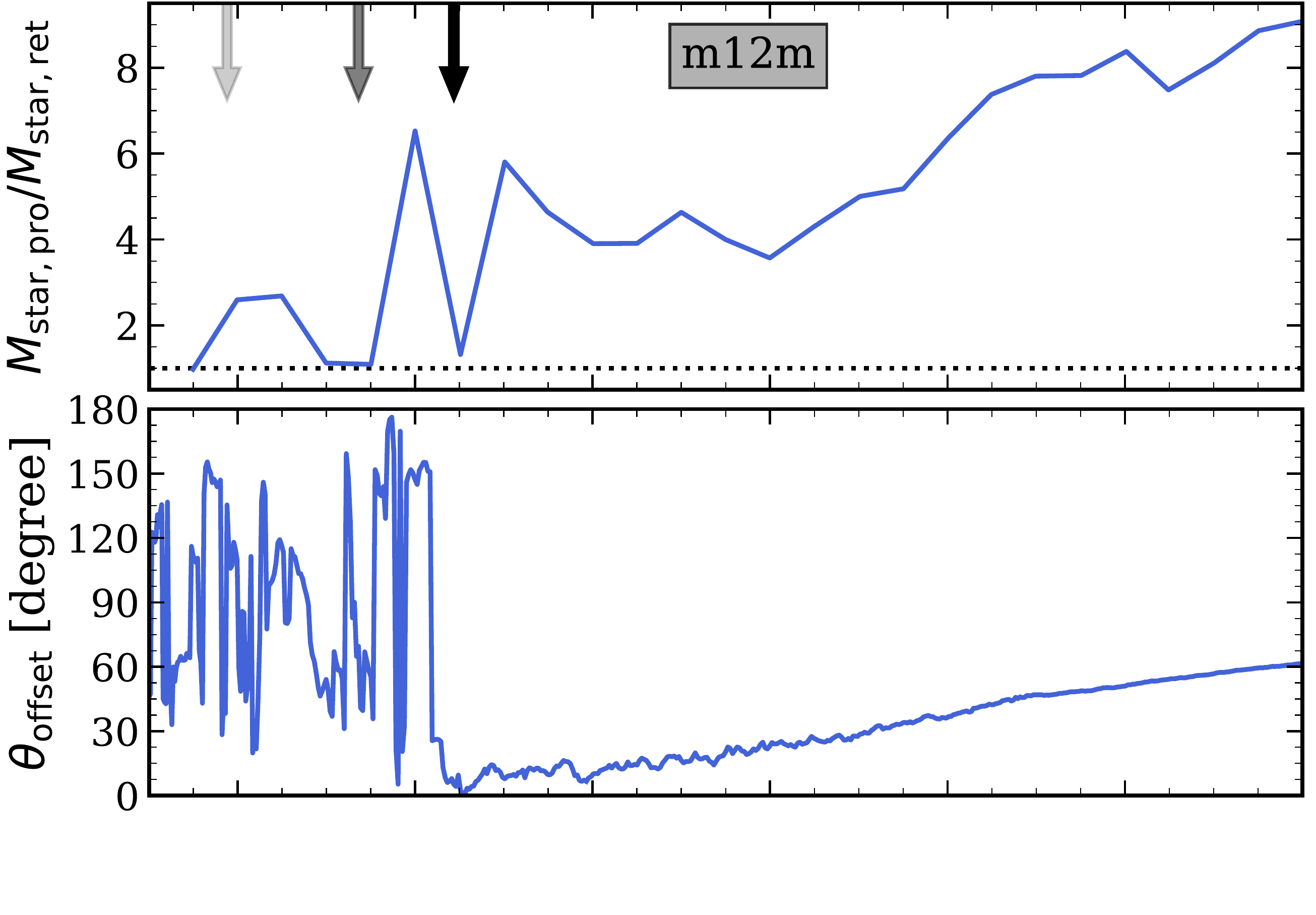}&
    \includegraphics[width=0.48\linewidth]{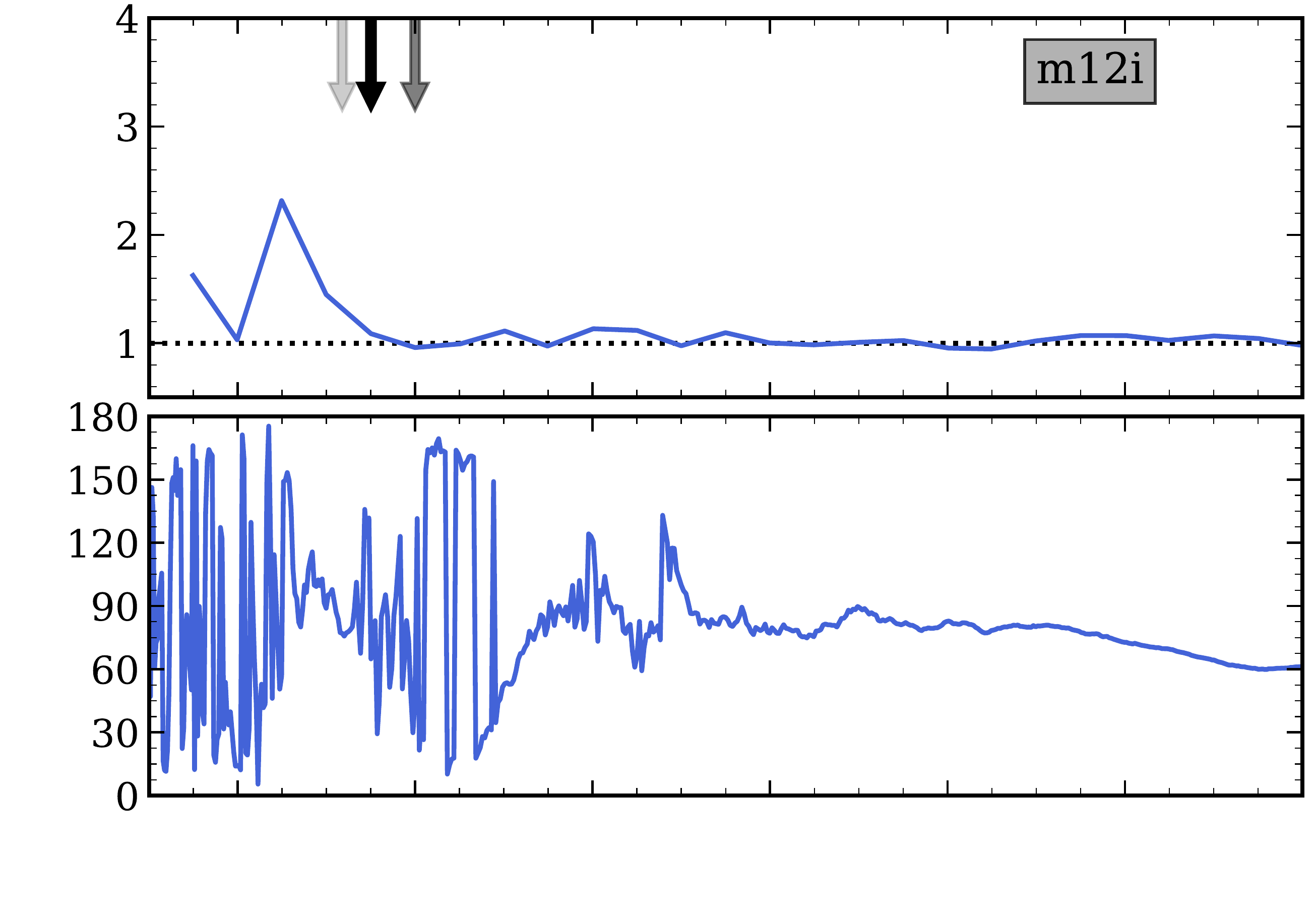}\\
    \includegraphics[width=0.48\linewidth]{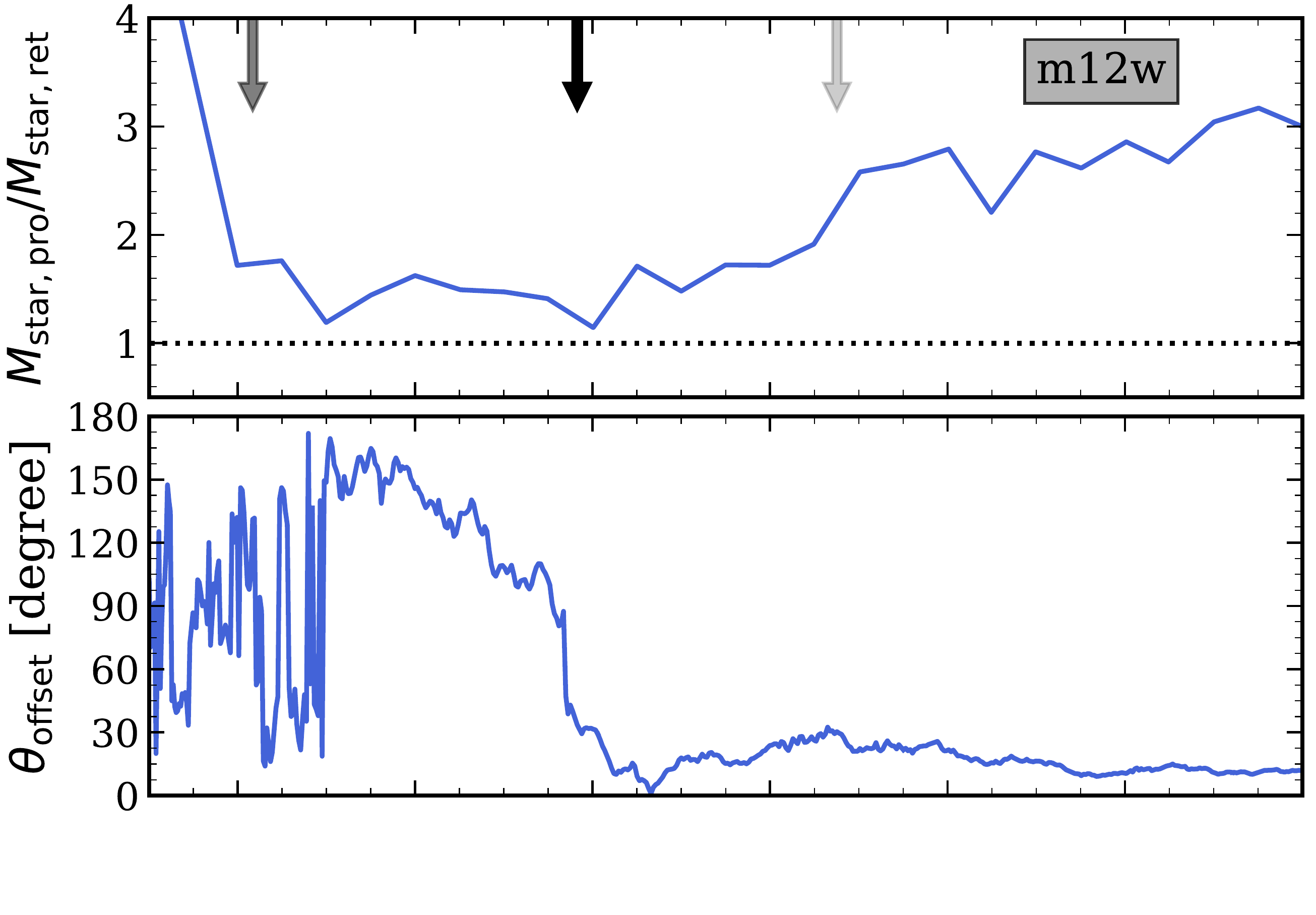}&
    \includegraphics[width=0.48\linewidth]{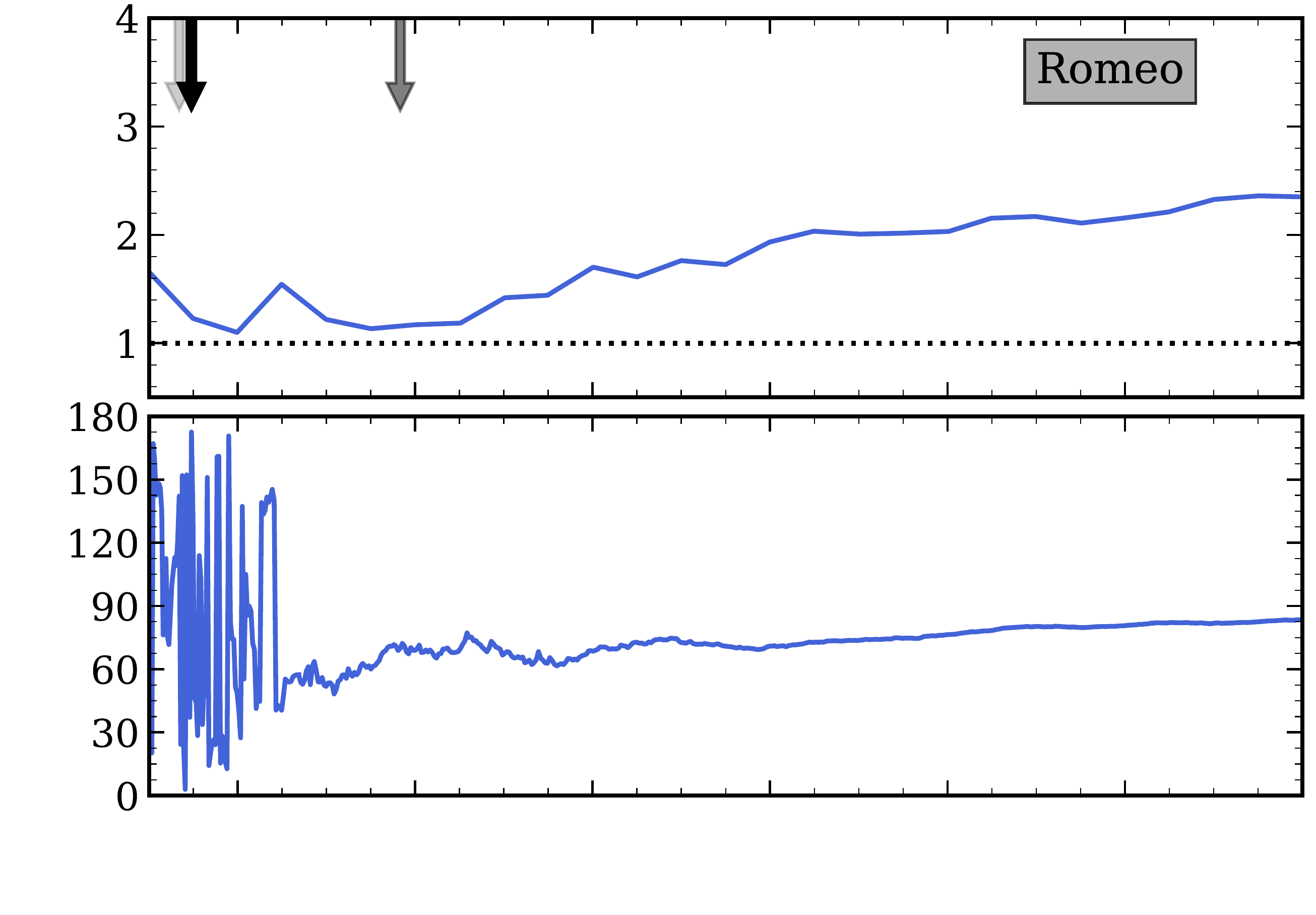}\\
    \includegraphics[width=0.48\linewidth]{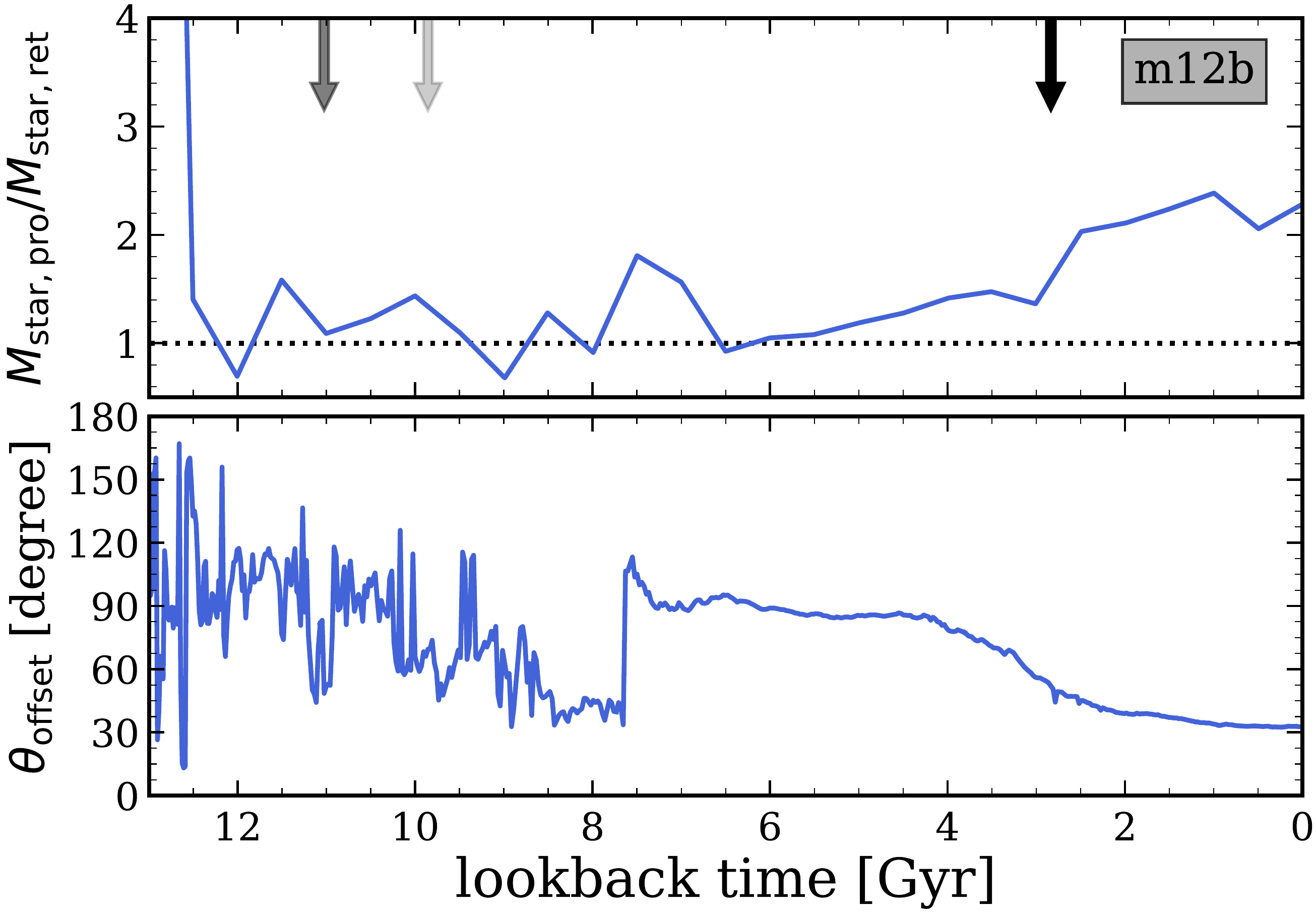}&
    \includegraphics[width=0.48\linewidth]{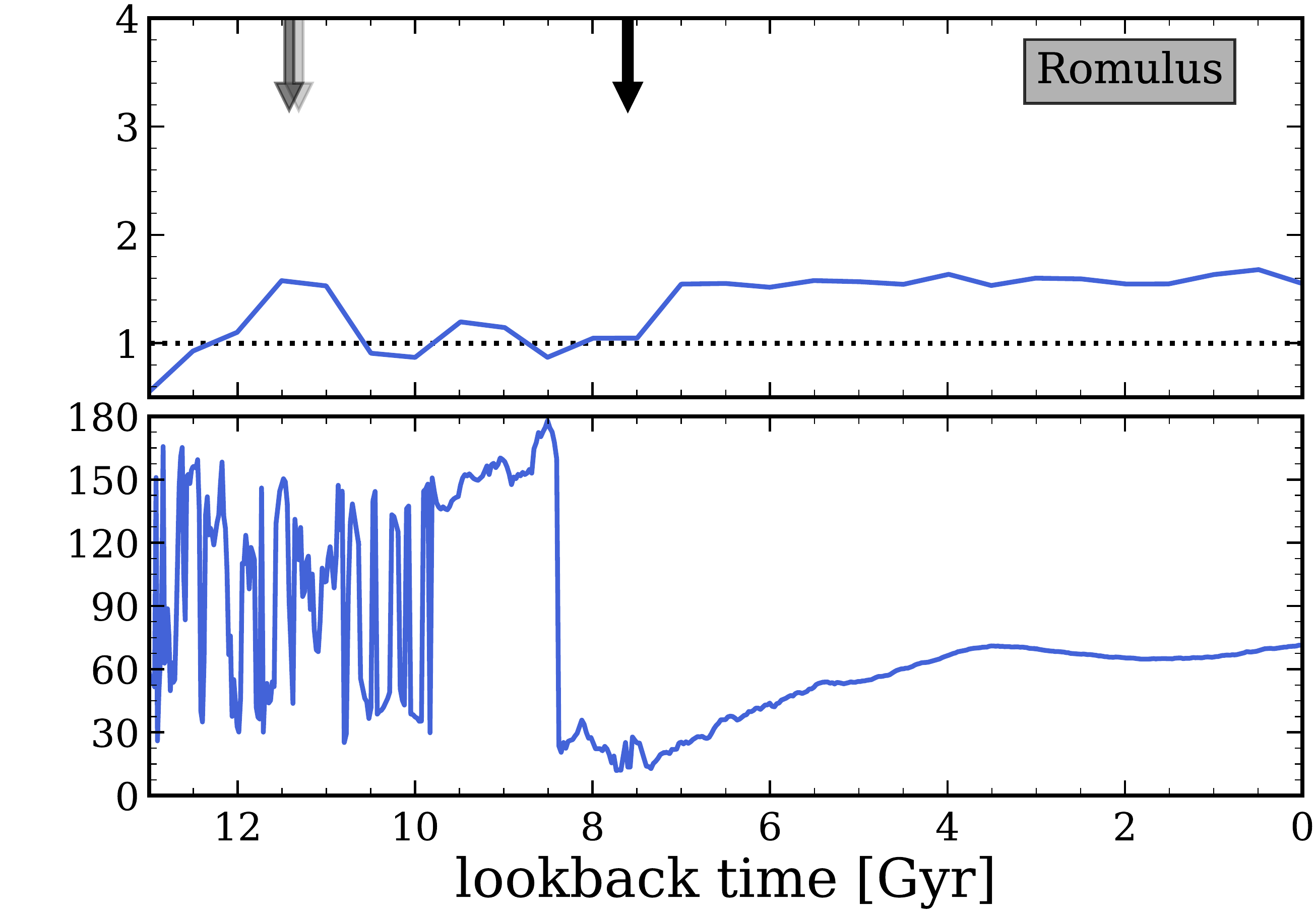}
\end{tabular}
\vspace{-2 mm}
\caption{
The formation histories for 6 galaxies in our sample.
We show half of our simulated galaxies, which represent the diverse origins of the prograde bias across our suite; the other galaxies have similar behavior.
Arrows indicate the times of the top three galaxy mergers, ranked by the mass that they contribute to prograde metal-poor ($\FeH < -2.5$) stars at $z = 0$, with arrow darkness indicating the rank, such that the primary merger is the darkest.
For each galaxy, the upper sub-panel shows the evolution of the prograde bias: the dotted horizontal line at 1 indicates no bias (equal mass in prograde and retrograde stars).
At early times, the prograde bias was typically near 1, though this reflects the fact that the disk had not formed yet, so prograde and retrograde are not well defined; see discussion below.
For all galaxies except m12i (top right), the prograde bias increased sharply following either the primary merger (as in m12m, m12w, m12b, Romulus) or the secondary merger (as in Romeo).
m12w also shows an additional increase after the tertiary merger.
For each galaxy, the lower sub-panel shows the evolution of the offset angle between the angular momentum vector of the galaxy's disk and the angular momentum vector of the primary merger's orbit, where we freeze the latter at its final orientation vector after the merger coalesces with the host (see Section~\ref{sec:mergers}).
At early times, the offset angle of the host's disk was un-aligned with the (future) merger, and it rapidly shifted given the high merger/accretion rates, and that the disk itself was only marginally defined.
However, the host's disk typically aligned itself with the orbital plane of the primary merger (as in m12m, m12w, m12b, Romulus) during/after the merger event.
These mergers were typically gas-rich and drove (via torquing and gas deposition) the formation and/or prograde orientation of the host's disk, seeding the origin of prograde metal-poor stars.
}
\label{fig:asym_offset}
\end{figure*}

We next investigate how the prograde bias evolved over time.
We calculate the prograde bias at each snapshot spaced $\approx 500 \Myr$, selecting stars based on the criteria in Section~\ref{sec:sample}.
We measured the orientation of the host's disk separately at each snapshot, defined according to the moment of inertia tensor of all star particles in the host.
We then calculated the prograde bias at each snapshot based on the host's rotational velocity, $\upsilon_{\rm rot}$, for normalizing $J_{\phi}$.
Finally, we do not limit this analysis across time to star particles that are confined to the plane of the disk throughout their lifetimes, only those that are within the action selection windows discussed above.

Figure~\ref{fig:asym_offset} (top sub-panels) shows the evolution of the prograde bias for 6 of our hosts as a function of lookback time.
These hosts show a wide range of prograde biases and histories, and they are representative of the other 6 hosts that we do not show.
The vertical arrows show when each primary, secondary, and tertiary merger took place (darkest to lightest, respectively).

Figure~\ref{fig:asym_offset} (top left) shows m12m, the host with the strongest prograde bias, $\approx 9$.
As with many hosts, the prograde bias at early times was highly variable, caused by rapid merger/accretion activity and no well-defined long-lived disk.
The prograde bias increased dramatically after the secondary and primary mergers, with continued gradual growth after the primary merger $\sim 6 \Gyr$ ago.

Figure~\ref{fig:asym_offset} (top right) shows our one host, m12i, that has no significant prograde bias over the last $10.5 \Gyr$.
The reason may be that all 3 mergers occurred at nearly the same time, possibly canceling out any coherent effects.
Both m12m and m12i highlight the extremes of the prograde biases found in our simulated sample.

Both m12w (middle left) and Romeo (middle right) are interesting because their prograde bias steadily increased across most of cosmic time, and their the most recent merger was not their primary merger.
The primary merger in m12w occurred $\sim 8.2 \Gyr$ ago and significantly increased the prograde bias, while the primary merger in Romeo occurred $\sim 12.5 \Gyr$ ago and had a more mild effect on the prograde bias, which subsequently decreased.
Interestingly, the tertiary merger in m12w continued to drive the prograde bias upward, and more importantly, the secondary merger corresponded to the main increase for Romeo.
Thus, secondary and tertiary mergers, which contributed fewer prograde metal-poor stars than the primary merger, still can effect the prograde bias.
Juliet, Louise, Remus, and m12f (not shown) are similar to Romeo, because they have gradually and monotonically increasing asymmetry biases over time.

m12b (bottom left) has the latest primary merger, $\sim 3 \Gyr$ ago, which increased the prograde bias to $\approx 2$.
Here, the (early) secondary and tertiary mergers did not have much of an effect.
m12b shows that a galaxy can have no significant long-term prograde bias throughout most of its history, until a late merger induces one.
Both Thelma and m12c (not shown) show similar behavior, with a negligible prograde bias until $\sim 6 - 7 \Gyr$ ago, which then continued to increase.
In Thelma, there was no prograde bias until $\sim5\Gyr$ ago, however, the primary merger occurred $\sim8.5\Gyr$ ago and had no lasting affect on its prograde bias.
The primary merger in m12c occurred $9\Gyr$ and did cause the prograde bias to spike to $\sim3.5$, however, the secondary merger caused the prograde bias to completely vanish around $0.5\Gyr$ later.
In both m12c and Thelma, it is likely that we do not track the mergers that cause their prograde biases to increase and become long-lived.

Finally, figure~\ref{fig:asym_offset} (bottom right) shows Romulus, which had two distinct periods of significant prograde bias.
The secondary and tertiary mergers induced the first period $\sim 11.5 \Gyr$ ago, but it quickly went away.
The primary merger then occurred $\sim 7.5 \Gyr$ ago, which drove the prograde bias to $\sim 1.5$, where it remained to $z = 0$.
This primary merger in Romulus contributed the highest fraction of prograde metal-poor stars compared to all other hosts (53 per cent).
Both m12b and Romulus show how the timing of these mergers affects the prograde bias.

We tested several variations in how to measure the prograde bias for each host across time, and we found similar overall trends using all of them.
First, we tested using the same $J_{\phi}$ selection region for each host at all snapshots, instead of scaling to each host's $\upsilon_{\rm rot}$ at each snapshot.
Second, while Figure~\ref{fig:asym_offset} shows the prograde bias using all metal-poor stars within the host's selection region at each snapshot, we tested following back only the star particles that are in the selection region at $z = 0$.
Finally, we examined how the prograde bias evolves selecting all metal-poor stars within a radius of 15 kpc from the host, as opposed to our fiducial `disk' spatial selection.
For all of these variations, we found the same general evolutionary trends, which reinforces that the prograde bias is a global feature of these MW/M31-mass galaxies.

\subsection{How did the primary merger affect the orientation of the host's disk?} %
\label{sec:mergers}                                      %
We next investigate the relation between the primary merger's orbit and the orientation of the stellar disk of the host galaxy.
In particular, we examine whether the merger was on an orbit that was aligned with the host's pre-existing disk, whether the merger torqued the direction of a pre-existing disk, and/or whether the merger helped to seed the formation and orientation of the disk.

The bottom sub-panels in Figure~\ref{fig:asym_offset} show the evolution of the offset angle, $\uptheta_{\rm offset}$, for each host, defined as the angle between the angular momentum of the primary merger's orbit and the angular momentum of the host galaxy's disk.
We store the final angular momentum vector of the primary merger just it merged into the host, and we compare it with the evolving angular momentum of the disk after the merger.

At early times, $\uptheta_{\rm offset}$ changed rapidly, caused by rapid accretion and mergers with many galaxies that built the host galaxy \citep[e.g.][]{Santistevan20}.
As a result, the angular momentum of the host's (often poorly defined) disk quickly changed.
All hosts in Figure~\ref{fig:asym_offset} except m12b show a rapid transition from a rapidly varying $\uptheta_{\rm offset}$ to a settled long-lived disk orientation.
This transition often coincides with one of the major mergers.

In both m12m and m12w (top left and middle left panels, respectively), just prior to the primary merger $\uptheta_{\rm offset}$ dipped significantly, and after the merger it went to near $0^{\circ}$.
This means the merging galaxy was aligned with the \textit{resultant} host's disk \textit{after the merger}, that is, the galaxy merged into the host, deposited the metal-poor stars that remain until $z = 0$, and defined the prograde direction of the host's disk.
These events occurred early and helped seed the formation of a stable long-lived stellar disk in the host.
After these merger events, the disk continued to gradually change its orientation, presumably from interaction with other galaxies and/or accretion events.
In m12m, the disk ends up tilted by $\approx 60^{\circ}$ with respect to the primary merger orbit.
However, in m12w, the disk did not rotate much, remaining at $\approx 10^{\circ}$ with respect to the primary merger orbit.

Romulus (bottom right panel) shows similar behavior as m12m and m12w; starting $\sim 500\Myr$ before the merger, the host's disk flipped from $\approx 180^{\circ}$ down to$ \approx 20^{\circ}$ and stayed there for $\sim 1 \Gyr$ during/after the merger, after which it gradually increased to $\approx 70^{\circ}$, similar to both Romeo and m12i (middle right and top right panels, respectively).
Both Juliet and m12c are similar to m12m, m12w, and Romulus, because there were temporary merging events that drove $\uptheta_{\rm offset}$ down, while subsequent merger events caused $\uptheta_{\rm offset}$ to rise/rotate again.

Romeo and m12i are different from the rest of the hosts shown here, because $\uptheta_{\rm offset}$ remained rapidly variable both before and after the primary merger, stabilizing only $\sim 1 - 2 \Gyr$ after.
The remaining hosts not shown here (m12f, Thelma, Louise, and Remus) reflect the same behavior as Romeo and m12i: rapid change in $\uptheta_{\rm offset}$ at early times, including before, during, and after the primary merger, but eventually the disk settleed and $\uptheta_{\rm offset}$ remained fairly constant at $30 - 140^{\circ}$.

Finally, m12b (bottom left panel) is unique among our suite, because the primary merger occurred late, $\sim 2.9 \Gyr$ ago.
By this time, the host's disk already formed and established itself, and because primary merger was only $< 20$ per cent of the stellar mass of the host, the merger only moderately torqued the orientation of the host's disk.
m12b provides our single example that a \textit{late-time} merger can drive significant prograde bias as well.

The amount of disk precession since it stabilized varies across all galaxies.
The change in offset angle defined above has no distinct trend in the direction of change, that is, in some hosts $\uptheta_{\rm offset}$ increased over time and in others $\uptheta_{\rm offset}$ decreased or remained constant.
The changes in $\uptheta_{\rm offset}$ range from $\sim0-115^{\circ}$, with a median of $25^{\circ}$ and 68 per cent scatter of $\sim15-35^{\circ}$.
We find comparable results for the change in the orientation of the disk from its $z = 0$ orientation to when it stabilized, ranging from $\sim5-130^{\circ}$, with a median of $15^{\circ}$ and 68 per cent scatter of $\sim10-45^{\circ}$.
Thus, we find moderate precession of the disk since its direction stabilized.

As we discuss in further detail in Section~\ref{sec:corr}, the strength of the prograde bias does not correlate with the gas or stellar masses of the primary mergers, nor the gas or stellar mass ratios of the primary mergers to their host galaxies.
At first glance, it may seem surprising that there is no correlation of these merger/host properties with the prograde bias because these mergers largely source the prograde metal-poor population, however, as we mentioned above, they also deposit gas that contributes to the formation of the host's disk and set the prograde direction.
These lacks of correlations hold true across different times (when the primary merger occurred, when the primary merger was at its peak stellar mass, $300\Myr$ before the primary merger) and across different ways we spatially and kinematically select gas particles in the galaxies.

We also checked whether the orbits of the primary, secondary, and tertiary mergers were aligned with each other.
Specifically, we computed the orbital angular momentum vector of each merging galaxy prior its merging into the host, and we computed the offset angles between the primary and secondary, the primary and tertiary, and the secondary and tertiary.
The majority of these offset angles (32 out of 36 pairs) were $\gtrsim 25^{\circ}$, that is, not particularly aligned.
We also checked if host galaxies with more aligned mergers have larger prograde biases, but we found no clear correlation between the alignment of the mergers and the strength of the prograde bias (p-values range from 0.21 - 0.56 for the three offset angles).

\subsection{Correlations with galaxy properties}
\label{sec:corr}

\begin{figure*}
\centering
\begin{tabular}{c @{\hspace{-0.1ex}} c}
\includegraphics[width=0.48\linewidth]{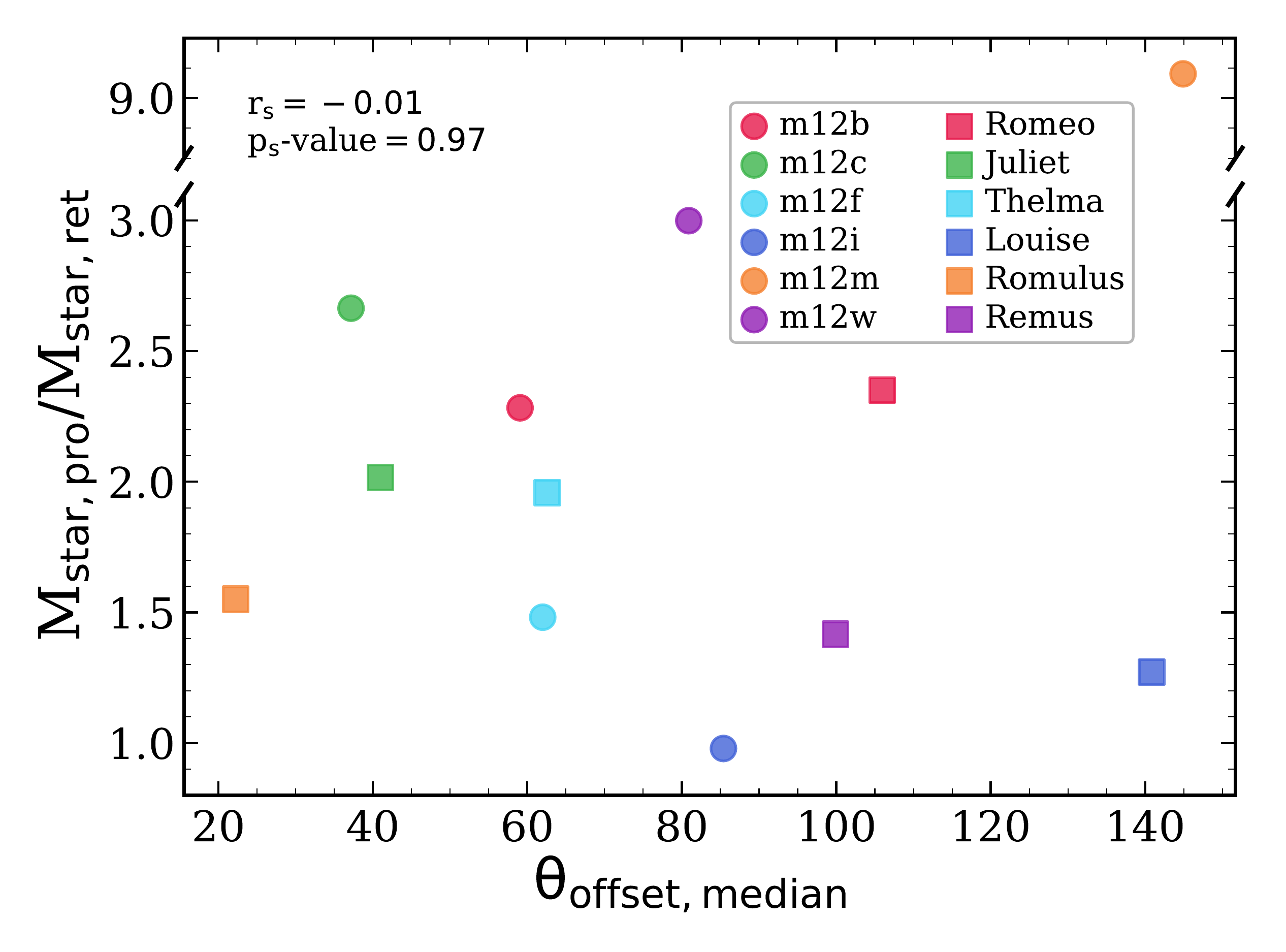}&
\includegraphics[width=0.48\linewidth]{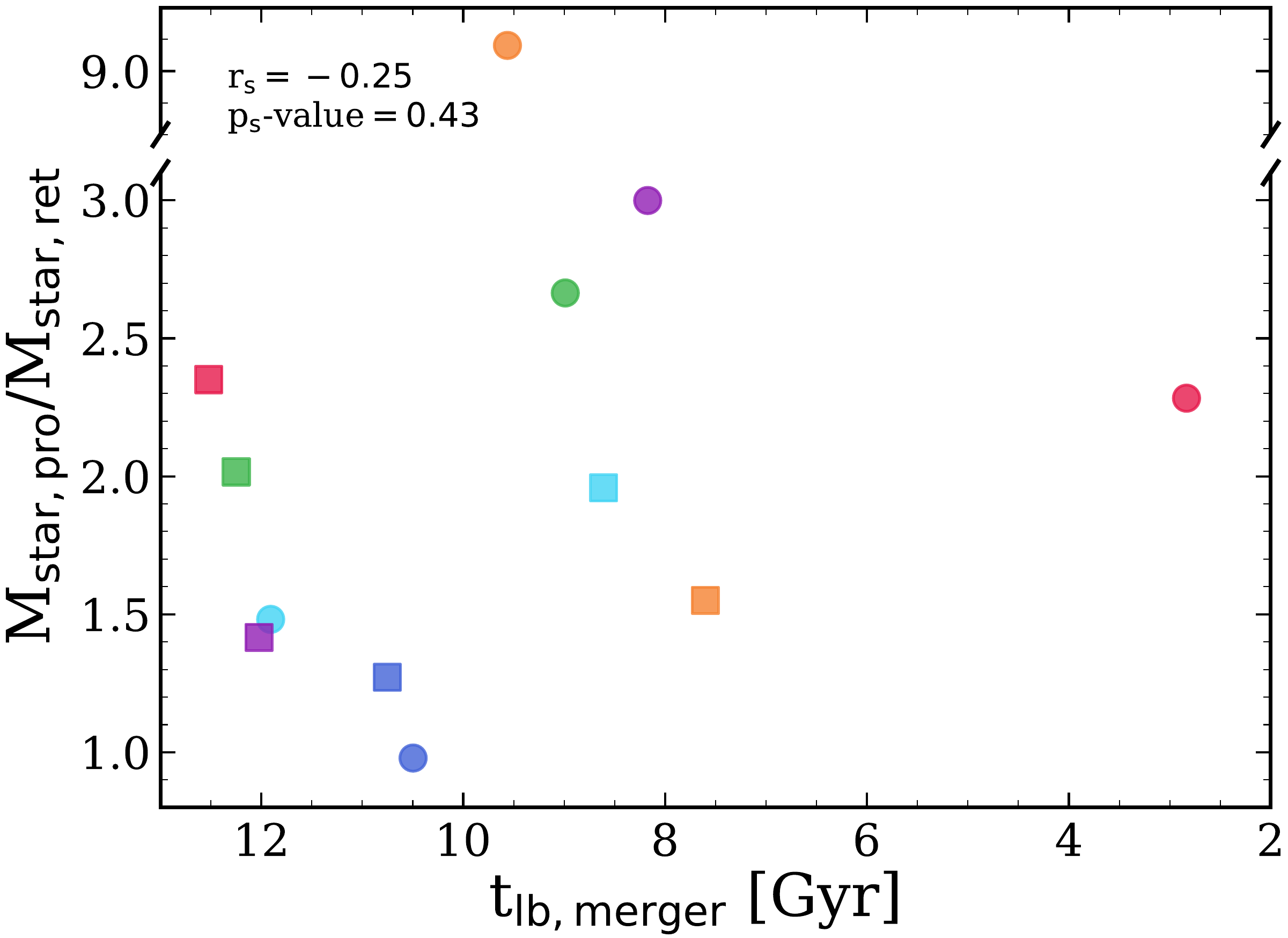}\\
\includegraphics[width=0.48\linewidth]{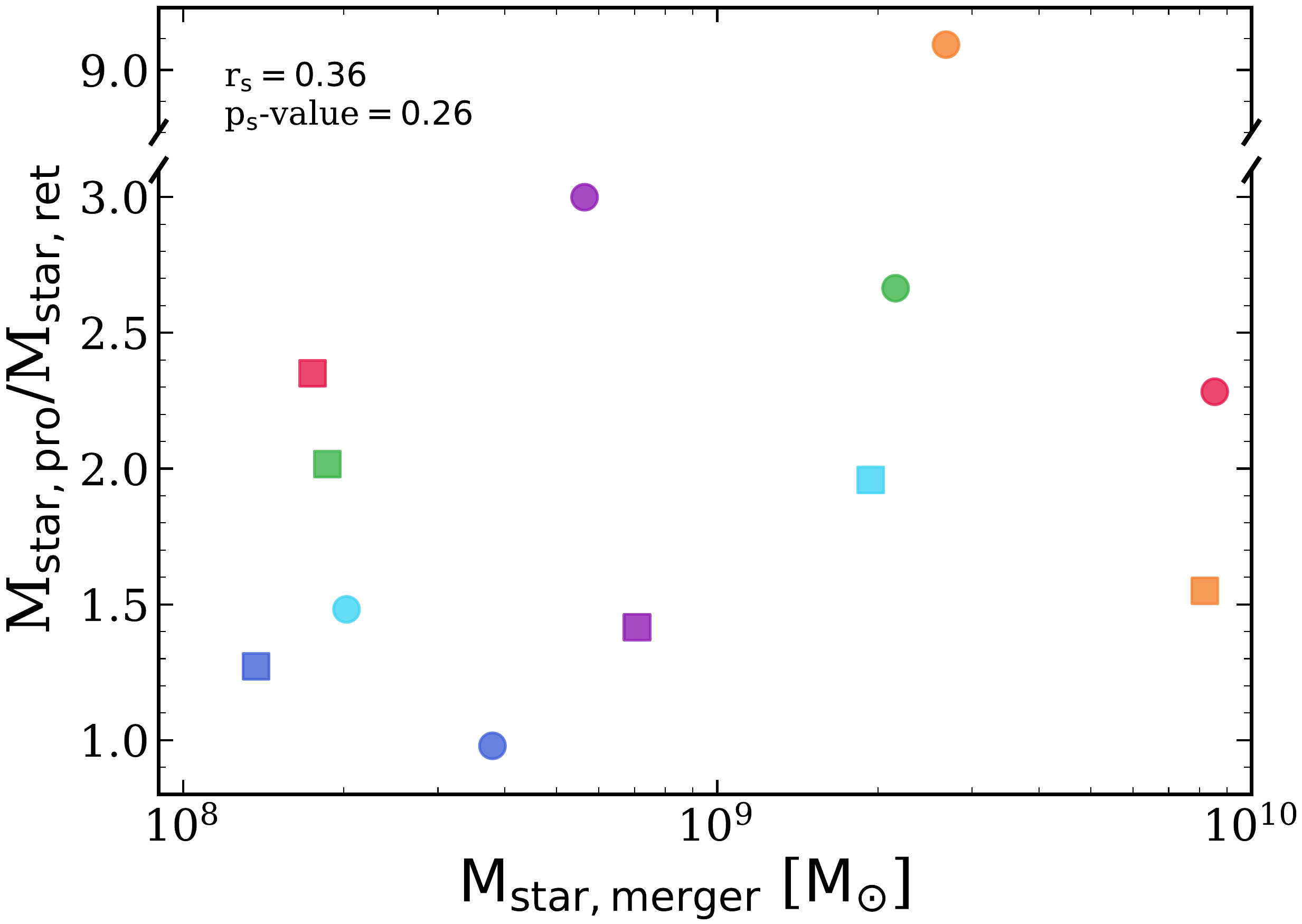}&
\includegraphics[width=0.48\linewidth]{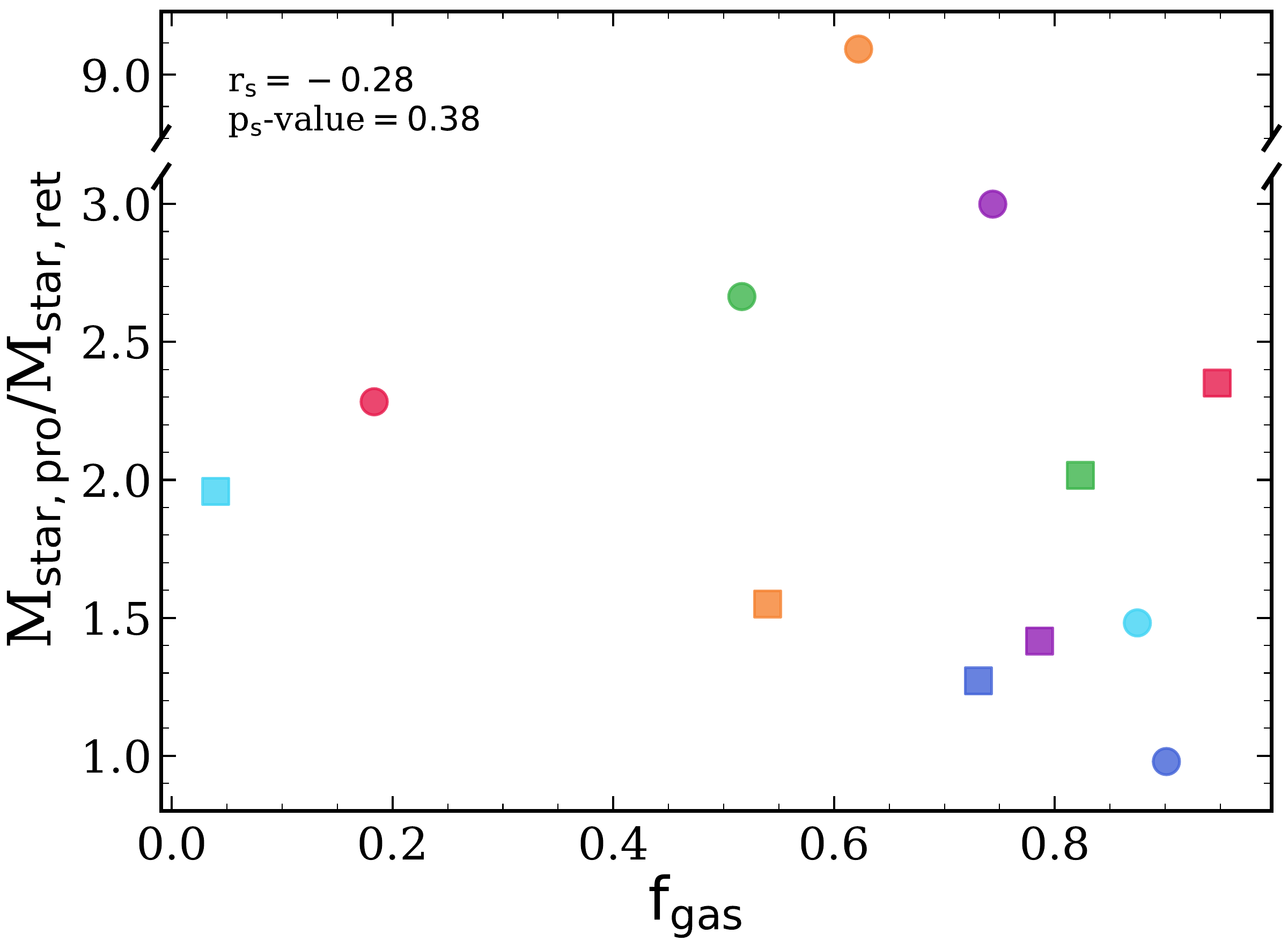}\\
\end{tabular}
\vspace{-2 mm}
\caption{
The prograde bias of metal-poor ($\FeH < -2.5$) stars in each MW/M31-mass galaxy at $z = 0$ versus various properties of its primary merger galaxy.
While we do not find strong correlations for these properties, these panels indicate the range of primary mergers to highlight their diversity.
\textbf{Top left}:
the median offset angle between the angular momentum vector of the host's disk and the angular momentum vector of the primary merger's orbit, $\uptheta_{\rm offset, median}$, over $500 \Myr$ prior to the merger.
$\uptheta_{\rm offset, median}$ indicates how prograde the merger's orbit was with respect to the host's existing disk, with 0$^{\circ}$ indicating complete alignment.
The prograde bias does not correlate with $\uptheta_{\rm offset, median}$; rather, the primary merger drove the orientation of the host's disk after the merger.
\textbf{Top right}:
the lookback time when the primary merger occurred, $t_{\rm lb,merger}$.
Most mergers occurred $7.5 - 12.5 \Gyr$ ago, although m12b is an example of a late merger, only $\approx 3 \Gyr$ ago.
We find a weak negative correlation with $t_{\rm lb,merger}$, such that earlier primary mergers cause a lower prograde biases, likely because the host galaxies have had more time to merge with other galaxies, which can phase-mix the stars.
\textbf{Bottom left}:
the stellar mass of the primary merger, $M_{\rm star,merger}$, which ranges from $10^8 - 10^{10} \Msun$, and which ranges from $0.1 - 1 \times$ the stellar mass of the host at the time of the merger.
The prograde bias increases weakly with $M_{\rm star,merger}$, as expected.
\textbf{Bottom right}:
the gas fraction of the primary merger, $f_{\rm gas,merger} = M_{\rm gas,merger} / \left( M_{\rm gas,merger} + M_{\rm star,merger} \right)$.
This ranges from $\approx 0.05 - 0.95$, though most (10 of 12) primary mergers have $f_{\rm gas,merger} > 0.5$.
The prograde bias decreases weakly with $f_{\rm gas,merger}$.
Each panel lists the Spearman correlation coefficient and p-value: we find no strong correlation between the prograde bias and these or almost any other property that we tested (see Table~\ref{tab:corr}).
Thus, while metal-poor stars nearly ubiquitously prefer prograde orbits in our simulations, the strength of this prograde bias has a complex dependence on formation/merger history.
}
\label{fig:merger_props}
\end{figure*}

Finally, we investigate correlations between properties of the MW/M31-mass host galaxies, or their galaxy mergers, and their prograde biases.
Figure~\ref{fig:merger_props} shows distributions of key properties of the primary mergers and lists the Spearman correlation coefficients and p-values with respect to the prograde bias.
Table~\ref{tab:corr} lists all of the correlations that we tested.

Given that both in-situ stars and the primary merger contribute significantly to the prograde stars, we explore whether hosts whose primary merger's orbit was more aligned with its disk orientation \textit{before the merger} show a stronger prograde bias.
Specifically, we examine the median $\uptheta_{\rm offset}$ over the $500 \Myr$ before the primary merger, and Figure~\ref{fig:merger_props} (top left) shows its relation to the prograde bias at $z = 0$ for all hosts.
$\uptheta_{\rm offset,median}$ ranges from $\sim 20 - 145^{\circ}$, with Romulus having the smallest and m12m having the largest offset.
In particular, m12m has both the largest pre-merger offset angle and the largest prograde bias.
m12w and m12i have values in the middle of the sample, even though they have the second-most and least prograde biases, respectively.
\textit{We thus conclude that the orientation of the primary merger's orbit with the host's pre-existing disk has no significant role in driving prograde bias for metal-poor stars.}

Figure~\ref{fig:merger_props} (top right) shows the prograde bias versus the lookback time of the primary merger.
Almost all primary mergers occurred $> 7 \Gyr$ ago, with the sole exception of m12b, whose merger occurred $\approx 3 \Gyr$ ago.
We find a weak correlation that hosts with an earlier primary merger tend to have a smaller prograde bias, although this correlation is weak.
This correlation may arise because hosts with earlier mergers have had more time to subsequently phase-mix their star via subsequent mergers, for example.

Figure~\ref{fig:merger_props} (bottom left) shows the prograde bias versus the stellar masses of the primary merger.
Because stars can start to be stripped from the merging galaxy well before coalescence, we measure the peak stellar mass of the merging galaxy throughout its history.
Among our sample, the peak stellar mass of the primary merger spans $10^8 - 10^{10} \Msun$, roughly evenly in log mass.
Given the stellar masses of the host galaxies at these merger times \citep{Santistevan20}, the primary merger galaxy was $\sim 7 - 95$ per cent as massive as the host during the merger.
As expected, the prograde bias increases with the stellar mass of the primary merger, but the strength of this correlation is weak, as is the correlation with the the ratio of stellar masses between the primary merger and the host (see Table~\ref{tab:corr}).

We also examine whether the gas content of the primary merger correlates with the prograde bias.
The prograde metal-poor stars formed $\gtrsim 12 \Gyr$ ago, before the primary merger occurred, so the gas content during the merger event did not contribute to this population.
Rather, the (more metal-rich) gas deposited by the merger could contribute to the formation/stabilization of the host's disk, because the merger deposited significant gas mass on a single angular momentum vector.
Thus, one might expect that higher gas content in the primary merger would drive stronger prograde-ness of the deposited metal-poor stars with respect to the resultant disk.
Because, gas can be stripped as the merger orbits around and coalesces with the host, we measure the gas mass of the merging galaxy $300 \Myr$ before merging.
We include all gas that is within $R_{\rm star,90}$, the radius that encloses 90 per cent of the stellar mass of the merging galaxy, and that has relative velocity $< 2 \times \sigma_{\rm star}$, the stellar velocity dispersion of the merging galaxy.
Most of these primary mergers were gas-rich, bringing in between $\approx 10^8 - 10^{10} \Msun$ of gas.
We then calculate their gas fractions as $f_{\rm gas,merger} = M_{\rm gas,merger} / (M_{\rm gas,merger} + M_{\rm star,merger})$.

Figure~\ref{fig:merger_props} (bottom right) shows the prograde bias versus the primary merger's $f_{\rm gas,merger}$.
Most (10 of 12) primary mergers were gas-rich, with $f_{\rm gas,merger} > 0.5$, and more than half had values $> 0.75$.
The two exceptions are m12b, which merged most recently, and Thelma, whose primary merger occurred $\sim 9 \Gyr$ ago but still was gas poor.
Perhaps surprisingly, we find that more gas-rich primary mergers correspond to slightly weaker prograde bias, though again this correlation is weak.
For example, m12i has the smallest prograde bias but also the second-highest $f_{\rm gas,merger}$ ($\approx 0.9$), while m12m has the largest prograde bias but also an intermediate $f_{\rm gas,merger}$.

\begin{table}
	\centering
	\begin{threeparttable}
	\caption{
Properties that we tested for a correlation with the prograde bias of metal-poor stars at $z = 0$.
Column list: 
property name;
Spearman correlation coefficient, r$_{\rm s}$;
Spearman p-value.
We calculate the mass properties/ratios at the following different times: present-day ($t_0$); at the time of the primary merger ($t_{\rm merger,1}$); at the time the primary galaxy merger had its peak stellar mass ($t_{\rm merger,\ peak}$); and 300 Myr before the primary merger ($t_{\rm 300}$).
}
\begin{tabular}{|l@{\hspace{-0.1 ex}}|c@{\hspace{-0.1ex}}|c|}
		\hline
		\hline
		Property & $\rm r_s$ & $\rm p_s$ value \\
		\hline
		$f_{\rm in-situ}$ & -0.11 & 0.73 \\
        $f_{\rm merger,1}$ & 0.31 & 0.33 \\
        \hline
        $M_{\rm star,\ host}(t_0)$ & 0.31 & 0.33 \\
        $M_{\rm halo, host}(t_0)$ & 0.10 & 0.76 \\
        $M_{\rm gas, host}(t_{\rm 300})$ & 0.23 & 0.47 \\
        $M_{\rm star, host}/  M_{\rm halo, host}(t_0)$ & 0.39 & 0.22 \\
        $M_{\rm star,\ merger}(t_{\rm merger,\ peak})$ & 0.36 & 0.26 \\
        $M_{\rm gas,\ merger}(t_{\rm 300})$ & -0.09 & 0.78 \\
        $M_{\rm star,\ merger} / M_{\rm star, host}(t_{\rm merger})$ & 0.20 & 0.53 \\
        $M_{\rm star, merger} / M_{\rm star, host}(t_{\rm merger, peak})$ & 0.06 & 0.85 \\
        $M_{\rm gas,\ merger}/M_{\rm gas,\ host}(t_{\rm 300})$ & -0.57 & 0.05 \\
        $M_{\rm gas,\ total}(t_{\rm 300})$ & 0.18 & 0.57 \\
        $f_{\rm gas, \ merger}$ & -0.28 & 0.38 \\
        \hline
        $t_{\rm lb,\ merger}$ & -0.25 & 0.43 \\
        $t_{\rm lb,\ 50}$ & -0.43 & 0.16 \\
        $t_{\rm lb,\ 90}$ & -0.28 & 0.38 \\
        $t_{\rm lb,\ form}$ & -0.34 & 0.27 \\
        \hline
        $N_{\rm sat}$ & 0.52 & 0.09 \\
        Primary merger ordering & 0.31 & 0.32 \\
        $\upsilon_{\rm rot}/\sigma$ & 0.29 & 0.35 \\
        $\uptheta_{\rm offset, median}$ & -0.01 & 0.97 \\
		\hline
		\hline
\end{tabular}
\label{tab:corr}
\end{threeparttable}
\end{table}

In total, we examined correlations of the prograde bias at $z = 0$ with 21 properties of the host (both today and near the time of the primary merger) or of the primary galaxy merger, which we show in Table~\ref{tab:corr}.
Because we measure each property in Table~\ref{tab:corr} at different times, we calculate the Spearman correlation coefficient ($r_s$) and p-value at present-day ($t_0$), at the time of the primary merger ($t_{\rm merger,1}$), at the time the primary galaxy merger had its peak stellar mass ($t_{\rm merger,peak}$), and $300\Myr$ before the primary merger ($t_{\rm 300}$).
We briefly address each below:

\begin{itemize}
    \item $f_{\rm in-situ}$ and $f_{\rm merger,1}$:
    While $f_{\rm merger,1}$ does positively correlate with the prograde bias, the strength of its correlation is weak, indicating additional dependence, such as time and gas-richness of the merger.
    Importantly, $f_{\rm in-situ}$ shows no meaningful correlation with the prograde bias.
    \item The ratio of stellar mass of the primary merger and the host does not correlate significantly with prograde bias.
    We tested measuring this ratio both at the time when the primary merger reached its peak stellar mass ($M_{\rm star, merger} / M_{\rm star, host}(t_{\rm merger, peak})$, and at the time immediately before merging into the host galaxy ($M_{\rm star, merger} / M_{\rm star, host}(t_{\rm merger})$.
%
%
    \item $M_{\rm gas, merger} / M_{\rm gas, host}(t_{\rm 300})$:
    the ratio of the gas mass in the primary merger to that in the host galaxy.
    We explored four ways of measuring gas masses, including $300 \Myr$ or $600 \Myr$ before the primary merger, using only cold gas ($T < 2 \times 10^4 \rm K$), or measuring gas out to $2 \times R_{\rm 90}$.
    Interestingly, this ratio $300 \Myr$ before the merger is the most significant correlation that we find (p-value = 0.05).
    However, we caution the reader to over-interpreted this result, because the strength of this correlation varies greatly with how we measure gas masses (p-values = 0.23-0.78).
    In particular, the gas mass of the primary merger is relatively robust across our methods, but the gas mass of the host galaxy varies more significantly, which may be because of ambiguity in mass association during or just before a merger.
    \item Host mass:
    $M_{\rm star,host}(t_0)$, $M_{\rm halo,host}(t_0)$, $M_{\rm gas,host}(t_{\rm 300})$, and $M_{\rm star,host} / M_{\rm halo,host}(t_0)$, correlate at most weakly with prograde bias.
    The strongest correlation is the current ratio of stellar mass to dark-matter halo mass.
%
    \item $M_{\rm gas,total}(t_{\rm 300})$:
    the total gas mass, summing that of the primary merger and the host galaxy, does not correlate significantly with prograde bias.
%
%
    \item $t_{\rm lb,50}$, $t_{\rm lb,90}$, and $t_{\rm lb, form}$:
    we investigated three metrics of host galaxy formation: the time when the galaxy formed 50 per cent of its final mass, the time when it reached 90 per cent of its final mass, and the time when it transitioned from having primarily ex-situ to primarily in-situ stellar mass.
    All of these correlate negatively with prograde bias, however, the significance of these correlations remains weak.
    $t_{\rm lb,50}$ correlates most strongly, perhaps because it correlates best with typical times of the most important mergers.
    Despite differences in the formation histories between galaxies in LG-like pairs, and galaxies in isolated environments mentioned in \citet{Santistevan20}, and the small positive correlation between the primary merger times ($t_{\rm lb, merger}$) and host galaxy formation times ($t_{\rm lb,form}$; p-value=0.05), we found no significant differences in the metal-poor, prograde biases between LG-like and isolated host galaxies, which suggests prograde bias does not correlate with host galaxy formation time.
    \item $N_{\rm sat}$:
    the number of satellite dwarf galaxies ($M_{\rm star} > 10^5\Msun$) within the host halo at $z = 0$ is curiously one of the strongest correlation that we find (p-value = 0.09).
    It remains unclear whether this is physically meaningful or a statistical fluke among our modest sample size.
    Taken at face value, it would imply that M31 has an even stronger prograde bias than the MW.
    \item Primary merger ordering:
    we compare the relative timing of the primary merger to the secondary or tertiary mergers, that is, whether the primary merger occurred first, second, or third, which we define as the primary merger ordering.
    We find only a weak correlation, such that hosts whose primary merger happened the most recently out of the 3 have a somewhat stronger prograde bias.
%
    \item $\upsilon_{\rm rot} / \sigma$: 
    the ratio of the rotational velocity to the velocity dispersion of stars in the host at $z = 0$, $\upsilon_{\rm rot} / \sigma$ is a kinematic metric of `diskiness'.
    We calculated both $\upsilon_{\rm rot}$ and $\sigma$ using all stars in the disk, however, we also calculated the ratio using only stars that are metal-poor (threshold from $-3.5 < \FeH < -0.5$).
    We find a marginal increase in prograde bias with $\upsilon_{\rm rot} / \sigma$, but it is weak.

\end{itemize}


\section{Summary \& Discussion} %
\label{sec:sumndisc}            %

\subsection{Summary} %
\label{sec:sum}      %
We investigated the prevalence and origin of the preference of metal-poor stars to be on prograde disk orbits using 12 MW/M31-mass galaxies from the FIRE-2 simulation suite.
We reiterate the questions that we articulated in the introduction, and we summarize our answers to them: \\


\begin{enumerate}
    \item \textit{How commonly do MW/M31-mass galaxies show a preference for prograde motions in their metal-poor stars, as observed in the MW, and what is the range in strength of this prograde bias?}
    \begin{itemize}
        \item 11 of 12 of our MW/M31-mass galaxies show a preference for prograde orbits among metal-poor stars (Figure~\ref{fig:selection}).
        \item Our simulations predict that this prograde bias is a general feature of MW/M31-mass galaxies throughout most of their history, and its presence does not depend significantly on the way that we spatially or kinematically select metal-poor stars (Figures~\ref{fig:metal}, \ref{fig:h3}, \ref{fig:age}).
        \item The present-day prograde biases for our sample range from $0.98 - 9.14$, with a median value of $2.0$ (Table~\ref{tab:hosts}).
        \item We find no significant differences in prograde bias  between LG-like and isolated host galaxies (Figure~\ref{fig:metal}).
    \end{itemize}
    
    \item \textit{How does this prograde bias depend on the metallicity, distance, and/or age of the stars?}
    \begin{itemize}
        \item We find little-to-no dependence of the prograde bias on the iron abundance of stars at $\FeH \lesssim -1$ ( Figure~\ref{fig:metal}).
        \item Our results broadly agree with recent observations of the MW from the Pristine and H3 surveys (Figures~\ref{fig:metal} \& \ref{fig:h3}).
        \item We find typically a similar prograde bias, $\sim 2 - 2.5$, for stars at distances $\gtrsim 20 \kpc$ in the stellar halo (Figure~\ref{fig:halo}).
        \item 11 of 12 of our galaxies have a prograde bias for arbitrarily old stars ($\gtrsim 13.5 \Gyr$) regardless of metallicity, and all hosts have a significant prograde bias for stars $\lesssim 12.5 \Gyr$ old (Figure~\ref{fig:age}).
    \end{itemize}
    
    \item \textit{What process(es) cause this prograde bias among metal-poor stars?}
    \begin{itemize}
        \item We find no large difference in the prograde bias for stars that formed in-situ versus ex-situ (Figure~\ref{fig:metal}).
        The sample of prograde metal-poor stars (both in-situ and ex-situ) almost entirely formed $\gtrsim 12 \Gyr$ ago, prior to the primary mergers.
        \item A \textit{single} galaxy merger is typically the most important cause of the prograde bias, with additional but sub-dominant contributions from in-situ star formation and lesser mergers (Figure~\ref{fig:merger_vs_insitu}).
        This primary merger typically contributed $\sim 24$ per cent of all prograde metal-poor stars, while the fraction that formed in-situ is $\sim 16$ per cent.
        Combining in-situ stars with those from the top 3 mergers accounts for $\sim 70$ per cent, on average (Table~\ref{tab:hosts}).
        \item We do not find any significant correlation in orientation between the primary merger's orbit and any pre-existing disk in the host (Figure~\ref{fig:asym_offset}).
        \item The primary merger occurred typically $7 - 12.5 \Gyr$ ago and typically was gas rich ($f_{\rm gas,merger} \gtrsim 0.5$).
        It thus deposited a significant population of old metal-poor stars and significant gas into the host on the same orbital vector, which typically seeded/shaped the formation of a long-lived disk in the host, giving rise to metal-poor stars preferentially on a prograde disk orbit (Figures~\ref{fig:asym_offset} \& \ref{fig:merger_props}).
        \item In 3 of our simulated hosts, the host stellar disk orientation became settled/long-lived within $\sim500\Myr$ of 1 of the 3 mergers we tracked, and in two cases, the orientation became settled, but it subsequently changed via other mergers (Figure~\ref{fig:asym_offset}).
        This suggests that mergers can be a dominant factor in the formation of stellar disks.
%
    \end{itemize}
    
    \item \textit{Do any properties of the MW/M31-mass galaxy or their galaxy mergers correlate with this prograde bias?}
    \begin{itemize}
        \item We explored correlations with 21 properties of the host galaxy or primary merger, but we find few clear correlations with the strength of the prograde bias (Table~\ref{tab:corr}).
        This indicates that the strength of the prograde bias has a complicated dependence on galaxy formation/merger history, which limits our immediate interpretation of how to translate recent observations of the prograde bias for the MW into a robust constraint on its formation history.

    \end{itemize}
\end{enumerate}

\subsection{Discussion} %
\label{sec:disc}         %

A few galaxy mergers predominantly drive the growth of the metal-poor component of MW/M31-mass galaxies/halos.
For example, using a suite of dark-matter-only simulations of MW-mass halos and applying abundance matching to assign stellar mass to (sub)halos, \citet{Deason16} found that most of the accreted stellar mass in the stellar halo comes from only 1-2 dwarf galaxies with $M_{\rm star} = 10^8 - 10^{10}\Msun$.
Similarly, in \citet{Santistevan20} we used the same 12 simulated galaxies in this paper and found that stars that form ex-situ (in galaxies other than the most massive progenitor) and accrete into the host galaxy largely come from the most massive progenitor galaxies, and that this is true for progenitors at all redshifts.
Because just a few galaxies are responsible for most of the accreted (metal-poor) population, they leave a kinematic imprint, as we have shown.
Accreted substructures populate our own galaxy, such as Gaia-Enceladus \citep{Helmi18, Belokurov18}, the Sagittarius stream \citep{Ibata94, Newberg03, Majewski03}, Sequoia \citep{Barba19, Myeong19, Matsuno19}, and a few others \citep[e.g. see][]{Naidu20}.

Although stars in Gaia-Enceladus are on primarily radial orbits, they may have some net rotation.
\citet{Helmi18}, \citet{Belokurov18}, \citet{Naidu20}, and \citet{Naidu21} all suggest that even though the mean rotational velocity of Gaia-Enceladus' stars is close to 0, it may exhibit a small degree of retrograde motion.
However, this is further complicated by the possible correlation of Gaia-Enceladus with other substructures that populate the MW's stellar halo.
\citet{Naidu20} lists the currently known substructures, and details 3 prograde structures (Sagittarius \& Helmi streams, Wukong) and 4 retrograde structures (Arjuna, Sequoia, I'itoi, Thamnos).
It remains unclear whether or not Thamnos, Sequoia, or I'itoi are associated with Gaia-Enceladus, or with each other.
Ignoring Gaia-Enceladus stars in their sample, \citet{Naidu20} report that the number of prograde to retrograde stars is $\sim 3/2$ within the stellar halo, which qualitatively agrees with our results, though is weaker than our median value in Figure~\ref{fig:halo}.
However, the Sagittarius stream dominates this signal in the MW, and when \citet{Naidu20} do not include Sagittarius stars, they find no prograde bias in the stellar halo.

Because the MW shows a clear prograde bias for metal-poor stars, at least near its disk region, then the results above, combined with our results, imply one or more of the following: (i) one or more of the mergers above imparted a prograde bias to the MW, but their stars that populate the stellar halo (beyond the disk region) have since radialized; (ii) the prograde bias in the MW originated from a yet-to-be discovered merger, perhaps whose remnant stars populate the inner halo; (iii) the prograde bias in the MW is entirely in-situ and not associated with any merger.
However, our results suggest that the latter scenario (alone) is rare.

In their analysis of the prograde bias in the H3 survey, \citet{Carter20} suggest that metal-poor stars likely came from either (1) a combination of in-situ and ex-situ star formation during the early assembly and formation of the MW, which is broadly similar to what we find, or (2) late accretion of dwarf galaxies on prograde orbits with respect to the disk.
However, because the stars in the MW have not been kinematically heated enough to change their prograde orbits, the authors claim that the angular momentum of the MW has been in place for at least 12 Gyr.
From the bottom sub-panels of Figure~\ref{fig:asym_offset}, we see find changes in $\uptheta_{\rm offset}$ at early times, with a sharp transition, after which the angle slowly precesses over time.
This transition marks when the stellar disk orientation stabilized in the host, and although the disk can precess afterwards, $\uptheta_{\rm offset}$ remains relatively stable.
We find a range of times when this transition occurs, across $\approx 5 - 11.5 \Gyr$ ago.
Romeo is the host galaxy that has the oldest transition akin to the MW, at $\sim 11.5 \Gyr$ ago.
However, in most of our hosts, this transition occurs more recently, which suggests that the MW may be uncommon in this regard if the angular momentum of the disk has been stable for the last 12 Gyr.
If instead we compare the $\hat{z}$ orientation of the disk at $z = 0$ to the orientation at early time, we see the same qualitative behavior in Figure~\ref{fig:asym_offset}: rapid changes likely driven by early assembly, then a sharp transition to a more gradual change from disk precession.
Changes in both $\uptheta_{\rm offset}$ and in the disk's orientation, from the time that these rapid changes settle to present-day, vary between $\sim0-130^{\circ}$ and have similar median values of $25^{\circ}$ and $15^{\circ}$, respectively.

\citet{Naidu21} recently aimed to reconstruct the merger of Gaia-Enceladus with the MW.
By comparing observational data from the H3 survey with $N$-body simulations, the authors constrained the orbital properties of the merger, such as the inclination, circularity, and direction of the orbit, as well as the stellar mass and size of the galaxy.
Interestingly, the authors conclude that Gaia-Enceladus was on an initial retrograde orbit and subsequently was radialized over time via dynamical friction, but they also show the mean angular momentum of Gaia-Enceladus stars as a function of galacto-centric distance.
Figure 15 of \citet{Naidu21} suggests that Gaia-Enceladus stars within $d_{\rm host} < 25\kpc$ are only slightly retrograde today, but beyond $25 \kpc$ the debris becomes increasingly more retrograde.
One caveat of their analysis is that the authors do not account for the amount of disk precession over time; we see precession between $5-130^{\circ}$ from the time the disk stabilizes to its $z=0$ orientation.
The results from \citet{Carter20} suggested that the stellar halo stars in the H3 survey show a prograde bias, similar to our results.
Our Figure~\ref{fig:halo} also shows that the stellar halo stars in our simulated galaxies have a \textit{prograde} bias that increases to $\sim2-2.5$ at $d_{\rm host} = 45\kpc$.
However, an important clarification to make is that we show \textit{all} stars in Figure~\ref{fig:halo} (and in Figures~\ref{fig:metal}-\ref{fig:age}), not just stars in a single merger.
This motivates future work to more carefully test our results (and those in \citet{Carter20}) in the context of individual merger remnants, as in \citet{Naidu21}.

Although mergers are violent processes, they can result in the formation of a disk \citep[e.g.][]{Springel05, Robertson06, Sparre17, Tapia17} or change the properties of existing disks \citep[e.g.][]{Toomre72, Bournaud04, Bournaud05, Hopkins09}.
For example, following a series of papers aimed at understanding disk formation through major mergers \citep{Athanassoula16, Rodionov17, Peschken17}, \citet{Peschken20} analyzed galaxies in the Illustris simulation \citep{Illustris} and showed that even though 38 of them experienced a $\gtrsim$ 4:1 merger since $z \sim 1.5$, they eventually formed disks by $z = 0$.
The most important factor in determining the final outcome of such a merger is the amount of gas, given that stars can form from this gas after the merger to build a disk, while the older pre-existing stars can be heated during the merger to form a thick-disk/halo/ellipsoidal component in the galaxy.
In our simulated sample, 300 Myr prior to the primary mergers, the gas mass ratios of the primary merging galaxy to the host span values from $\sim 0.002$ in Thelma, to as high as $\sim 1.6$ in Louise.
Similarly, the total gas masses (primary merging galaxy + host) 300 Myr prior to the primary mergers span values from $\sim 5 \times 10^8 - 4 \times 10^{10} \Msun$.
In 3 of our hosts (m12m, m12w, and Romulus), the orientation of the disk settles from the rapid variations caused by early accretion/mergers within $\approx 500 \Myr$ of the three mergers we investigated here, suggesting that these mergers drove the formation and prograde orientation of the disk.
In 2 other hosts (m12c and Juliet), 1 of the 3 mergers caused the orientation of the disk to settle, but other subsequent mergers caused the disk to then change orientation again.
Our results qualitatively agree with \citet{Peschken20} and suggest that mergers were a dominant mode of promoting the transition to thin-disk morphology, with much of the gas mass mentioned above likely contributing to subsequent star formation in the disk region.

Recently, \citet{Sestito_new} performed a similar study as ours using 5 of the NIHAO-UHD cosmological zoom-in simulations of MW-mass galaxies \citep{NIHAO_UHD}.
Building on their previous work, they found that the simulated galaxies also show a prevalence for a prograde bias.
The authors also showed that a rotating spheroid of metal-poor stars, akin to a halo-like or pressure supported distribution of stars from the early accretion phase, cannot explain the origin of the prograde bias in a galaxy.
Rather, the early accretion and mergers of progenitor galaxies is the dominant source of metal-poor planar stars, with a smaller additional contribution from later accretion of satellite galaxies on prograde orbits that get dragged and disrupted into the proto-disk.
Furthermore, they suggest that $\sim 88 - 93$ per cent of metal-poor stars ($\FeH < -2$) in those simulations are older than $\sim 12 \Gyr$, where the retrograde planar population traces the early accretion of the main galaxy and the prograde planar population better traces the full history.
Our results qualitatively agree with this picture, and we discuss the similarities and differences below.

Both our analysis and \citet{Sestito_new} find that this population of metal-poor stars within the disks of MW-mass galaxies has a preference of prograde motion as opposed to retrograde.
\citet{Sestito_new} find a prograde bias in all 5 NIHAO-UHD simulated galaxies, and we find a prograde bias in 11 of our 12 galaxies.
This reinforces that the prograde bias for metal-poor stars is a nearly (but not completely) ubiquitous feature of MW/M31-mass galaxies.
Our sample of 12 host galaxies is larger than that in \citet{Sestito_new}, which is likely why we find one host without a prograde bias.
The strengths of the prograde biases differ significantly between our analysis and theirs.
\citet{Sestito_new} find prograde biases of $\sim 3.5 - 9.1$, higher than the MW, while our simulations have $0.98 - 9.1$, with a median value nearly identical to the MW.
In particular, 11 of our hosts have prograde biases $\lesssim 3$, below the smallest prograde bias in \citet{Sestito_new}, which may indicate the importance on these results from differences in physical and/or numerical implementations between our FIRE-2 and the NIHAO-UHD simulations.


We qualitatively agree about the origins of prograde metal-poor stars.
These star particles primarily came from a number of mergers, both from the early accretion and late growth phases.
Even though a non-negligible fraction of these stars form in-situ ($5 - 30$ per cent) in our simulations, they are too old to have formed in a disk-like component within the host.
In Figure 5 of \citet{Sestito_new}, massive satellite galaxies ($M_{\rm star} \gtrsim 10^8\Msun$) merged into the host on orbits aligned with the angular momentum of the stellar disk around the time the host galaxy formed 25 per cent of its final mass ($t_{25}$).
At these times, they report that $\sim 37 - 90$ per cent of the prograde planar stars are already in the host galaxy from the early assembly phase.
However, from the bottom sub-panels in Figure~\ref{fig:asym_offset}, we show that the orientation of the proto-disk is not well defined until the primary merger (or another lesser merger) occurs, implying that the primary merger helped to seed the formation and set the orientation of the host's disk by depositing a significant supply of gas on a coherent angular momentum vector.
Thus, this means that the primary merger determined the prograde direction in the host, and that these progenitor galaxies do not merge into the host on orbits aligned with the disk because the disk was not well-defined before this.
Our answer to the origin of prograde metal-poor stars in the disks of MW-mass galaxies is then most similar to explanation (iii) from \citet{Sestito19}: these stars formed inside one or more progenitor galaxies that merged into the MW-mass host as it was forming.

As a final comparison, we note important differences between our analysis and \citet{Sestito_new}.
The authors do not explore the stability of the stellar disk orientations in their Figure 5 at $t_{25}$.
Therefore, it is unclear what effect, if any, the satellite progenitor galaxies had on the orientation of the host.
Although they quantify the growth of the accreted stellar mass of the host galaxy, they do not examine the gas mass of the merging galaxies, which plays a large role in shaping the disk.
In this paper, we only focus on the evolution of the ratio of the stellar mass of prograde to retrograde stars, but \citet{Sestito_new} suggests that $74 - 90$ per cent of the retrograde planar population are primarily accreted before $t_{25}$, while the prograde population is accreted throughout most of cosmic time.
This paper also investigated several properties of the host galaxy to look for possible correlations with the strength of its prograde bias.
Finally, we do not analyze other regions in kinematic space, such as the high eccentricity feature in \citet{Sestito_new}.

Evidence for the prograde, metal-poor population in the MW disk is further supported by a recent study involving stars located in the stellar halo and thick disk from the \textit{SkyMapper Survey for extremely metal-poor stars} \citep{SkyMapper}.
In a sample of 475 stars with iron abundances in the range of $-6.5 \lesssim \FeH \lesssim -2$, \citet{Cordoni20} reported that $\sim21$ per cent of their sample have disk-like orbits confined to $3\kpc$ of the disk's midplane.
The authors suggest that this sub-sample of metal-poor stars with disk-like orbits consists of a high-eccentricity and low-eccentricity population, where the low-eccentricity population are on primarily prograde orbits (defined as $J_{\phi}>0$) with a prograde bias of $\sim3.1$.
\citet{Cordoni20} postulate that the prograde and retrograde low-eccentricity stars confined to $|Z|<3\kpc$ likely have different origins: the prograde stars possibly formed in the thick disk of the MW, while the retrograde were accreted from a disrupted satellite galaxy.
These results broadly agree with what we presented here, however, it is likely that some of the prograde, low-eccentricity stars formed both in-situ \textit{and} were accreted early in the MW's formation history.

Similar studies focused on old, metal-poor stars in the MW also halo have investigated their origin and kinematics.
By selecting stars within $3 \kpc$ of the Sun with halo-like kinematics, \citet{Bonaca17} found that there are an excess of metal-rich halo stars ($\FeH > -1$) with prograde orbits compared to the metal-poor halo stars ($\FeH < -1$) which have no orbital preference.
The authors compared this with one of the simulations in our sample, m12i, and found similar results.
They found that the two halo populations also have different origins, with the metal-rich stars primarily forming in-situ, and the metal-poor population accreting from mergers (ex-situ).
A similar study using the APOSTLE simulations by \citet{Starkenburg17} suggests that the fractions of metal-poor ($\FeH < -2.5$) and old ($t_{\rm form} < 0.8 \Gyr$) stars increase for larger galactocentric radii, with $\sim 60$ per cent of the stars orbiting outside of the solar circle.
We did not examine the prograde bias of the halo region in our simulations, but rather, we compared how the prograde bias evolution changed when selecting \textit{all} metal-poor ($\FeH < -2.5$) stars within $15 \kpc$, radially.
We saw the same qualitative evolution of the prograde biases as in the top sub-panels of Figure~\ref{fig:asym_offset}.
This is not a strict `halo selection' of stars, but given that we showed that the prograde bias is a general feature, independent of the kinematic and spatial selection windows we examined, we do not expect our results to change significantly by making a more strict selection of halo stars.
Similarly, \citet{ElBadry18} investigated the spatial distribution and dynamics of the oldest stars ($\rm z_{form} \gtrsim 5$) in MW-mass galaxies and found that the majority formed ex-situ.
The authors found that metal-rich stars ($\FeH > -1$) typically have disk-like orbits, while the more metal-poor stars reflect an isotropic distribution.
Furthermore, when only selecting the oldest stars ($z_{\rm form} \gtrsim 5$), the in-situ and ex-situ populations in m12f have nearly identical velocity distributions, however, both populations do show a slight prograde bias, which are centered at $\upsilon_{\phi}>0\kmsi$.

Instead of selecting stars based on their distributions in velocity space, as in a Toomre diagram, one recent study tested a Gaussian mixture model to deconstruct the different components of the MW using the star's velocities and iron abundances \citep{Nikakhtar21}.
The authors used data from APOGEE \citep{APOGEE} and Gaia DR2 \citep{GaiaDR2}, and created mock catalogs using 3 of the simulations in our study (m12f, m12i, m12m), and showed that the MW is best described by five different components: the stellar halo, the thin disk, the metal-rich thick disk, and a 2-component metal-poor thick disk.
\citet{Nikakhtar21} suggest that the 3-component thick disk originates from a combination of kinematic heating (via gravitational interactions) and radial migration of early forming stars: we show that a significant fraction of the most metal-poor stars in the disk come from multiple mergers during the galaxy's early assembly.
In a different approach, using the same 12 simulated galaxies as in our sample, \citet{Yu21} differentiated thin disk and thick disk stars based on their circularity (angular momenta relative to the angular momentum of a circular orbit) and suggest that the transition from thick disk to thin disk formation correlates with the transition from early, bursty star formation to steady, near constant star formation.
This transition occurred some time between $\sim 0.6 - 6.5 \Gyr$ ago, and galaxies that had earlier transitions often have older thick disk stars and more prominent thin disks.

One interesting avenue for future work is to investigate the correlation of the prograde bias with larger-scale structure.
To first order, we found no distinct differences in the metal-poor prograde bias between isolated and LG-like environments, which suggests that the prograde bias is not inherent in LG-like environments only, as in the MW.



\section*{Acknowledgements}

We greatly appreciate interesting and fruitful discussions with both Federico Sestito and Nicolas Martin, as well as Courtney Carter and Charlie Conroy.
We also express our gratitude for their sharing of observational data.
Finally, we wish to thank the reviewer, Ted Mackereth, for offering many great suggestions to strengthen the paper.

IBS, AW, and JS received support from NASA through ATP grants 80NSSC18K1097 and 80NSSC20K0513; HST grants GO-14734, AR-15057, AR-15809, and GO-15902 from STScI; a Scialog Award from the Heising-Simons Foundation; and a Hellman Fellowship.
AW performed this work in part at KITP, supported by NSF grant PHY-1748958.
CAFG was supported by NSF through grants AST-1715216 and CAREER award AST-1652522; by NASA through grant 17-ATP17-0067; by STScI through grant HST-AR-16124.001-A; and by a Cottrell Scholar Award and a Scialog Award from the Research Corporation for Science Advancement.
RES acknowledges support from NASA grant 19-ATP19-0068, NSF grant AST-2009828, and HST-AR-15809 from the Space Telescope Science Institute (STScI), which is operated by AURA, Inc., under NASA contract NAS5-26555.
We ran simulations using: XSEDE, supported by NSF grant ACI-1548562; Blue Waters, supported by the NSF; Pleiades, via the NASA HEC program through the NAS Division at Ames Research Center.

This paper used various python packages including \textsc{NumPy} \citep{Numpy}, \textsc{SciPy} \citep{SciPy}, and \textsc{Matplotlib} \citep{Matplotlib}, as well as NASA's Astrophysics Data System.


\section*{Data availability}
Full simulation snapshots at $z = 0$ are available for m12i, m12f, and m12m at ananke.hub.yt.
The python code used to analyze these data is available at \url{https://bitbucket.org/isantis/iron\_poor\_stars}, which uses the publicly available packages \url{https://bitbucket.org/awetzel/gizmo\_analysis}, \url{https://bitbucket.org/awetzel/halo\_analysis}, and \url{https://bitbucket.org/awetzel/utilities}.
Finally, data values in each figure are available at \url{https://ibsantistevan.wixsite.com/mysite/publications}.


\bibliographystyle{mnras}
\bibliography{metal_poor}

\begin{thebibliography}{}
\makeatletter
\relax
\def\mn@urlcharsother{\let\do\@makeother \do\$\do\&\do\#\do\^\do\_\do\%\do\~}
\def\mn@doi{\begingroup\mn@urlcharsother \@ifnextchar [ {\mn@doi@}
  {\mn@doi@[]}}
\def\mn@doi@[#1]#2{\def\@tempa{#1}\ifx\@tempa\@empty \href
  {http://dx.doi.org/#2} {doi:#2}\else \href {http://dx.doi.org/#2} {#1}\fi
  \endgroup}
\def\mn@eprint#1#2{\mn@eprint@#1:#2::\@nil}
\def\mn@eprint@arXiv#1{\href {http://arxiv.org/abs/#1} {{\tt arXiv:#1}}}
\def\mn@eprint@dblp#1{\href {http://dblp.uni-trier.de/rec/bibtex/#1.xml}
  {dblp:#1}}
\def\mn@eprint@#1:#2:#3:#4\@nil{\def\@tempa {#1}\def\@tempb {#2}\def\@tempc
  {#3}\ifx \@tempc \@empty \let \@tempc \@tempb \let \@tempb \@tempa \fi \ifx
  \@tempb \@empty \def\@tempb {arXiv}\fi \@ifundefined
  {mn@eprint@\@tempb}{\@tempb:\@tempc}{\expandafter \expandafter \csname
  mn@eprint@\@tempb\endcsname \expandafter{\@tempc}}}

\bibitem[\protect\citeauthoryear{{Athanassoula}, {Rodionov}, {Peschken}  \&
  {Lambert}}{{Athanassoula} et~al.}{2016}]{Athanassoula16}
{Athanassoula} E.,  {Rodionov} S.~A.,  {Peschken} N.,   {Lambert} J.~C.,  2016,
  \mn@doi [\apj] {10.3847/0004-637X/821/2/90}, \href
  {https://ui.adsabs.harvard.edu/abs/2016ApJ...821...90A} {821, 90}

\bibitem[\protect\citeauthoryear{{Barb{\'a}}, {Minniti}, {Geisler},
  {Alonso-Garc{\'\i}a}, {Hempel}, {Monachesi}, {Arias}  \&
  {G{\'o}mez}}{{Barb{\'a}} et~al.}{2019}]{Barba19}
{Barb{\'a}} R.~H.,  {Minniti} D.,  {Geisler} D.,  {Alonso-Garc{\'\i}a} J.,
  {Hempel} M.,  {Monachesi} A.,  {Arias} J.~I.,   {G{\'o}mez} F.~A.,  2019,
  \mn@doi [\apjl] {10.3847/2041-8213/aaf811}, \href
  {https://ui.adsabs.harvard.edu/abs/2019ApJ...870L..24B} {870, L24}

\bibitem[\protect\citeauthoryear{{Behroozi}, {Wechsler}  \& {Wu}}{{Behroozi}
  et~al.}{2013a}]{Behroozi13a}
{Behroozi} P.~S.,  {Wechsler} R.~H.,   {Wu} H.-Y.,  2013a, \mn@doi [\apj]
  {10.1088/0004-637X/762/2/109}, \href
  {https://ui.adsabs.harvard.edu/\#abs/2013ApJ...762..109B} {762, 109}

\bibitem[\protect\citeauthoryear{{Behroozi}, {Wechsler}, {Wu}, {Busha},
  {Klypin}  \& {Primack}}{{Behroozi} et~al.}{2013b}]{Behroozi13b}
{Behroozi} P.~S.,  {Wechsler} R.~H.,  {Wu} H.-Y.,  {Busha} M.~T.,  {Klypin}
  A.~A.,   {Primack} J.~R.,  2013b, \mn@doi [\apj]
  {10.1088/0004-637X/763/1/18}, \href
  {https://ui.adsabs.harvard.edu/\#abs/2013ApJ...763...18B} {763, 18}

\bibitem[\protect\citeauthoryear{{Belokurov}, {Erkal}, {Evans}, {Koposov}  \&
  {Deason}}{{Belokurov} et~al.}{2018}]{Belokurov18}
{Belokurov} V.,  {Erkal} D.,  {Evans} N.~W.,  {Koposov} S.~E.,   {Deason}
  A.~J.,  2018, \mn@doi [Monthly Notices of the Royal Astronomical Society]
  {10.1093/mnras/sty982}, \href
  {https://ui.adsabs.harvard.edu/abs/2018MNRAS.478..611B} {478, 611}

\bibitem[\protect\citeauthoryear{{Bird}, {Kazantzidis}, {Weinberg}, {Guedes},
  {Callegari}, {Mayer}  \& {Madau}}{{Bird} et~al.}{2013}]{Bird13}
{Bird} J.~C.,  {Kazantzidis} S.,  {Weinberg} D.~H.,  {Guedes} J.,  {Callegari}
  S.,  {Mayer} L.,   {Madau} P.,  2013, \mn@doi [\apj]
  {10.1088/0004-637X/773/1/43}, \href
  {https://ui.adsabs.harvard.edu/abs/2013ApJ...773...43B} {773, 43}

\bibitem[\protect\citeauthoryear{{Bird}, {Loebman}, {Weinberg}, {Brooks},
  {Quinn}  \& {Christensen}}{{Bird} et~al.}{2021}]{Bird21}
{Bird} J.~C.,  {Loebman} S.~R.,  {Weinberg} D.~H.,  {Brooks} A.~M.,  {Quinn}
  T.~R.,   {Christensen} C.~R.,  2021, \mn@doi [\mnras]
  {10.1093/mnras/stab289}, \href
  {https://ui.adsabs.harvard.edu/abs/2021MNRAS.503.1815B} {503, 1815}

\bibitem[\protect\citeauthoryear{{Bonaca}, {Conroy}, {Wetzel}, {Hopkins}  \&
  {Kere{\v s}}}{{Bonaca} et~al.}{2017}]{Bonaca17}
{Bonaca} A.,  {Conroy} C.,  {Wetzel} A.,  {Hopkins} P.~F.,   {Kere{\v s}} D.,
  2017, \mn@doi [\apj] {10.3847/1538-4357/aa7d0c}, \href
  {http://adsabs.harvard.edu/abs/2017ApJ...845..101B} {845, 101}

\bibitem[\protect\citeauthoryear{{Bournaud}, {Combes}  \& {Jog}}{{Bournaud}
  et~al.}{2004}]{Bournaud04}
{Bournaud} F.,  {Combes} F.,   {Jog} C.~J.,  2004, \mn@doi [\aap]
  {10.1051/0004-6361:20040114}, \href
  {https://ui.adsabs.harvard.edu/abs/2004A&A...418L..27B} {418, L27}

\bibitem[\protect\citeauthoryear{{Bournaud}, {Jog}  \& {Combes}}{{Bournaud}
  et~al.}{2005}]{Bournaud05}
{Bournaud} F.,  {Jog} C.~J.,   {Combes} F.,  2005, \mn@doi [\aap]
  {10.1051/0004-6361:20042036}, \href
  {https://ui.adsabs.harvard.edu/abs/2005A&A...437...69B} {437, 69}

\bibitem[\protect\citeauthoryear{{Brook}, {Kawata}, {Scannapieco}, {Martel}  \&
  {Gibson}}{{Brook} et~al.}{2007}]{Brook07}
{Brook} C.~B.,  {Kawata} D.,  {Scannapieco} E.,  {Martel} H.,   {Gibson} B.~K.,
   2007, \mn@doi [\apj] {10.1086/511514}, \href
  {https://ui.adsabs.harvard.edu/abs/2007ApJ...661...10B} {661, 10}

\bibitem[\protect\citeauthoryear{{Buck}, {Obreja}, {Macci{\`o}}, {Minchev},
  {Dutton}  \& {Ostriker}}{{Buck} et~al.}{2020}]{NIHAO_UHD}
{Buck} T.,  {Obreja} A.,  {Macci{\`o}} A.~V.,  {Minchev} I.,  {Dutton} A.~A.,
  {Ostriker} J.~P.,  2020, \mn@doi [\mnras] {10.1093/mnras/stz3241}, \href
  {https://ui.adsabs.harvard.edu/abs/2020MNRAS.491.3461B} {491, 3461}

\bibitem[\protect\citeauthoryear{{Bullock} \& {Johnston}}{{Bullock} \&
  {Johnston}}{2005}]{Bullock05}
{Bullock} J.~S.,  {Johnston} K.~V.,  2005, \mn@doi [\apj] {10.1086/497422},
  \href {https://ui.adsabs.harvard.edu/abs/2005ApJ...635..931B} {635, 931}

\bibitem[\protect\citeauthoryear{{Bullock}, {Kravtsov}  \&
  {Weinberg}}{{Bullock} et~al.}{2001}]{Bullock01}
{Bullock} J.~S.,  {Kravtsov} A.~V.,   {Weinberg} D.~H.,  2001, \mn@doi [\apj]
  {10.1086/318681}, \href
  {https://ui.adsabs.harvard.edu/abs/2001ApJ...548...33B} {548, 33}

\bibitem[\protect\citeauthoryear{{Carney}, {Laird}, {Latham}  \&
  {Aguilar}}{{Carney} et~al.}{1996}]{Carney96}
{Carney} B.~W.,  {Laird} J.~B.,  {Latham} D.~W.,   {Aguilar} L.~A.,  1996,
  \mn@doi [\aj] {10.1086/118042}, \href
  {https://ui.adsabs.harvard.edu/abs/1996AJ....112..668C} {112, 668}

\bibitem[\protect\citeauthoryear{{Carter} et~al.,}{{Carter}
  et~al.}{2021}]{Carter20}
{Carter} C.,  et~al., 2021, \mn@doi [\apj] {10.3847/1538-4357/abcda4}, \href
  {https://ui.adsabs.harvard.edu/abs/2021ApJ...908..208C} {908, 208}

\bibitem[\protect\citeauthoryear{{Chiba} \& {Beers}}{{Chiba} \&
  {Beers}}{2000}]{Chiba00}
{Chiba} M.,  {Beers} T.~C.,  2000, \mn@doi [\aj] {10.1086/301409}, \href
  {https://ui.adsabs.harvard.edu/abs/2000AJ....119.2843C} {119, 2843}

\bibitem[\protect\citeauthoryear{{Conroy} et~al.,}{{Conroy}
  et~al.}{2019}]{H3SURVEY}
{Conroy} C.,  et~al., 2019, \mn@doi [\apj] {10.3847/1538-4357/ab38b8}, \href
  {https://ui.adsabs.harvard.edu/abs/2019ApJ...883..107C} {883, 107}

\bibitem[\protect\citeauthoryear{{Cordoni} et~al.,}{{Cordoni}
  et~al.}{2020}]{Cordoni20}
{Cordoni} G.,  et~al., 2020, \mn@doi [\mnras] {10.1093/mnras/staa3417}, \href
  {https://ui.adsabs.harvard.edu/abs/2020MNRAS.tmp.3219C} {}

\bibitem[\protect\citeauthoryear{{Da Costa} et~al.,}{{Da Costa}
  et~al.}{2019}]{SkyMapper}
{Da Costa} G.~S.,  et~al., 2019, \mn@doi [\mnras] {10.1093/mnras/stz2550},
  \href {https://ui.adsabs.harvard.edu/abs/2019MNRAS.489.5900D} {489, 5900}

\bibitem[\protect\citeauthoryear{{De Silva} et~al.,}{{De Silva}
  et~al.}{2015}]{GALAH}
{De Silva} G.~M.,  et~al., 2015, \mn@doi [\mnras] {10.1093/mnras/stv327}, \href
  {https://ui.adsabs.harvard.edu/abs/2015MNRAS.449.2604D} {449, 2604}

\bibitem[\protect\citeauthoryear{{Deason}, {Belokurov}  \& {Evans}}{{Deason}
  et~al.}{2011}]{Deason11}
{Deason} A.~J.,  {Belokurov} V.,   {Evans} N.~W.,  2011, \mn@doi [\mnras]
  {10.1111/j.1365-2966.2010.17785.x}, \href
  {https://ui.adsabs.harvard.edu/abs/2011MNRAS.411.1480D} {411, 1480}

\bibitem[\protect\citeauthoryear{{Deason}, {Mao}  \& {Wechsler}}{{Deason}
  et~al.}{2016}]{Deason16}
{Deason} A.~J.,  {Mao} Y.-Y.,   {Wechsler} R.~H.,  2016, \mn@doi [\apj]
  {10.3847/0004-637X/821/1/5}, \href
  {https://ui.adsabs.harvard.edu/abs/2016ApJ...821....5D} {821, 5}

\bibitem[\protect\citeauthoryear{{Di Matteo}, {Spite}, {Haywood}, {Bonifacio},
  {G{\'o}mez}, {Spite}  \& {Caffau}}{{Di Matteo} et~al.}{2020}]{DiMatteo20}
{Di Matteo} P.,  {Spite} M.,  {Haywood} M.,  {Bonifacio} P.,  {G{\'o}mez} A.,
  {Spite} F.,   {Caffau} E.,  2020, \mn@doi [\aap]
  {10.1051/0004-6361/201937016}, \href
  {https://ui.adsabs.harvard.edu/abs/2020A&A...636A.115D} {636, A115}

\bibitem[\protect\citeauthoryear{{Eggen}, {Lynden-Bell}  \& {Sandage}}{{Eggen}
  et~al.}{1962}]{Eggen62}
{Eggen} O.~J.,  {Lynden-Bell} D.,   {Sandage} A.~R.,  1962, \mn@doi [\apj]
  {10.1086/147433}, \href
  {https://ui.adsabs.harvard.edu/abs/1962ApJ...136..748E} {136, 748}

\bibitem[\protect\citeauthoryear{{El-Badry} et~al.,}{{El-Badry}
  et~al.}{2018a}]{ElBadry18_gas}
{El-Badry} K.,  et~al., 2018a, \mn@doi [\mnras] {10.1093/mnras/stx2482}, \href
  {https://ui.adsabs.harvard.edu/abs/2018MNRAS.473.1930E} {473, 1930}

\bibitem[\protect\citeauthoryear{{El-Badry} et~al.,}{{El-Badry}
  et~al.}{2018b}]{ElBadry18}
{El-Badry} K.,  et~al., 2018b, \mn@doi [Monthly Notices of the Royal
  Astronomical Society] {10.1093/mnras/sty1864}, \href
  {https://ui.adsabs.harvard.edu/abs/2018MNRAS.480..652E} {480, 652}

\bibitem[\protect\citeauthoryear{{Escala} et~al.,}{{Escala}
  et~al.}{2018}]{Escala18}
{Escala} I.,  et~al., 2018, \mn@doi [\mnras] {10.1093/mnras/stx2858}, \href
  {https://ui.adsabs.harvard.edu/abs/2018MNRAS.474.2194E} {474, 2194}

\bibitem[\protect\citeauthoryear{{Fall} \& {Efstathiou}}{{Fall} \&
  {Efstathiou}}{1980}]{Fall80}
{Fall} S.~M.,  {Efstathiou} G.,  1980, \mn@doi [\mnras]
  {10.1093/mnras/193.2.189}, \href
  {https://ui.adsabs.harvard.edu/abs/1980MNRAS.193..189F} {193, 189}

\bibitem[\protect\citeauthoryear{{Faucher-Gigu{\`e}re}, {Lidz}, {Zaldarriaga}
  \& {Hernquist}}{{Faucher-Gigu{\`e}re} et~al.}{2009}]{FaucherGiguere09}
{Faucher-Gigu{\`e}re} C.-A.,  {Lidz} A.,  {Zaldarriaga} M.,   {Hernquist} L.,
  2009, \mn@doi [\apj] {10.1088/0004-637X/703/2/1416}, \href
  {https://ui.adsabs.harvard.edu/abs/2009ApJ...703.1416F} {703, 1416}

\bibitem[\protect\citeauthoryear{{Freeman} \& {Bland-Hawthorn}}{{Freeman} \&
  {Bland-Hawthorn}}{2002}]{Freeman02}
{Freeman} K.,  {Bland-Hawthorn} J.,  2002, \mn@doi [\araa]
  {10.1146/annurev.astro.40.060401.093840}, \href
  {https://ui.adsabs.harvard.edu/abs/2002ARA&A..40..487F} {40, 487}

\bibitem[\protect\citeauthoryear{{Gaia Collaboration} et~al.,}{{Gaia
  Collaboration} et~al.}{2018}]{GaiaDR2}
{Gaia Collaboration} et~al., 2018, \mn@doi [\aap]
  {10.1051/0004-6361/201833051}, \href
  {https://ui.adsabs.harvard.edu/abs/2018A&A...616A...1G} {616, A1}

\bibitem[\protect\citeauthoryear{{Gallart}, {Bernard}, {Brook}, {Ruiz-Lara},
  {Cassisi}, {Hill}  \& {Monelli}}{{Gallart} et~al.}{2019}]{Gallart19}
{Gallart} C.,  {Bernard} E.~J.,  {Brook} C.~B.,  {Ruiz-Lara} T.,  {Cassisi} S.,
   {Hill} V.,   {Monelli} M.,  2019, \mn@doi [Nature Astronomy]
  {10.1038/s41550-019-0829-5}, \href
  {https://ui.adsabs.harvard.edu/abs/2019NatAs...3..932G} {3, 932}

\bibitem[\protect\citeauthoryear{{Garrison-Kimmel} et~al.,}{{Garrison-Kimmel}
  et~al.}{2017}]{GarrisonKimmel17}
{Garrison-Kimmel} S.,  et~al., 2017, \mn@doi [\mnras] {10.1093/mnras/stx1710},
  \href {https://ui.adsabs.harvard.edu/abs/2017MNRAS.471.1709G} {471, 1709}

\bibitem[\protect\citeauthoryear{{Garrison-Kimmel} et~al.,}{{Garrison-Kimmel}
  et~al.}{2018}]{GarrisonKimmel18}
{Garrison-Kimmel} S.,  et~al., 2018, \mn@doi [\mnras] {10.1093/mnras/sty2513},
  \href {https://ui.adsabs.harvard.edu/abs/2018MNRAS.481.4133G} {481, 4133}

\bibitem[\protect\citeauthoryear{{Garrison-Kimmel} et~al.,}{{Garrison-Kimmel}
  et~al.}{2019a}]{GarrisonKimmel19b}
{Garrison-Kimmel} S.,  et~al., 2019a, \mn@doi [\mnras] {10.1093/mnras/stz1317},
  \href {https://ui.adsabs.harvard.edu/abs/2019MNRAS.487.1380G} {487, 1380}

\bibitem[\protect\citeauthoryear{{Garrison-Kimmel} et~al.,}{{Garrison-Kimmel}
  et~al.}{2019b}]{GarrisonKimmel19a}
{Garrison-Kimmel} S.,  et~al., 2019b, \mn@doi [\mnras] {10.1093/mnras/stz2507},
  \href {https://ui.adsabs.harvard.edu/abs/2019MNRAS.489.4574G} {489, 4574}

\bibitem[\protect\citeauthoryear{{Gilmore} \& {Reid}}{{Gilmore} \&
  {Reid}}{1983}]{Gilmore83}
{Gilmore} G.,  {Reid} N.,  1983, \mn@doi [\mnras] {10.1093/mnras/202.4.1025},
  \href {https://ui.adsabs.harvard.edu/abs/1983MNRAS.202.1025G} {202, 1025}

\bibitem[\protect\citeauthoryear{{Grand}, {Springel}, {G{\'o}mez}, {Marinacci},
  {Pakmor}, {Campbell}  \& {Jenkins}}{{Grand} et~al.}{2016}]{Grand16}
{Grand} R. J.~J.,  {Springel} V.,  {G{\'o}mez} F.~A.,  {Marinacci} F.,
  {Pakmor} R.,  {Campbell} D. J.~R.,   {Jenkins} A.,  2016, \mn@doi [\mnras]
  {10.1093/mnras/stw601}, \href
  {https://ui.adsabs.harvard.edu/abs/2016MNRAS.459..199G} {459, 199}

\bibitem[\protect\citeauthoryear{{Griffen}, {Dooley}, {Ji}, {O'Shea},
  {G{\'o}mez}  \& {Frebel}}{{Griffen} et~al.}{2018}]{Griffen18}
{Griffen} B.~F.,  {Dooley} G.~A.,  {Ji} A.~P.,  {O'Shea} B.~W.,  {G{\'o}mez}
  F.~A.,   {Frebel} A.,  2018, \mn@doi [\mnras] {10.1093/mnras/stx2749}, \href
  {https://ui.adsabs.harvard.edu/abs/2018MNRAS.474..443G} {474, 443}

\bibitem[\protect\citeauthoryear{{Guedes}, {Callegari}, {Madau}  \&
  {Mayer}}{{Guedes} et~al.}{2011}]{ERIS}
{Guedes} J.,  {Callegari} S.,  {Madau} P.,   {Mayer} L.,  2011, \mn@doi [\apj]
  {10.1088/0004-637X/742/2/76}, \href
  {https://ui.adsabs.harvard.edu/abs/2011ApJ...742...76G} {742, 76}

\bibitem[\protect\citeauthoryear{{Hahn} \& {Abel}}{{Hahn} \&
  {Abel}}{2011}]{Hahn11}
{Hahn} O.,  {Abel} T.,  2011, \mn@doi [\mnras]
  {10.1111/j.1365-2966.2011.18820.x}, \href
  {https://ui.adsabs.harvard.edu/abs/2011MNRAS.415.2101H} {415, 2101}

\bibitem[\protect\citeauthoryear{Harris et~al.,}{Harris et~al.}{2020}]{Numpy}
Harris C.~R.,  et~al., 2020, \mn@doi [Nature] {10.1038/s41586-020-2649-2}, 585,
  357

\bibitem[\protect\citeauthoryear{{Haywood}, {Di Matteo}, {Lehnert}, {Snaith},
  {Khoperskov}  \& {G{\'o}mez}}{{Haywood} et~al.}{2018}]{Haywood18}
{Haywood} M.,  {Di Matteo} P.,  {Lehnert} M.~D.,  {Snaith} O.,  {Khoperskov}
  S.,   {G{\'o}mez} A.,  2018, \mn@doi [\apj] {10.3847/1538-4357/aad235}, \href
  {https://ui.adsabs.harvard.edu/abs/2018ApJ...863..113H} {863, 113}

\bibitem[\protect\citeauthoryear{{Helmi}, {Babusiaux}, {Koppelman}, {Massari},
  {Veljanoski}  \& {Brown}}{{Helmi} et~al.}{2018}]{Helmi18}
{Helmi} A.,  {Babusiaux} C.,  {Koppelman} H.~H.,  {Massari} D.,  {Veljanoski}
  J.,   {Brown} A. G.~A.,  2018, \mn@doi [Nature] {10.1038/s41586-018-0625-x},
  \href {https://ui.adsabs.harvard.edu/abs/2018Natur.563...85H} {563, 85}

\bibitem[\protect\citeauthoryear{{Hopkins}}{{Hopkins}}{2015}]{Hopkins15}
{Hopkins} P.~F.,  2015, \mn@doi [\mnras] {10.1093/mnras/stv195}, \href
  {https://ui.adsabs.harvard.edu/abs/2015MNRAS.450...53H} {450, 53}

\bibitem[\protect\citeauthoryear{{Hopkins}}{{Hopkins}}{2016}]{Hopkins16}
{Hopkins} P.~F.,  2016, \mn@doi [\mnras] {10.1093/mnras/stv2226}, \href
  {https://ui.adsabs.harvard.edu/abs/2016MNRAS.455...89H} {455, 89}

\bibitem[\protect\citeauthoryear{{Hopkins}, {Cox}, {Younger}  \&
  {Hernquist}}{{Hopkins} et~al.}{2009}]{Hopkins09}
{Hopkins} P.~F.,  {Cox} T.~J.,  {Younger} J.~D.,   {Hernquist} L.,  2009,
  \mn@doi [\apj] {10.1088/0004-637X/691/2/1168}, \href
  {https://ui.adsabs.harvard.edu/abs/2009ApJ...691.1168H} {691, 1168}

\bibitem[\protect\citeauthoryear{{Hopkins} et~al.,}{{Hopkins}
  et~al.}{2018}]{Hopkins18}
{Hopkins} P.~F.,  et~al., 2018, \mn@doi [\mnras] {10.1093/mnras/sty1690}, \href
  {http://adsabs.harvard.edu/abs/2018MNRAS.480..800H} {480, 800}

\bibitem[\protect\citeauthoryear{{Hunter}}{{Hunter}}{2007}]{Matplotlib}
{Hunter} J.~D.,  2007, Computing in Science Engineering, 9, 90

\bibitem[\protect\citeauthoryear{{Ibata}, {Gilmore}  \& {Irwin}}{{Ibata}
  et~al.}{1994}]{Ibata94}
{Ibata} R.~A.,  {Gilmore} G.,   {Irwin} M.~J.,  1994, \mn@doi [\nat]
  {10.1038/370194a0}, \href
  {https://ui.adsabs.harvard.edu/abs/1994Natur.370..194I} {370, 194}

\bibitem[\protect\citeauthoryear{{Kroupa}}{{Kroupa}}{2001}]{Kroupa01}
{Kroupa} P.,  2001, \mn@doi [\mnras] {10.1046/j.1365-8711.2001.04022.x}, \href
  {https://ui.adsabs.harvard.edu/abs/2001MNRAS.322..231K} {322, 231}

\bibitem[\protect\citeauthoryear{{Krumholz} \& {Gnedin}}{{Krumholz} \&
  {Gnedin}}{2011}]{Krumholz11}
{Krumholz} M.~R.,  {Gnedin} N.~Y.,  2011, \mn@doi [\apj]
  {10.1088/0004-637X/729/1/36}, \href
  {https://ui.adsabs.harvard.edu/abs/2011ApJ...729...36K} {729, 36}

\bibitem[\protect\citeauthoryear{{Leitherer} et~al.,}{{Leitherer}
  et~al.}{1999}]{Leitherer99}
{Leitherer} C.,  et~al., 1999, \mn@doi [\apjs] {10.1086/313233}, \href
  {https://ui.adsabs.harvard.edu/abs/1999ApJS..123....3L} {123, 3}

\bibitem[\protect\citeauthoryear{{Li}, {Zhao}, {Christlieb}, {Wang}, {Wang},
  {Zhang}, {Hou}  \& {Yuan}}{{Li} et~al.}{2015}]{LAMOST}
{Li} H.-N.,  {Zhao} G.,  {Christlieb} N.,  {Wang} L.,  {Wang} W.,  {Zhang} Y.,
  {Hou} Y.,   {Yuan} H.,  2015, \mn@doi [\apj] {10.1088/0004-637X/798/2/110},
  \href {https://ui.adsabs.harvard.edu/abs/2015ApJ...798..110L} {798, 110}

\bibitem[\protect\citeauthoryear{{Li}, {Tan}  \& {Zhao}}{{Li}
  et~al.}{2018}]{Li18}
{Li} H.,  {Tan} K.,   {Zhao} G.,  2018, \mn@doi [\apjs]
  {10.3847/1538-4365/aada4a}, \href
  {https://ui.adsabs.harvard.edu/abs/2018ApJS..238...16L} {238, 16}

\bibitem[\protect\citeauthoryear{{Loebman}, {Ro{\v{s}}kar}, {Debattista},
  {Ivezi{\'c}}, {Quinn}  \& {Wadsley}}{{Loebman} et~al.}{2011}]{Loebman11}
{Loebman} S.~R.,  {Ro{\v{s}}kar} R.,  {Debattista} V.~P.,  {Ivezi{\'c}}
  {\v{Z}}.,  {Quinn} T.~R.,   {Wadsley} J.,  2011, \mn@doi [\apj]
  {10.1088/0004-637X/737/1/8}, \href
  {https://ui.adsabs.harvard.edu/abs/2011ApJ...737....8L} {737, 8}

\bibitem[\protect\citeauthoryear{{Ma}, {Hopkins}, {Wetzel}, {Kirby},
  {Angl{\'e}s-Alc{\'a}zar}, {Faucher-Gigu{\`e}re}, {Kere{\v{s}}}  \&
  {Quataert}}{{Ma} et~al.}{2017}]{Ma17}
{Ma} X.,  {Hopkins} P.~F.,  {Wetzel} A.~R.,  {Kirby} E.~N.,
  {Angl{\'e}s-Alc{\'a}zar} D.,  {Faucher-Gigu{\`e}re} C.-A.,  {Kere{\v{s}}} D.,
    {Quataert} E.,  2017, \mn@doi [\mnras] {10.1093/mnras/stx273}, \href
  {https://ui.adsabs.harvard.edu/abs/2017MNRAS.467.2430M} {467, 2430}

\bibitem[\protect\citeauthoryear{{Majewski}, {Skrutskie}, {Weinberg}  \&
  {Ostheimer}}{{Majewski} et~al.}{2003}]{Majewski03}
{Majewski} S.~R.,  {Skrutskie} M.~F.,  {Weinberg} M.~D.,   {Ostheimer} J.~C.,
  2003, \mn@doi [\apj] {10.1086/379504}, \href
  {https://ui.adsabs.harvard.edu/abs/2003ApJ...599.1082M} {599, 1082}

\bibitem[\protect\citeauthoryear{{Majewski} et~al.,}{{Majewski}
  et~al.}{2017}]{APOGEE}
{Majewski} S.~R.,  et~al., 2017, \mn@doi [\aj] {10.3847/1538-3881/aa784d},
  \href {https://ui.adsabs.harvard.edu/abs/2017AJ....154...94M} {154, 94}

\bibitem[\protect\citeauthoryear{{Matsuno}, {Aoki}  \& {Suda}}{{Matsuno}
  et~al.}{2019}]{Matsuno19}
{Matsuno} T.,  {Aoki} W.,   {Suda} T.,  2019, \mn@doi [\apjl]
  {10.3847/2041-8213/ab0ec0}, \href
  {https://ui.adsabs.harvard.edu/abs/2019ApJ...874L..35M} {874, L35}

\bibitem[\protect\citeauthoryear{{Mo}, {Mao}  \& {White}}{{Mo}
  et~al.}{1998}]{Mo98}
{Mo} H.~J.,  {Mao} S.,   {White} S. D.~M.,  1998, \mn@doi [\mnras]
  {10.1046/j.1365-8711.1998.01227.x}, \href
  {https://ui.adsabs.harvard.edu/abs/1998MNRAS.295..319M} {295, 319}

\bibitem[\protect\citeauthoryear{{Myeong}, {Vasiliev}, {Iorio}, {Evans}  \&
  {Belokurov}}{{Myeong} et~al.}{2019}]{Myeong19}
{Myeong} G.~C.,  {Vasiliev} E.,  {Iorio} G.,  {Evans} N.~W.,   {Belokurov} V.,
  2019, \mn@doi [\mnras] {10.1093/mnras/stz1770}, \href
  {https://ui.adsabs.harvard.edu/abs/2019MNRAS.488.1235M} {488, 1235}

\bibitem[\protect\citeauthoryear{{Naidu}, {Conroy}, {Bonaca}, {Johnson},
  {Ting}, {Caldwell}, {Zaritsky}  \& {Cargile}}{{Naidu} et~al.}{2020}]{Naidu20}
{Naidu} R.~P.,  {Conroy} C.,  {Bonaca} A.,  {Johnson} B.~D.,  {Ting} Y.-S.,
  {Caldwell} N.,  {Zaritsky} D.,   {Cargile} P.~A.,  2020, \mn@doi [\apj]
  {10.3847/1538-4357/abaef4}, \href
  {https://ui.adsabs.harvard.edu/abs/2020ApJ...901...48N} {901, 48}

\bibitem[\protect\citeauthoryear{{Naidu} et~al.,}{{Naidu}
  et~al.}{2021}]{Naidu21}
{Naidu} R.~P.,  et~al., 2021, arXiv e-prints, \href
  {https://ui.adsabs.harvard.edu/abs/2021arXiv210303251N} {p. arXiv:2103.03251}

\bibitem[\protect\citeauthoryear{{Newberg} et~al.,}{{Newberg}
  et~al.}{2003}]{Newberg03}
{Newberg} H.~J.,  et~al., 2003, \mn@doi [\apjl] {10.1086/379316}, \href
  {https://ui.adsabs.harvard.edu/abs/2003ApJ...596L.191N} {596, L191}

\bibitem[\protect\citeauthoryear{{Nikakhtar} et~al.,}{{Nikakhtar}
  et~al.}{2021}]{Nikakhtar21}
{Nikakhtar} F.,  et~al., 2021, arXiv e-prints, \href
  {https://ui.adsabs.harvard.edu/abs/2021arXiv210408394N} {p. arXiv:2104.08394}

\bibitem[\protect\citeauthoryear{{Nissen} \& {Schuster}}{{Nissen} \&
  {Schuster}}{2010}]{Nissen10}
{Nissen} P.~E.,  {Schuster} W.~J.,  2010, \mn@doi [\aap]
  {10.1051/0004-6361/200913877}, \href
  {https://ui.adsabs.harvard.edu/abs/2010A&A...511L..10N} {511, L10}

\bibitem[\protect\citeauthoryear{{Norris} \& {Ryan}}{{Norris} \&
  {Ryan}}{1989}]{Norris89}
{Norris} J.~E.,  {Ryan} S.~G.,  1989, \mn@doi [\apjl] {10.1086/185351}, \href
  {https://ui.adsabs.harvard.edu/abs/1989ApJ...336L..17N} {336, L17}

\bibitem[\protect\citeauthoryear{{Ostriker} \& {Tremaine}}{{Ostriker} \&
  {Tremaine}}{1975}]{Ostriker75}
{Ostriker} J.~P.,  {Tremaine} S.~D.,  1975, \mn@doi [\apjl] {10.1086/181992},
  \href {https://ui.adsabs.harvard.edu/abs/1975ApJ...202L.113O} {202, L113}

\bibitem[\protect\citeauthoryear{{Peschken}, {Athanassoula}  \&
  {Rodionov}}{{Peschken} et~al.}{2017}]{Peschken17}
{Peschken} N.,  {Athanassoula} E.,   {Rodionov} S.~A.,  2017, \mn@doi [\mnras]
  {10.1093/mnras/stx481}, \href
  {https://ui.adsabs.harvard.edu/abs/2017MNRAS.468..994P} {468, 994}

\bibitem[\protect\citeauthoryear{{Peschken}, {{\L}okas}  \&
  {Athanassoula}}{{Peschken} et~al.}{2020}]{Peschken20}
{Peschken} N.,  {{\L}okas} E.~L.,   {Athanassoula} E.,  2020, \mn@doi [\mnras]
  {10.1093/mnras/staa299}, \href
  {https://ui.adsabs.harvard.edu/abs/2020MNRAS.493.1375P} {493, 1375}

\bibitem[\protect\citeauthoryear{{Planck Collaboration} et~al.,}{{Planck
  Collaboration} et~al.}{2020}]{Planck18}
{Planck Collaboration} et~al., 2020, \mn@doi [\aap]
  {10.1051/0004-6361/201833910}, \href
  {https://ui.adsabs.harvard.edu/abs/2020A&A...641A...6P} {641, A6}

\bibitem[\protect\citeauthoryear{{Rees} \& {Ostriker}}{{Rees} \&
  {Ostriker}}{1977}]{Rees77}
{Rees} M.~J.,  {Ostriker} J.~P.,  1977, \mn@doi [\mnras]
  {10.1093/mnras/179.4.541}, \href
  {https://ui.adsabs.harvard.edu/abs/1977MNRAS.179..541R} {179, 541}

\bibitem[\protect\citeauthoryear{{Robertson}, {Bullock}, {Cox}, {Di Matteo},
  {Hernquist}, {Springel}  \& {Yoshida}}{{Robertson}
  et~al.}{2006}]{Robertson06}
{Robertson} B.,  {Bullock} J.~S.,  {Cox} T.~J.,  {Di Matteo} T.,  {Hernquist}
  L.,  {Springel} V.,   {Yoshida} N.,  2006, \mn@doi [\apj] {10.1086/504412},
  \href {https://ui.adsabs.harvard.edu/abs/2006ApJ...645..986R} {645, 986}

\bibitem[\protect\citeauthoryear{{Rodionov}, {Athanassoula}  \&
  {Peschken}}{{Rodionov} et~al.}{2017}]{Rodionov17}
{Rodionov} S.~A.,  {Athanassoula} E.,   {Peschken} N.,  2017, \mn@doi [\aap]
  {10.1051/0004-6361/201628319}, \href
  {https://ui.adsabs.harvard.edu/abs/2017A&A...600A..25R} {600, A25}

\bibitem[\protect\citeauthoryear{{Samuel} et~al.,}{{Samuel}
  et~al.}{2020}]{Samuel20}
{Samuel} J.,  et~al., 2020, \mn@doi [\mnras] {10.1093/mnras/stz3054}, \href
  {https://ui.adsabs.harvard.edu/abs/2020MNRAS.491.1471S} {491, 1471}

\bibitem[\protect\citeauthoryear{{Samuel}, {Wetzel}, {Chapman}, {Tollerud},
  {Hopkins}, {Boylan-Kolchin}, {Bailin}  \& {Faucher-Gigu{\`e}re}}{{Samuel}
  et~al.}{2021}]{Samuel20_Planes}
{Samuel} J.,  {Wetzel} A.,  {Chapman} S.,  {Tollerud} E.,  {Hopkins} P.~F.,
  {Boylan-Kolchin} M.,  {Bailin} J.,   {Faucher-Gigu{\`e}re} C.-A.,  2021,
  \mn@doi [\mnras] {10.1093/mnras/stab955}, \href
  {https://ui.adsabs.harvard.edu/abs/2021MNRAS.504.1379S} {504, 1379}

\bibitem[\protect\citeauthoryear{{Sanderson} et~al.,}{{Sanderson}
  et~al.}{2018}]{Sanderson18_halo}
{Sanderson} R.~E.,  et~al., 2018, \mn@doi [\apj] {10.3847/1538-4357/aaeb33},
  \href {https://ui.adsabs.harvard.edu/abs/2018ApJ...869...12S} {869, 12}

\bibitem[\protect\citeauthoryear{{Sanderson} et~al.,}{{Sanderson}
  et~al.}{2020}]{Sanderson18_gaia}
{Sanderson} R.~E.,  et~al., 2020, \mn@doi [\apjs] {10.3847/1538-4365/ab5b9d},
  \href {https://ui.adsabs.harvard.edu/abs/2020ApJS..246....6S} {246, 6}

\bibitem[\protect\citeauthoryear{{Santistevan}, {Wetzel}, {El-Badry},
  {Bland-Hawthorn}, {Boylan-Kolchin}, {Bailin}, {Faucher-Gigu{\`e}re}  \&
  {Benincasa}}{{Santistevan} et~al.}{2020}]{Santistevan20}
{Santistevan} I.~B.,  {Wetzel} A.,  {El-Badry} K.,  {Bland-Hawthorn} J.,
  {Boylan-Kolchin} M.,  {Bailin} J.,  {Faucher-Gigu{\`e}re} C.-A.,
  {Benincasa} S.,  2020, \mn@doi [\mnras] {10.1093/mnras/staa1923}, \href
  {https://ui.adsabs.harvard.edu/abs/2020MNRAS.497..747S} {497, 747}

\bibitem[\protect\citeauthoryear{{Scannapieco}, {Kawata}, {Brook}, {Schneider},
  {Ferrara}  \& {Gibson}}{{Scannapieco} et~al.}{2006}]{Scannapieco06}
{Scannapieco} E.,  {Kawata} D.,  {Brook} C.~B.,  {Schneider} R.,  {Ferrara} A.,
    {Gibson} B.~K.,  2006, \mn@doi [\apj] {10.1086/508487}, \href
  {https://ui.adsabs.harvard.edu/abs/2006ApJ...653..285S} {653, 285}

\bibitem[\protect\citeauthoryear{{Searle} \& {Zinn}}{{Searle} \&
  {Zinn}}{1978}]{Searle78}
{Searle} L.,  {Zinn} R.,  1978, \mn@doi [\apj] {10.1086/156499}, \href
  {https://ui.adsabs.harvard.edu/abs/1978ApJ...225..357S} {225, 357}

\bibitem[\protect\citeauthoryear{{Sestito} et~al.,}{{Sestito}
  et~al.}{2019}]{Sestito19}
{Sestito} F.,  et~al., 2019, \mn@doi [\mnras] {10.1093/mnras/stz043}, \href
  {https://ui.adsabs.harvard.edu/abs/2019MNRAS.484.2166S} {484, 2166}

\bibitem[\protect\citeauthoryear{{Sestito} et~al.,}{{Sestito}
  et~al.}{2020}]{Sestito20}
{Sestito} F.,  et~al., 2020, \mn@doi [\mnras] {10.1093/mnrasl/slaa022}, \href
  {https://ui.adsabs.harvard.edu/abs/2020MNRAS.497L...7S} {497, L7}

\bibitem[\protect\citeauthoryear{{Sestito} et~al.,}{{Sestito}
  et~al.}{2021}]{Sestito_new}
{Sestito} F.,  et~al., 2021, \mn@doi [\mnras] {10.1093/mnras/staa3479}, \href
  {https://ui.adsabs.harvard.edu/abs/2021MNRAS.500.3750S} {500, 3750}

\bibitem[\protect\citeauthoryear{{Sparre} \& {Springel}}{{Sparre} \&
  {Springel}}{2017}]{Sparre17}
{Sparre} M.,  {Springel} V.,  2017, \mn@doi [\mnras] {10.1093/mnras/stx1516},
  \href {https://ui.adsabs.harvard.edu/abs/2017MNRAS.470.3946S} {470, 3946}

\bibitem[\protect\citeauthoryear{{Springel} \& {Hernquist}}{{Springel} \&
  {Hernquist}}{2005}]{Springel05}
{Springel} V.,  {Hernquist} L.,  2005, \mn@doi [\apjl] {10.1086/429486}, \href
  {https://ui.adsabs.harvard.edu/abs/2005ApJ...622L...9S} {622, L9}

\bibitem[\protect\citeauthoryear{{Starkenburg}, {Oman}, {Navarro}, {Crain},
  {Fattahi}, {Frenk}, {Sawala}  \& {Schaye}}{{Starkenburg}
  et~al.}{2017a}]{Starkenburg17}
{Starkenburg} E.,  {Oman} K.~A.,  {Navarro} J.~F.,  {Crain} R.~A.,  {Fattahi}
  A.,  {Frenk} C.~S.,  {Sawala} T.,   {Schaye} J.,  2017a, \mn@doi [\mnras]
  {10.1093/mnras/stw2873}, \href
  {https://ui.adsabs.harvard.edu/abs/2017MNRAS.465.2212S} {465, 2212}

\bibitem[\protect\citeauthoryear{{Starkenburg} et~al.,}{{Starkenburg}
  et~al.}{2017b}]{PRISTINE}
{Starkenburg} E.,  et~al., 2017b, \mn@doi [\mnras] {10.1093/mnras/stx1068},
  \href {https://ui.adsabs.harvard.edu/abs/2017MNRAS.471.2587S} {471, 2587}

\bibitem[\protect\citeauthoryear{{Steinmetz} et~al.,}{{Steinmetz}
  et~al.}{2006}]{RAVE}
{Steinmetz} M.,  et~al., 2006, \mn@doi [\aj] {10.1086/506564}, \href
  {https://ui.adsabs.harvard.edu/abs/2006AJ....132.1645S} {132, 1645}

\bibitem[\protect\citeauthoryear{{Su}, {Hopkins}, {Hayward},
  {Faucher-Gigu{\`e}re}, {Kere{\v{s}}}, {Ma}  \& {Robles}}{{Su}
  et~al.}{2017}]{Su17}
{Su} K.-Y.,  {Hopkins} P.~F.,  {Hayward} C.~C.,  {Faucher-Gigu{\`e}re} C.-A.,
  {Kere{\v{s}}} D.,  {Ma} X.,   {Robles} V.~H.,  2017, \mn@doi [\mnras]
  {10.1093/mnras/stx1463}, \href
  {https://ui.adsabs.harvard.edu/abs/2017MNRAS.471..144S} {471, 144}

\bibitem[\protect\citeauthoryear{{Tapia}, {Eliche-Moral}, {Aceves},
  {Rodr{\'\i}guez-P{\'e}rez}, {Borlaff}  \& {Querejeta}}{{Tapia}
  et~al.}{2017}]{Tapia17}
{Tapia} T.,  {Eliche-Moral} M.~C.,  {Aceves} H.,  {Rodr{\'\i}guez-P{\'e}rez}
  C.,  {Borlaff} A.,   {Querejeta} M.,  2017, \mn@doi [\aap]
  {10.1051/0004-6361/201628821}, \href
  {https://ui.adsabs.harvard.edu/abs/2017A&A...604A.105T} {604, A105}

\bibitem[\protect\citeauthoryear{{Toomre} \& {Toomre}}{{Toomre} \&
  {Toomre}}{1972}]{Toomre72}
{Toomre} A.,  {Toomre} J.,  1972, \mn@doi [\apj] {10.1086/151823}, \href
  {https://ui.adsabs.harvard.edu/abs/1972ApJ...178..623T} {178, 623}

\bibitem[\protect\citeauthoryear{{Venn} et~al.,}{{Venn} et~al.}{2020}]{Venn20}
{Venn} K.~A.,  et~al., 2020, \mn@doi [\mnras] {10.1093/mnras/stz3546}, \href
  {https://ui.adsabs.harvard.edu/abs/2020MNRAS.492.3241V} {492, 3241}

\bibitem[\protect\citeauthoryear{{Villalobos} \& {Helmi}}{{Villalobos} \&
  {Helmi}}{2008}]{Villalobos08}
{Villalobos} {\'A}.,  {Helmi} A.,  2008, \mn@doi [\mnras]
  {10.1111/j.1365-2966.2008.13979.x}, \href
  {https://ui.adsabs.harvard.edu/abs/2008MNRAS.391.1806V} {391, 1806}

\bibitem[\protect\citeauthoryear{{Virtanen} et~al.,}{{Virtanen}
  et~al.}{2020}]{SciPy}
{Virtanen} P.,  et~al., 2020, \mn@doi [Nature Methods]
  {https://doi.org/10.1038/s41592-019-0686-2}, \href {https://rdcu.be/b08Wh}
  {17, 261}

\bibitem[\protect\citeauthoryear{{Vogelsberger} et~al.,}{{Vogelsberger}
  et~al.}{2014}]{Illustris}
{Vogelsberger} M.,  et~al., 2014, \mn@doi [\mnras] {10.1093/mnras/stu1536},
  \href {https://ui.adsabs.harvard.edu/abs/2014MNRAS.444.1518V} {444, 1518}

\bibitem[\protect\citeauthoryear{{Walker}, {Mihos}  \& {Hernquist}}{{Walker}
  et~al.}{1996}]{Walker96}
{Walker} I.~R.,  {Mihos} J.~C.,   {Hernquist} L.,  1996, \mn@doi [\apj]
  {10.1086/176956}, \href
  {https://ui.adsabs.harvard.edu/abs/1996ApJ...460..121W} {460, 121}

\bibitem[\protect\citeauthoryear{{Wetzel} \& {Garrison-Kimmel}}{{Wetzel} \&
  {Garrison-Kimmel}}{2020a}]{HaloAnalysis}
{Wetzel} A.,  {Garrison-Kimmel} S.,  2020a, {HaloAnalysis: Read and analyze
  halo catalogs and merger trees} (\mn@eprint {ascl} {2002.014})

\bibitem[\protect\citeauthoryear{{Wetzel} \& {Garrison-Kimmel}}{{Wetzel} \&
  {Garrison-Kimmel}}{2020b}]{GizmoAnalysis}
{Wetzel} A.,  {Garrison-Kimmel} S.,  2020b, {GizmoAnalysis: Read and analyze
  Gizmo simulations} (\mn@eprint {ascl} {2002.015})

\bibitem[\protect\citeauthoryear{{Wetzel}, {Hopkins}, {Kim},
  {Faucher-Gigu{\`e}re}, {Kere{\v s}}  \& {Quataert}}{{Wetzel}
  et~al.}{2016}]{Wetzel16}
{Wetzel} A.~R.,  {Hopkins} P.~F.,  {Kim} J.-h.,  {Faucher-Gigu{\`e}re} C.-A.,
  {Kere{\v s}} D.,   {Quataert} E.,  2016, \mn@doi [\apjl]
  {10.3847/2041-8205/827/2/L23}, \href
  {http://adsabs.harvard.edu/abs/2016ApJ...827L..23W} {827, L23}

\bibitem[\protect\citeauthoryear{{White} \& {Frenk}}{{White} \&
  {Frenk}}{1991}]{White91}
{White} S. D.~M.,  {Frenk} C.~S.,  1991, \mn@doi [\apj] {10.1086/170483}, \href
  {https://ui.adsabs.harvard.edu/abs/1991ApJ...379...52W} {379, 52}

\bibitem[\protect\citeauthoryear{{White} \& {Rees}}{{White} \&
  {Rees}}{1978}]{White78}
{White} S.~D.~M.,  {Rees} M.~J.,  1978, \mn@doi [\mnras]
  {10.1093/mnras/183.3.341}, \href
  {https://ui.adsabs.harvard.edu/abs/1978MNRAS.183..341W} {183, 341}

\bibitem[\protect\citeauthoryear{{Yu} et~al.,}{{Yu} et~al.}{2021}]{Yu21}
{Yu} S.,  et~al., 2021, arXiv e-prints, \href
  {https://ui.adsabs.harvard.edu/abs/2021arXiv210303888Y} {p. arXiv:2103.03888}

\bibitem[\protect\citeauthoryear{{Zinn}}{{Zinn}}{1993}]{Zinn93}
{Zinn} R.,  1993, in {Smith} G.~H.,  {Brodie} J.~P.,  eds,  Astronomical
  Society of the Pacific Conference Series Vol. 48, The Globular Cluster-Galaxy
  Connection. p.~38

\makeatother
\end{thebibliography}


\bsp	
\label{lastpage}
\end{document}